\documentclass[journal=jacsat,manuscript=article]{achemso}
\SectionNumbersOn
\usepackage{color,soul}
\usepackage{xspace}
\usepackage{xcolor}
\usepackage{tablefootnote}

\usepackage[version=3]{mhchem} % Formula subscripts using \ce{}
\usepackage{mathptmx}
\usepackage{booktabs}

\usepackage{siunitx}
\usepackage{physics}

\author{Thomas J. Longo}%
\affiliation{Institute for Physical Science and Technology, University of Maryland, College Park, MD 20742, USA}

\author{Sergey V. Buldyrev}%
\affiliation{ Department of Physics, Yeshiva University, New York, NY 10033, USA}
\alsoaffiliation{Department of Physics, Boston University, MA 02215, USA}

\author{Mikhail A. Anisimov}%
\affiliation{Institute for Physical Science and Technology, University of Maryland, College Park, MD 20742, USA}
\alsoaffiliation{Department of Chemical and Biomolecular Engineering, University of Maryland, College Park, MD 20742, USA}

\author{Fr\'ed\'eric Caupin}
\email{frederic.caupin@univ-lyon1.fr}
\affiliation{Institut Lumi\`ere Mati\`ere, Universit\'e de Lyon, Universit\'e Claude Bernard Lyon 1, CNRS, Institut universitaire de France, F-69622 Villeurbanne, France}

\title{Interfacial Properties of Fluids Exhibiting Liquid Polyamorphism and Water-Like Anomalies}

\abbreviations{IR,NMR,UV}
\keywords{American Chemical Society, \LaTeX}

\begin{document}


%\begin{tocentry}

%Some journals require a graphical entry for the Table of Contents. This should be laid out ``print ready'' so that the sizing of the text is correct.

%Inside the \texttt{tocentry} environment, the font used is Helvetica
%8\,pt, as required by \emph{Journal of the American Chemical

%The surrounding frame is 9\,cm by 3.5\,cm, which is the maximum
%permitted for  \emph{Journal of the American Chemical Society}
%graphical table of content entries. The box will not resize if the
%content is too big: instead it will overflow the edge of the box.

%This box and the associated title will always be printed on a
%separate page at the end of the document.

%\end{tocentry}

\begin{abstract}
  It has been hypothesized that liquid polyamorphism, the existence of multiple amorphous states in a single component substance, may be caused by molecular or supramolecular interconversion. A simple microscopic model [Caupin and Anisimov, \textit{Phys. Rev. Lett.}, \textbf{127}, 185701, (2021)] introduces interconversion in a compressible binary lattice to generate various thermodynamic scenarios for fluids that exhibit liquid polyamorphism and/or water-like anomalies. Using this model,  we demonstrate the dramatic effects of interconversion on the interfacial properties. In particular, we find that the liquid-vapor surface tension exhibits either an inflection point or two extrema in its temperature dependence. Correspondingly, we observe anomalous behavior of the interfacial thickness and a significant shift in the location of the concentration profile with respect to the location of the density profile.
\end{abstract}

\section{Introduction}
Typically, pure substances may be found with only one gaseous or liquid state, while their solid state may exist in various polymorphic crystalline states. The existence of two distinct liquid forms in a single component substance is more unusual since liquids lack the long-range order common to crystals. Yet, the existence of multiple amorphous liquid states in a single component substance, a phenomenon known as ``liquid polyamorphism,''\cite{Debenedetti_Water_1998,Stanley_Liquid_2013,Anisimov_Polyamorphism_2018,Tanaka_Liquid_2020} has been observed or predicted in a wide variety of substances, such as superfluid helium \cite{Vollhardt_He_1990,Schmitt_He_2015}, high-pressure hydrogen \cite{Ohta_H_2015,Zaghoo_H_2016,McWilliams_H_2016,Norman_Review_2021,Fried_Hydrogen_2022}, sulfur \cite{Henry_Sulfur_2020}, phosphorous \cite{Katayama_Phos_2000,Katayama_Phos_2004}, carbon \cite{Glosli_Liquid_1999}, silicon \cite{Sastry_Silicon_2003,Beye_Silicon_2010,Vasisht_Solicon_2011,Sciorino_Silicon_2011}, silica \cite{Saika_Silica_2000,Lascaris_Silica_2014}, selenium and tellurium \cite{Tsuchiya_SeTe_1982,Brazhkin_SeTe_1999}, and cerium \cite{Cadien_Ce_2013}. Liquid polyamorphism is also highly plausible in deeply supercooled liquid water\cite{Angell_TwoState_1971,Angell_Amorphous_2004,Stanley_Liquid_2013,Anisimov_Polyamorphism_2018,Tanaka_Liquid_2020,Poole_Water_1992,Debenedetti_Water_1998,Holten_Water_2012,Holten_Water_2014,Gallo_Water_2016,Biddle_Water_2017,Caupin_Thermodynamics_2019,Duska_Water_2020}. 

In addition to the hypothesized existence of a liquid-liquid phase transition in supercooled water, other anomalies in water's thermodynamic properties have been reported, namely a maximum in the temperature dependence of its isothermal compressibility~\cite{Holten_Compressibility_2017,Kim_Maxima_2017} and a maximum in the isobaric heat capacity~\cite{Pathak_Enhancement_2021}. The possibility of anomalous behavior of the liquid-vapor surface tension, $\sigma_\mathrm{LV}$, of supercooled water has been a topic of long-standing interest. In 1951, an inflection point in the temperature dependence of $\sigma_\mathrm{LV}$ was reported to occur near \SI{0}{\celsius}~\cite{Hacker_experimental_1951}, but later studies, showing larger uncertainties, cast doubts on the early measurements~\cite{Floriano_Surface_1990,Trinh_Measurement_1995}. Only recently, the highly accurate studies by Hruby and coworkers became available \cite{Hruby_Surface_2014,Vins_Surface_2015,Vins_Surface_2017,Vins_Surface_2020}. Initially, in Refs.~\cite{Hruby_Surface_2014,Vins_Surface_2015,Vins_Surface_2017} it was concluded that no anomaly occurred in $\sigma_\mathrm{LV}(T)$ down to \SI{-26}{\celsius}; however, the results of the latest experiment~\cite{Vins_Surface_2020}, reaching \SI{-31.4}{\celsius}, suggest that an inflection point might be possible. Theoretical studies support the existence of anomalies in liquid-vapor surface tension of supercooled water\cite{Feeney_Theoretical_2003,Hruby_Twostructure_2004,Lu_Second_2006,Wang_Second_2019,Rogers_Possible_2016,Malek_Surface_2019,Gorfer_Highdensity_2022}. Using two closely related microscopic models of water-like associating fluids, Feeney and Debenedetti\cite{Feeney_Theoretical_2003} predicted  either an inflection point or a maximum, depending on the details and assumptions of the approach. Hruby and Holten\cite{Hruby_Twostructure_2004} proposed a two-state model able to generate an inflection point in the liquid-vapor surface tension of water. The inflection point has also been predicted by molecular dynamics simulations with several water potentials, such as SPC/E~\cite{Lu_Second_2006,Wang_Second_2019}, WAIL~\cite{Rogers_Possible_2016}, and TIP4P/2005~\cite{Malek_Surface_2019,Wang_Second_2019,Gorfer_Highdensity_2022}.

In the case of a fluid with a liquid-liquid phase transition, the liquid-liquid surface tension is also of particular interest. It controls the nucleation of the second liquid phase and the possibility of observing liquid-liquid coexistence in confined systems~\cite{Singh_Thermodynamic_2019}. For their models of water, Feeney and Debenedetti\cite{Feeney_Theoretical_2003} found a liquid-liquid surface tension two orders of magnitude lower than the liquid-vapor one. They attributed this phenomenon to the significant difference in the corresponding range of densities spanned by each transition.

In this work, inspired by the ideas of Feeney and Debenedetti, we investigate the interfacial properties in a simple microscopic model for liquid polyamorphism caused by the molecular interconversion between two species \cite{Caupin_Polyamorphism_2021}. Our approach provides a first-principle derivation of the surface tension and the density and concentration interfacial profiles for the liquid-vapor interface, as well as for the corresponding liquid-liquid interfacial properties, if liquid polyamorphism takes place. We demonstrate that, depending on the interaction parameters between the two interconverting species, an inflection point or two extrema, may emerge in the temperature dependence of the liquid-vapor surface tension.

\section{Methods}
In this section, we describe the  model developed by Caupin and Anisimov\cite{Caupin_Polyamorphism_2021}, referred to as the ``blinking-checkers'' lattice model, and its application in the density-gradient theory (DGT) to calculate the interfacial properties.

\subsection{Blinking-Checkers Lattice Model}

We consider a compressible binary-lattice of fixed total volume, $V$, where each of the $N$ lattice sites can either be empty or occupied by one particle of two types (1 and 2). The numbers of particles of type 1 and 2 are $N_1$ and $N_2$, respectively. The number density is $\rho = (N_1 + N_2)/N$ and the fraction of particles of type 1 in the mixture is $x = N_1/(N_1 + N_2)$. The interactions of each particle with its $z$ nearest neighbors is given by interaction parameters between each particle type of the form, $\omega_{11} = -z\epsilon_{11}/2$, $\omega_{22}=-z\epsilon_{22}/2$, and $\omega_{12}=-z\epsilon_{12}/2$, where the epsilons represent the energy of the pair interactions. There is no interaction with empty sites. The Helmholtz energy per lattice site, $f = F/N$, in the non-reacting version of the blinking-checkers model is given by
\begin{equation}\label{Eqn_Binary_Helmholtz_Energy}
    \begin{split}
        f(T,\rho,x) =& \rho\varphi_2^\circ  + \rho x \varphi_{12}^\circ - \rho^2[\omega_{11}x^2 + \omega_{22}(1-x)^2 + 2\omega_{12} x (1-x)]\\ &+ T\left[\rho x \ln{x} + \rho(1-x)\ln(1-x)\right] + T\left[\rho\ln\rho + (1-\rho)\ln(1-\rho)\right]
    \end{split}
\end{equation}
where $\varphi_{12}^\circ = \varphi_1^\circ -\varphi_2^\circ$, in which $\varphi_1^\circ=\varphi_1^\circ(T)$ and $\varphi_2^\circ=\varphi_2^\circ(T)$ are functions of temperature only, containing the arbitrary zero points of energy and entropy\cite{Anisimov_Polyamorphism_2018}. In Eq.~(\ref{Eqn_Binary_Helmholtz_Energy}), we adopt Boltzmann's constant, $k_\text{B}$, as $k_\text{B}=1$. For the units of the various quantities in this work, see Sec. S1 in the supplemental material (SM). The three terms in square brackets in Eq.~(\ref{Eqn_Binary_Helmholtz_Energy}) describe the contribution to the free energy from the energy of interactions, the entropy of mixing of the two species, and the entropy of mixing of the occupied and empty sites.

\begin{figure}[t]
    \centering
    \includegraphics[width=0.49\linewidth]{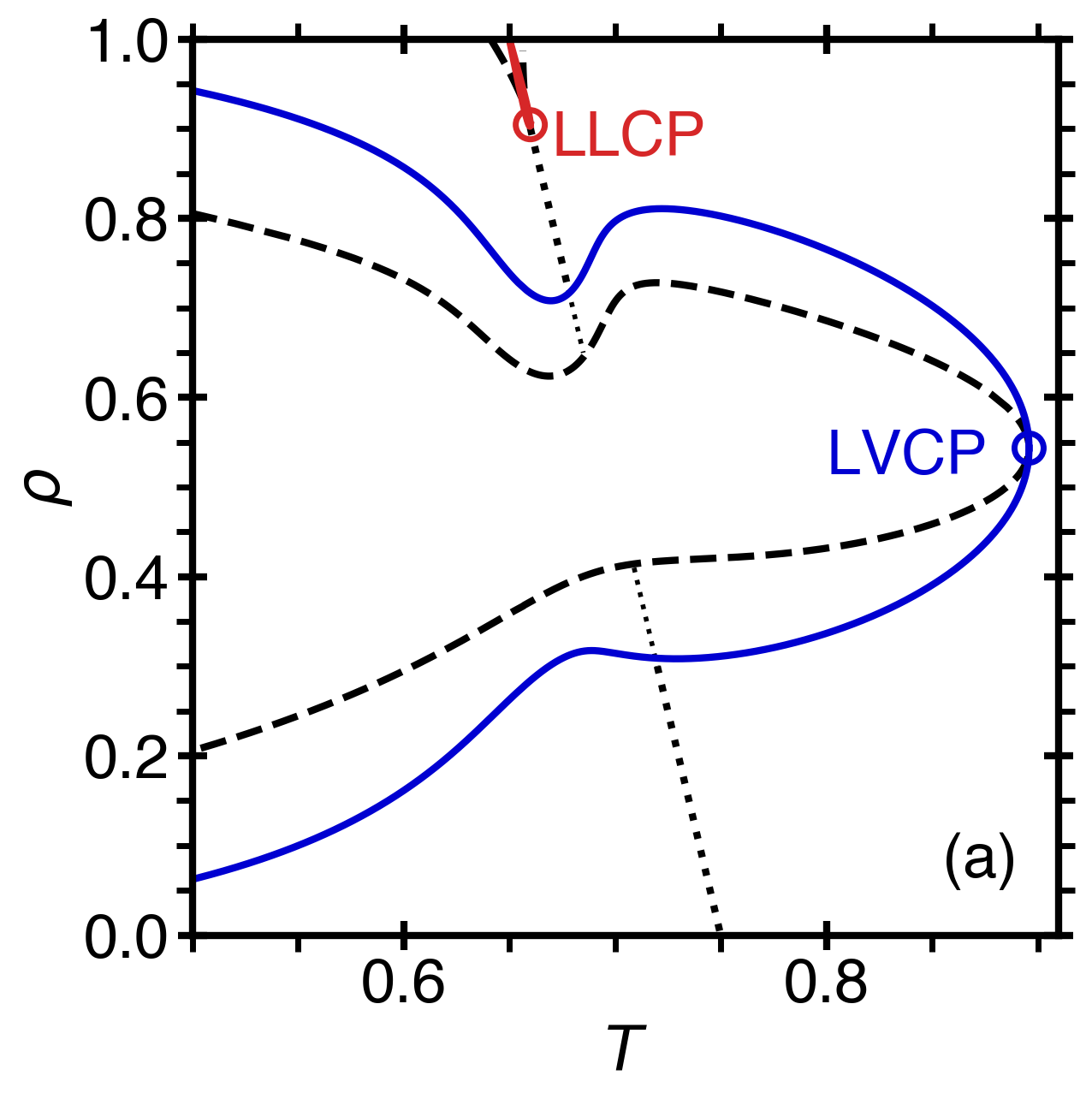}
    \includegraphics[width=0.49\linewidth]{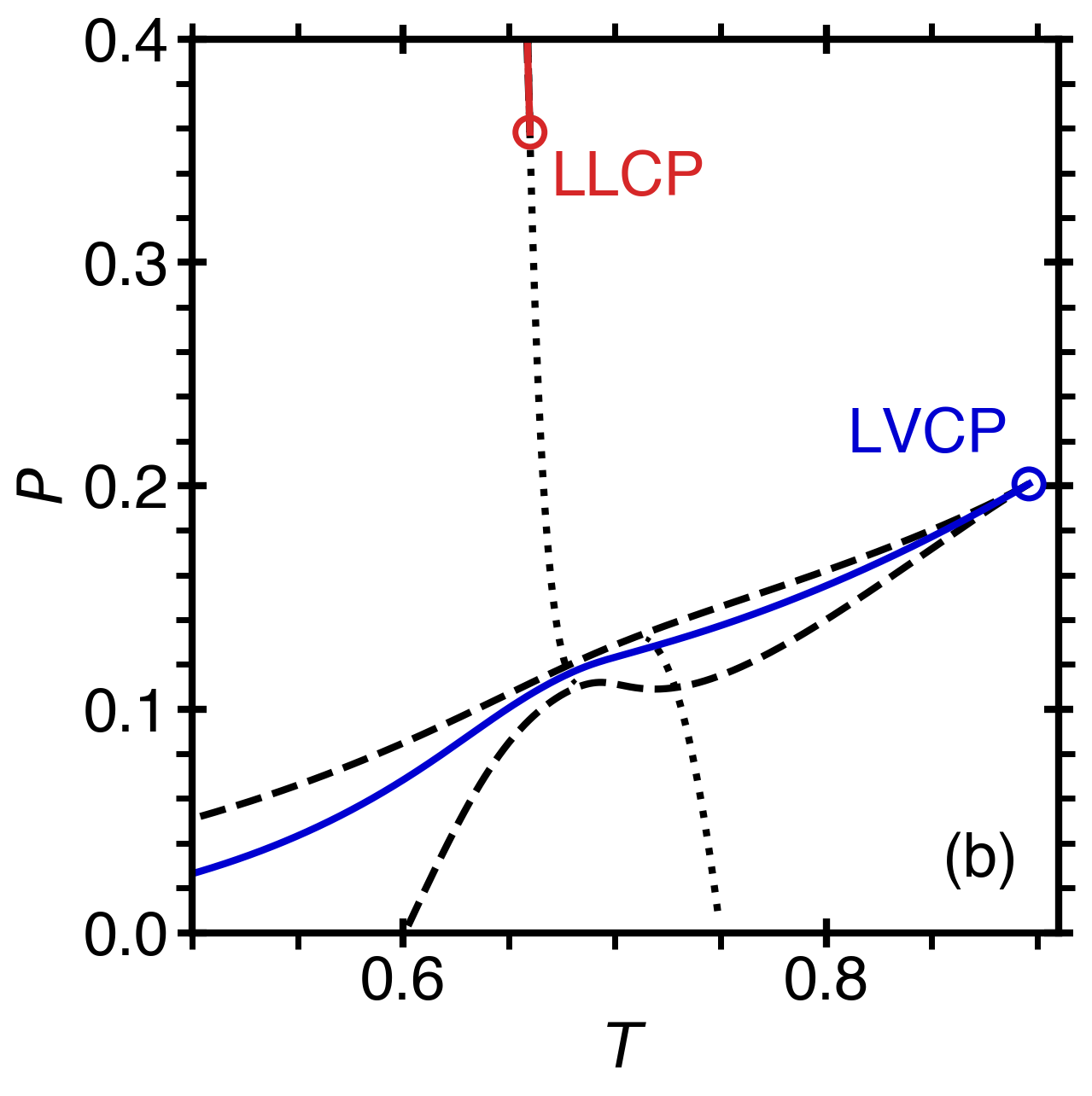}
    \caption{Density-temperature and pressure-temperature phase diagrams for the blinking-checkers model with $\omega_{11}=1.6$, $\omega_{22}=2.0$, $\omega_{12}=1.08$, $e=3$, and $s=4$. The liquid-vapor coexistence (blue curves) terminates at the liquid-vapor critical point, LVCP. The liquid-liquid coexistence (red curves) terminates at the liquid-liquid critical point, LLCP. The limits of thermodynamic stability (spinodals) are given by the dashed curves. In (a,b), the dotted line corresponds to the condition, $x=1/2$, which qualitatively separates the regions enriched either by species 1 (at low temperatures and low densities) or species 2 (at high temperatures and high densities).}
    \label{Fig_w12_108_Comparison}
\end{figure}

Since it has been hypothesized that the molecular interconversion of species could be a generic cause of liquid polyamorphism \cite{Holten_Water_2014,Anisimov_Polyamorphism_2018,Caupin_Thermodynamics_2019}, species 1 and 2 are allowed to interconvert via a simple reaction of the form, $\ce{1 <=> 2}$. The blinking-checkers lattice is a generic model, which has been used to generate liquid polyamorphism and reproduce a variety of the anomalies in the thermodynamic properties of supercooled water \cite{Caupin_Polyamorphism_2021}. Phenomenologically, species 1 and 2 could also represent supramolecular states 1 and 2, which enables one to use this model, via a coarse-grained approach, to mimic the different scenarios for the polyamorphic phase behavior considered for supercooled water. The most important aspect of mapping the blinking-checkers model to describe the phase behavior of polyamorphic substances is to assign the appropriate energy and entropy of reaction for the system~\cite{Anisimov_Polyamorphism_2018,Caupin_Polyamorphism_2021}. Applying the condition for chemical-reaction equilibrium, $\partial f/\partial x|_{T,\rho} =\mu_{12}=0$, where $\mu_{12}=\mu_{1}-\mu_{2}$ is the difference between the two chemical potentials of each species in the interconverting mixture, makes the temperature-dependent function, $\varphi_{12}^\circ(T)$, well defined. We emphasize that unlike previous models for water, which introduce a ``local'' density difference by changing the occupancy~\cite{Ciach_Model_2008} or volume~\cite{Sastry_SingularityFree_1996,Stokely_Effect_2010,Cerdeirina_Water_2019,Harvey_Improving_2023} of each cell, the blinking-checkers model assumes that there is no volume change of reaction. This indicates that the nonideality of the mixture is the primary ingredient for the anomalous behavior of the thermodynamic properties in fluids exhibiting polyamorphic or water-like behavior~\cite{Caupin_Polyamorphism_2021}. We specify this function through the energy of reaction, $e$, and the entropy of reaction, $s$, as $\varphi_{12}^0 = -(e-Ts)$, the simplest linear approximation of this function\cite{Anisimov_Polyamorphism_2018,Caupin_Polyamorphism_2021}. The condition for chemical-reaction equilibrium, $\mu_{12}=0$, reduces the number of thermodynamic degrees of freedom by one and defines the equilibrium concentrations (molecular fractions) of species 1, $x=x_\text{e}(T,\rho)$. Therefore, the concentration is no longer an independent variable, and thermodynamically, the interconverting binary mixture thermodynamically behaves as a single component fluid. 

\begin{figure}[ht!]
    \centering
    \includegraphics[width=0.49\linewidth]{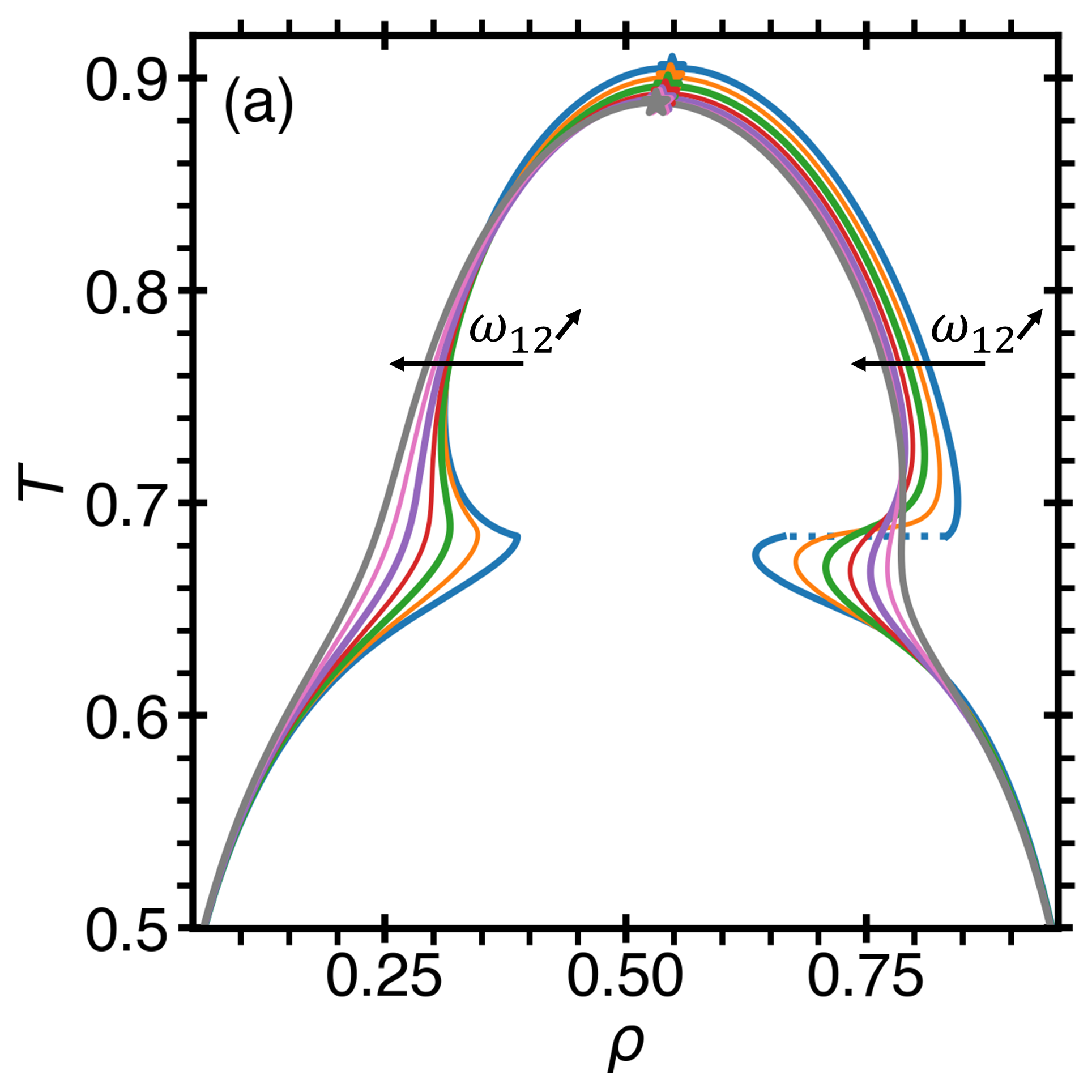}
    \includegraphics[width=0.49\linewidth]{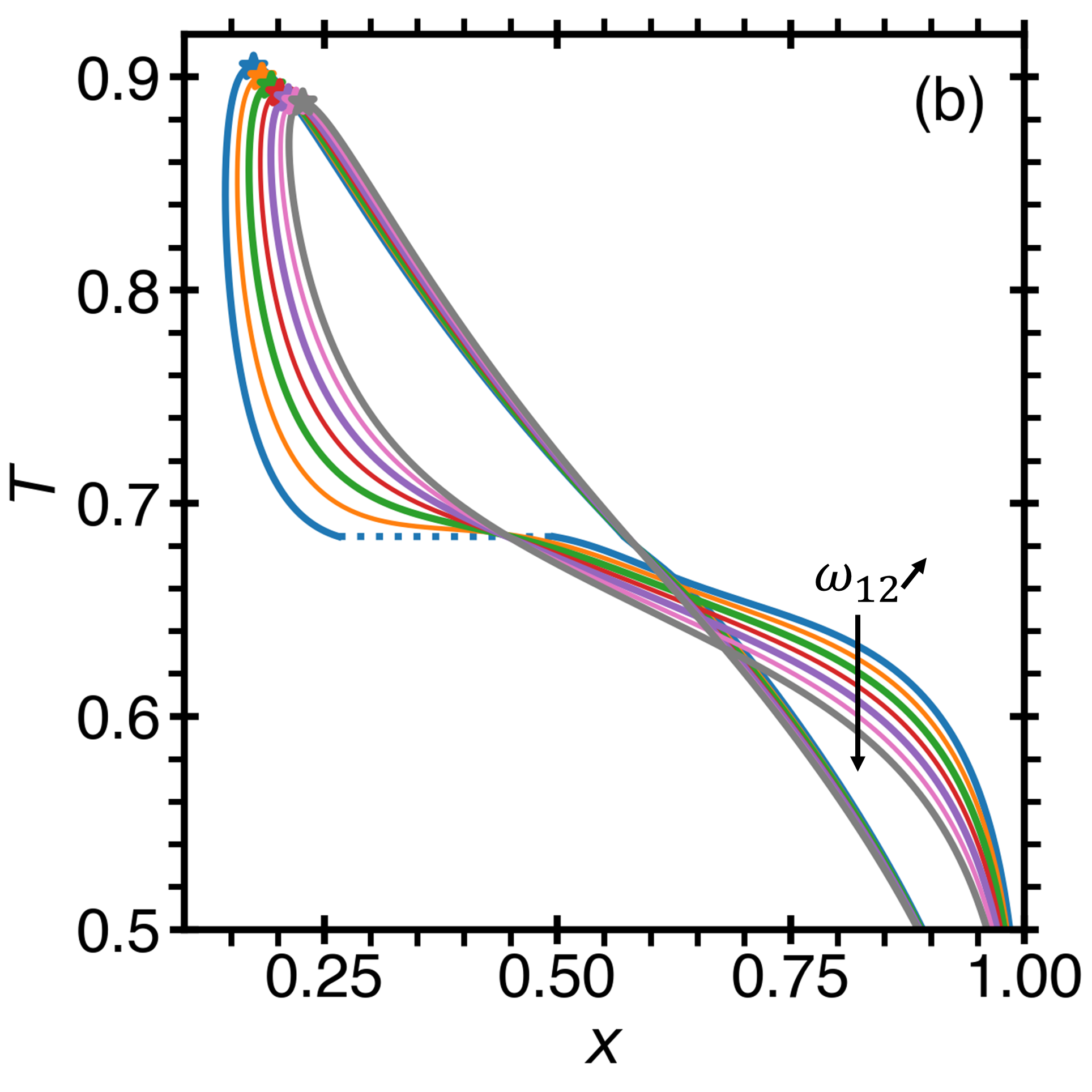}
    \includegraphics[width=0.49\linewidth]{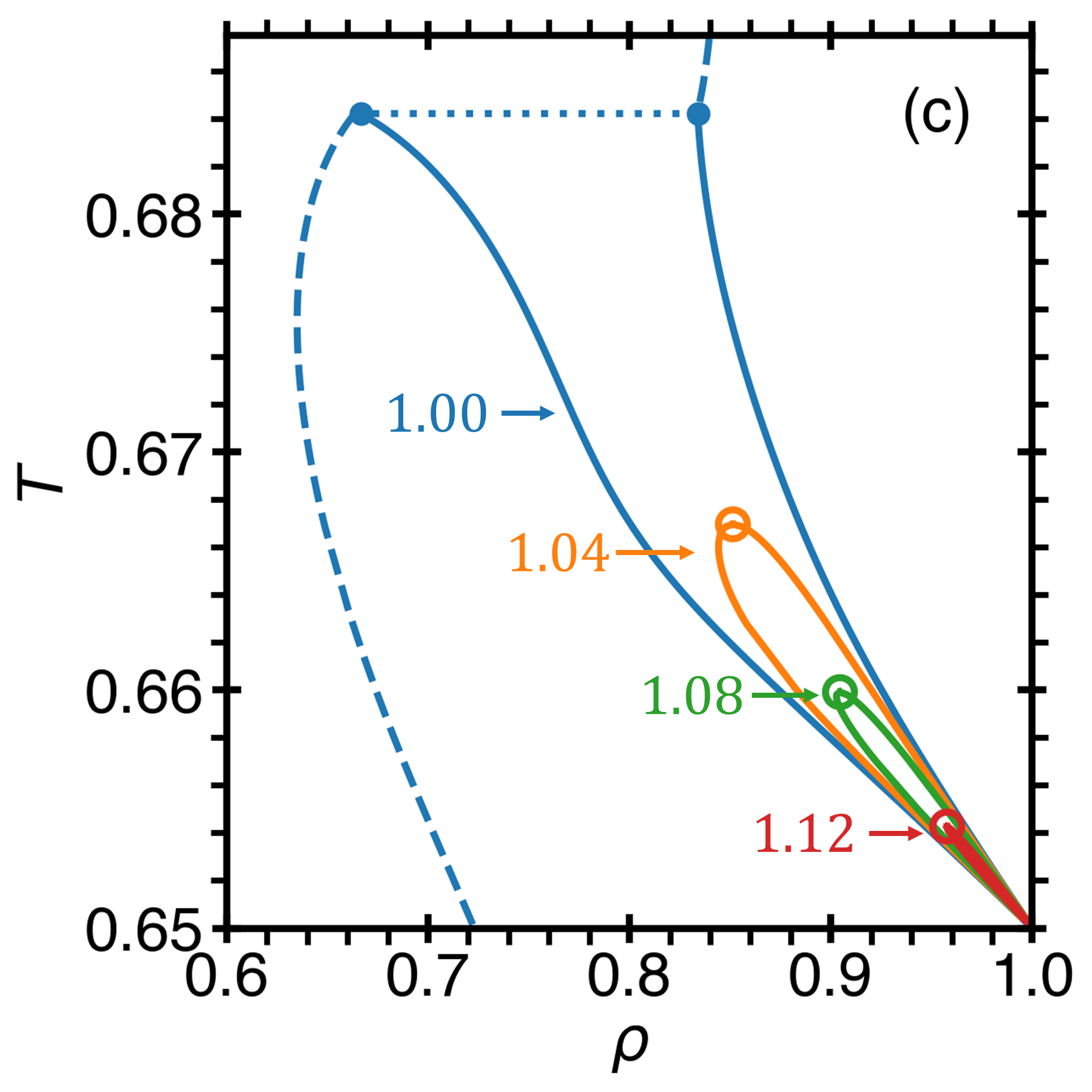}
    \includegraphics[width=0.49\linewidth]{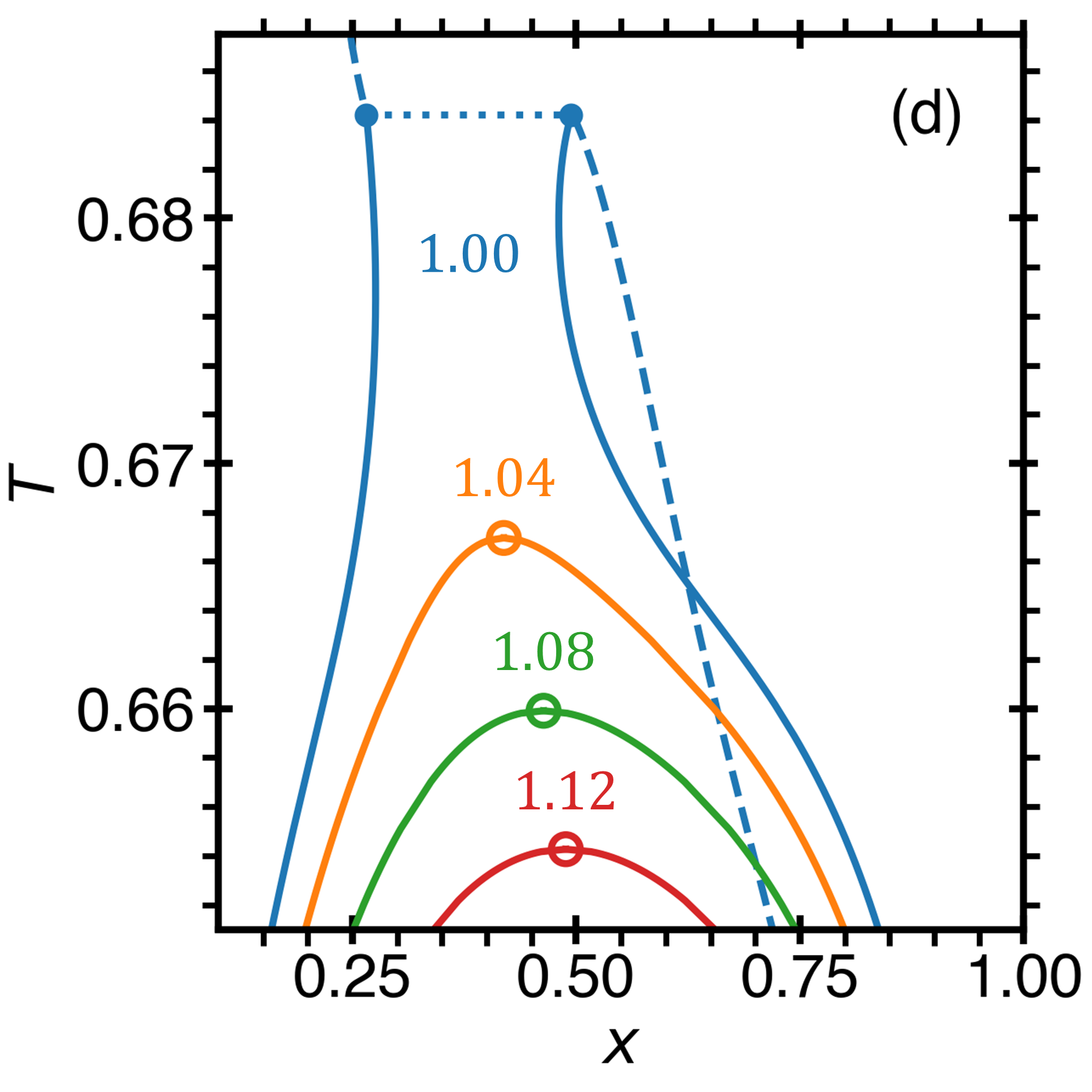}
    \caption{Liquid-vapor (a,b) and liquid-liquid (c,d) coexistence curves for the seven systems with $\omega_{11}=1.6$, $\omega_{22}=2.0$, $e=3$, $s=4$, and with various values of $\omega_{12}$: $\omega_{12}=1.00$ (blue), $\omega_{12}=1.04$ (orange), $\omega_{12}=1.08$ (green), $\omega_{12}=1.12$ (red), $\omega_{12}=1.16$ (purple), $\omega_{12}=1.20$ (pink), and $\omega_{12}=1.24$ (gray). The critical points, indicated by the stars in (a,b) and open circles in (c,d), are the unique liquid-vapor (LVCP) or liquid-liquid (LLCP) critical points in the interconverting system, referred to as ``actual'' critical points. In (a,b) the arrows indicate the direction of increasing $\omega_{12}$. As also  indicated in Fig.~\ref{Fig_w12_108_Comparison}, species 1 is enriched in the low-density, low-temperature region, while species 2 is enriched in the high-density, high-temperature region. For the system with $\omega_{12}=1.00$, in (c,d), the dashed blue curves indicate the liquid-vapor coexistence, while in (a-d) the dotted line indicates the discontinuity at the triple point. For more details, see SM Sec. S2.}
    \label{Fig_AllCoex}
\end{figure}

Figure~\ref{Fig_w12_108_Comparison}(a) and (b) illustrates the $\rho$-$T$ and $P$-$T$ phase diagram of the blinking-checkers lattice for an example set of interaction parameters ($\omega_{11}=1.6$, $\omega_{22}=2.0$, and $\omega_{12}=1.08$) and interconversion-reaction parameters ($e=3$ and $s=4$). For this choice of the energy and the entropy of reaction, we obtain a negative slope for the liquid-liquid phase transition, similar to that predicted for supercooled water. The line that qualitatively separates the region enriched by species 1, at low temperatures and low densities (referred to as ``L1''), from the region enriched by species 2, at high temperatures and high densities (referred to as ``L2''), is indicated by the dotted line in Fig.~\ref{Fig_w12_108_Comparison}(a,b). Furthermore, for this set of interaction parameters, two critical points and a ``bottleneck'' in the liquid-vapor coexistence are observed. The $\rho$-$T$ and $x$-$T$ phase diagrams for seven systems are illustrated in Fig.~\ref{Fig_AllCoex}(a,b), see more details in SM Secs. S1 and S2. We note that for different sets of parameters, one may obtain multiple fluid-fluid critical points, representing the more complex phase behavior of polyamorphic fluids.

In the treatment of the blinking-checkers model, we utilize the meanfield approximation, which is more accurate in the region away from the critical point, where the correlation length of concentration or density fluctuations is not significantly larger than the distance between molecules (the Ginzburg criterion\cite{LL_Stat_Phys_II}). To estimate the effect of the critical fluctuations on the phase behavior, we also conducted exact Monte Carlo (MC) simulations of the blinking checkers model. The preliminary results of these simulations, presented in SM Secs. S3 and S12, demonstrate a qualitatively similar bottleneck anomaly of the liquid-vapor coexistence as well as interfacial profile behavior near the minimum of this bottleneck. Also, these simulations have confirmed that the phase transitions in the interconverting blinking-checkers model belong to the three-dimensional Ising-model universality class\cite{fisher_scaling_1983}.

\subsection{Virtual Critical Points\label{Sec_Virtual_CPs}}
If interconversion does not occur, the blinking-checkers model describes a compressible binary mixture, which may exhibit liquid-vapor and liquid-liquid coexistence, as well as the corresponding critical lines~\cite{Caupin_Polyamorphism_2021}. Consider a point on a critical line with temperature, $T_\mathrm{c}$, density, $\rho_\mathrm{c}$, and type 1 particles' molecular fraction, $x_\mathrm{c}$. In a fixed volume, $V$, the corresponding critical isochore, at fixed composition, contains a fixed number of particles 1 and 2, given by $\rho_\mathrm{c} x_\mathrm{c} V$ and $\rho_\mathrm{c} (1- x_\mathrm{c}) V$, respectively. At temperatures below $T_\mathrm{c}$, the system will separate in two phases, $\alpha$ and $\beta$ (which could be liquid and vapor or liquid and liquid). For phase $i=\alpha$ or $\beta$, let $\rho_i$, $x_i$, and $V_i$ be the density, type 1 fraction, and volume, respectively. At each $T \leq T_\mathrm{c}$, the six values $(\rho_i,x_i,V_i)_{i=\alpha,\beta}$ are fully determined by three conservation equations (one for volume and two for mass), and three equilibrium conditions (two for the chemical potentials and one for the pressure).

% or along a selected path (\textit{e.g.} isobars, isotherms, or isochores) in the one phase region

As explained in the previous section, when interconversion takes place, the system (in terms of the Gibbs Phase Rule \cite{LL_Stat_Phys_II}) thermodynamically behaves as a single component fluid, following the given paths along liquid-vapor or liquid-liquid coexistence in the two phase region. Therefore, for each point along the interconversion path, there is a corresponding a critical point of the non-reacting binary mixture, which is connected to this point on the path by a critical isochore for the non-reacting mixture at fixed composition. We refer to this corresponding binary-mixture critical point as the ``virtual'' (\textit{i.e.} invisible along the interconversion path) critical point, while we refer to the interconverting system's unique liquid-vapor critical point (LVCP) as the ``actual'' LVCP. Similarly, for systems exhibiting interconversion, we refer to the single liquid-liquid critical point (LLCP), as the actual LLCP. We emphasize that not only the phase diagram of the interconverting mixture is characterized by unique fluid-fluid critical points (like that of a single-component fluid), but the response functions, second derivatives of the free energy at $\mu_{12}=0$, also exhibit the singularities characteristic of single-component fluids.

%We emphasize that after the interconverting system reaches equilibrium, the thermodynamic properties, being state functions, are the same as a non-reacting binary mixture with the same equilibrium composition, temperature, and pressure.

\begin{figure}[t]
    \centering
    \includegraphics[width=0.49\linewidth]{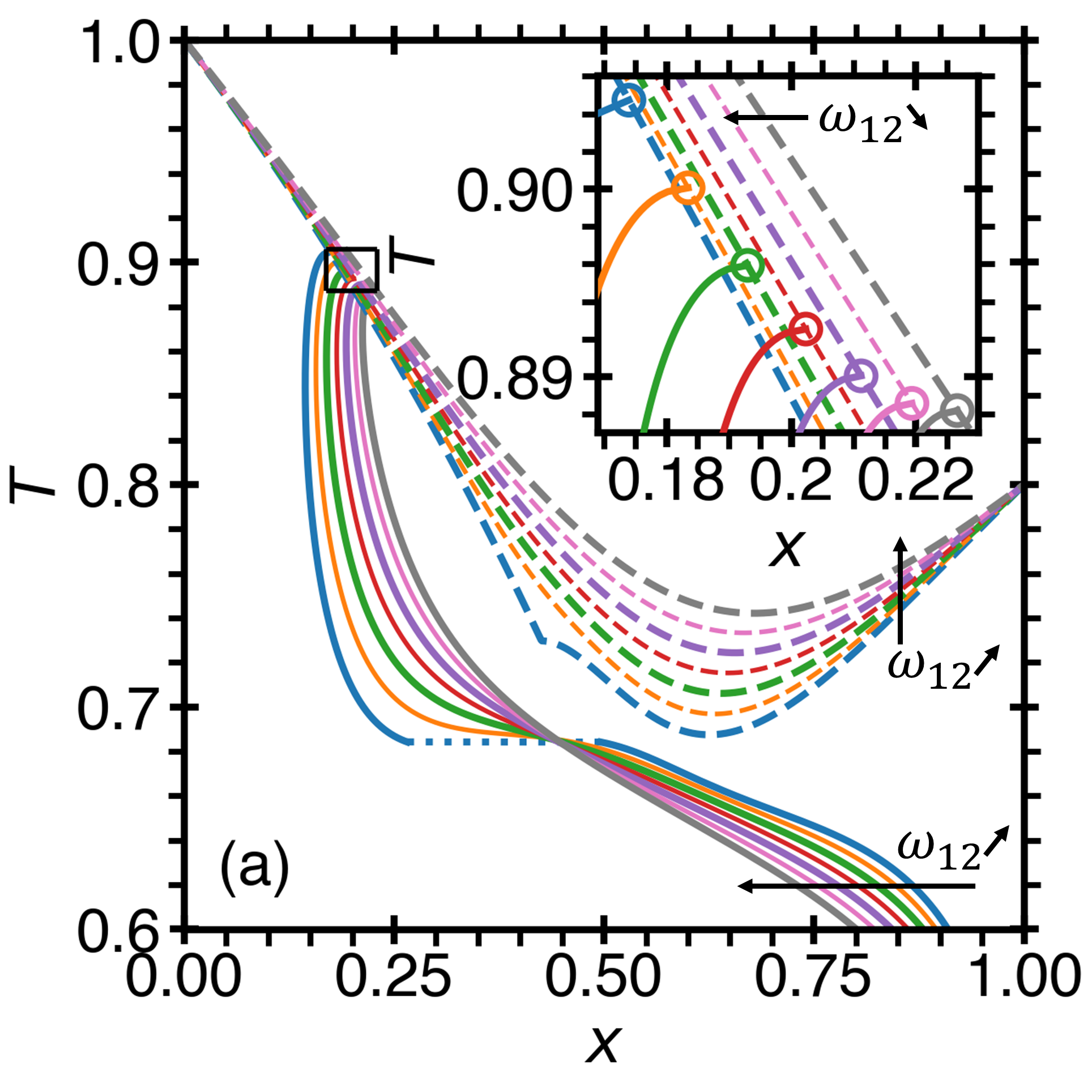}
    \includegraphics[width=0.49\linewidth]{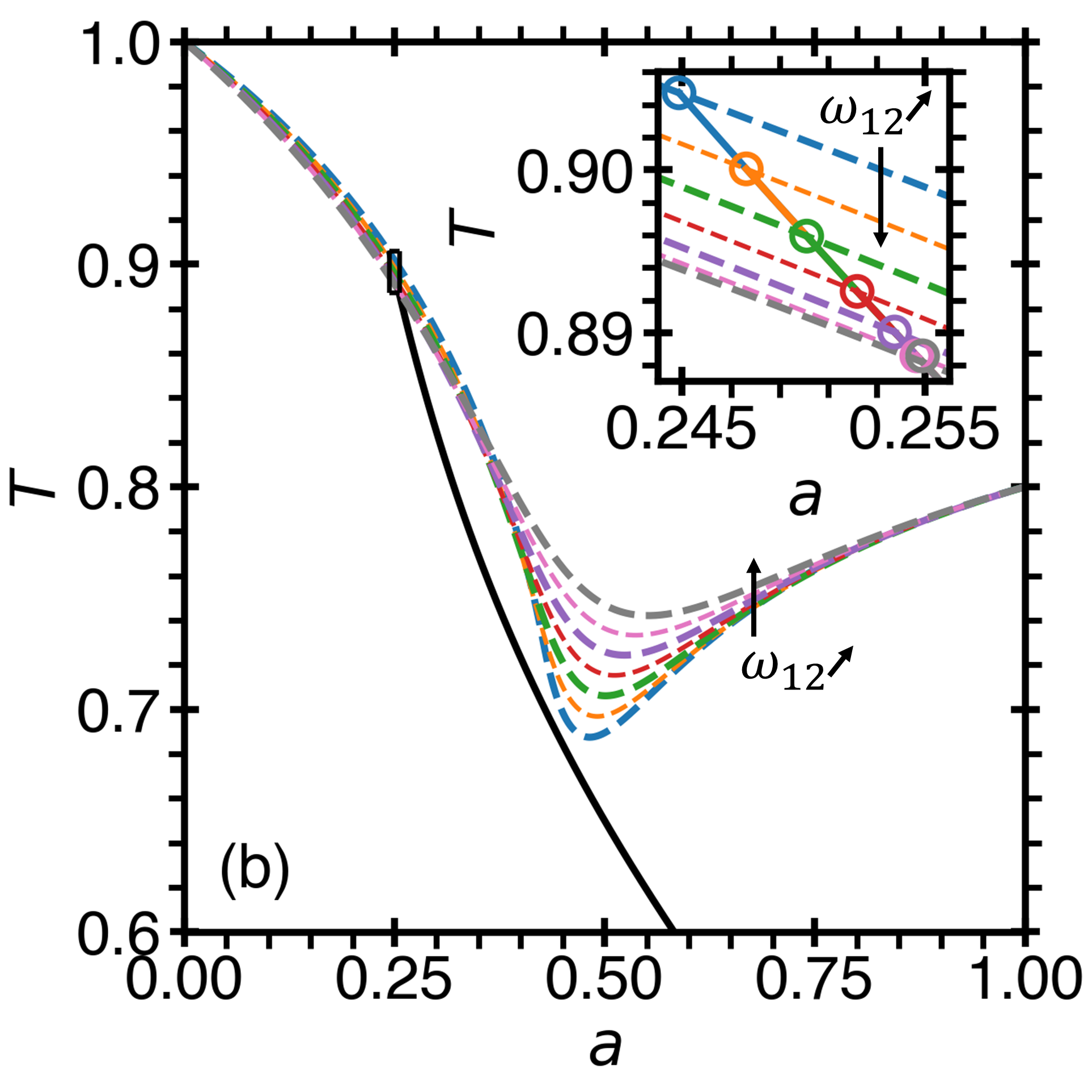}
    \caption{Illustration of the thermodynamic path selected by the interconversion reaction for $e=3$ and $s=4$ at constant volume, $V$, and at constant number of occupied lattice sites, $N_1 + N_2$, represented through (a) the liquid branch of the liquid-vapor temperature-concentration coexistence (see Fig.~\ref{Fig_AllCoex}b), and (b) the activity, $a = 1/[1+e^{-\Delta\mu_{12}/k_\mathrm{B}T}]$, for the systems with different interaction parameters: $\omega_{12}=1.00$ (blue), $\omega_{12}=1.04$ (orange), $\omega_{12}=1.08$ (green), $\omega_{12}=1.12$ (red), $\omega_{12}=1.16$ (purple), $\omega_{12}=1.20$ (pink), and $\omega_{12}=1.24$ (gray). For each system, the liquid-vapor critical line (LVCL) is shown by the dashed curves, while the collapsed coexistence, in (a), is illustrated by the black curve. The insets show the LV critical points for each scenario. In (a,b), the black arrow indicates the direction of increasing (or decreasing) $\omega_{12}$. Note that for the system with $\omega_{12} = 1.00$, the thermodynamic path crosses through the triple point, indicated by the dotted line in (a). For more details, ref.\cite{Caupin_Polyamorphism_2021} and SM Sec. S2.}
    \label{Fig_Thermo_Path}
\end{figure}

An illustration of the thermodynamic path along liquid-vapor equilibrium in the interconverting fluid for seven different sets of interaction parameters, $\omega_{12}$, is shown in Fig.~\ref{Fig_Thermo_Path}(a). In the coexisting liquid and vapor phases, there are two branches of the density and concentration, given by $\rho_\mathrm{L}^\mathrm{cxc}(T)$, $\rho_\mathrm{V}^\mathrm{cxc}(T)$, $x_\mathrm{L}^\mathrm{cxc}(T)$, and $x_\mathrm{V}^\mathrm{cxc}(T)$, see Fig.~\ref{Fig_AllCoex}(a,b) for details. For simplicity, in Fig.~\ref{Fig_Thermo_Path}(a) we show only the liquid branch of the liquid-vapor coexistence, see Fig.~\ref{Fig_AllCoex}(b) for both branches. The liquid-vapor critical lines for the seven binary mixtures with the same interaction parameters are also shown  (dashed lines). Fig.~\ref{Fig_Thermo_Path}(b) displays the same, but replacing the abscissa, $x$, by the ``activity'', $a = 1/[1+e^{-\Delta\mu_{12}/k_\mathrm{B}T}]$, \cite{Leung_Properties_1973,Jun_Fluids_1992,Anisimov_Hidden_1992} where $\Delta\mu_{12} = \mu_{12} - \varphi_{12}^0 $, see SM Sec. S1 for details. Since, due to the chemical reaction equilibrium condition, $\mu_{12}=0$ and $\varphi_{12}^\circ = -(e-Ts)$, $\Delta\mu_{12}$ is given by
\begin{equation}\label{Eq_mu12}
    \Delta\mu_{12} = -\varphi_{12}^\circ = e - Ts
\end{equation}
(see SM Sec. S1 for details). Thus, for each point along the thermodynamic path selected by interconversion, the activity is restricted by Eq.~(\ref{Eq_mu12}), such that for the seven systems considered in this work, the activities collapse into the line shown in Fig.~\ref{Fig_Thermo_Path}(b), as only $\omega_{12}$ is varied in each system.

%Thus, in the seven systems considered in this work, the activity is a straight line, as only $\omega_{12}$ is varied in each system, see Fig.~\ref{Fig_Thermo_Path}(b).

The proximity of the virtual critical line affects the properties along coexistence in the interconverting fluid, causing, in particular, the bottlenecked shape of the $\rho$-$T$ liquid-vapor coexistence (Fig.~\ref{Fig_AllCoex}). For our selection of the interaction parameters, this particular effect is pronounced because the difference between $\omega_{11}$ and $\omega_{22}$ is significant ($\omega_{11}=1.6$ and $\omega_{22}=2.0$). In addition, the asymmetry of the liquid-vapor coexistence occurs due to the existence of the liquid-liquid critical point, and even occurs in the singularity-free scenario where the liquid-liquid critical point is moving to indefinite pressure (see ref.\cite{Caupin_Polyamorphism_2021} and SM Sec. S2 for more details).

\subsection{Interfacial Properties via Density Gradient Theory}

To model the fluid interfaces, we consider density gradient theory (DGT) \cite{Debye_Dissymmetry_1959,Yang_Molecular_1976,Lu_Density_1985,Kahl_Interfacial_2002,Stephan_Interfaces_2020}, in which the free energy of the system is expanded in a Taylor series up to second-order in terms of derivatives of the concentration and density with respect to the coordinate perpendicular to the interface \cite{Stephan_Interfaces_2020}. Following the ideas presented by van der Waals \cite{VDW_Thermodynamics_1893,Rowlinson_Capillarity_1982}, and later elaborated by Cahn and Hilliard\cite{Cahn_Hillaird_1958}, the interfacial tension of a binary fluid may be expressed as\cite{Poser_Surface_1979,Poser_Interfacial_1981,Kahl_Interfacial_2002}
\begin{equation}\label{Eq_sigma}
    \sigma = \int\left[\Delta\Omega(x,\rho, T) + \frac{1}{2}c_x(\rho)|\nabla x|^2 + \frac{1}{2}c_\rho(x)|\nabla \rho|^2 + c_{\rho,x}(\rho,x)\nabla\rho\cdot\nabla x\right]\dd{V}
\end{equation}
where $c_x$, $c_\rho$, and $c_{\rho,x}$ are the microscopic ``influence'' coefficients. $\Delta\Omega$ is the excess grand potential per lattice site, given by\cite{Kahl_Interfacial_2002}
\begin{equation}\label{Eq_Grand_Potential}
    \Delta\Omega(x,\rho, T) = f(x,\rho, T) - \rho x \mu_2^\text{cxc} - \rho(1-x)\mu_1^\text{cxc} + P^\text{cxc}
\end{equation}
where $\mu_1$, $\mu_2$, and $P$ are the chemical potentials of species 1 and 2 per lattice site in solution and the pressure per lattice site, respectively, while the superscript ``cxc'' indicates that the quantity is evaluated along the phase coexistence. We note that $\Delta\Omega$ may be calculated for a non-reactive mixture, as in Eq.~(\ref{Eq_Grand_Potential}), or it may be calculated for a reactive mixture. Importantly, in either case, due to the chemical-reaction equilibrium condition, the grand thermodynamic potential is the same for the non-reacting and for the reacting cases of the blinking-checkers model (see SM Sec. S5 for details).

Assuming planar fluid interfaces in the lattice, we find that the three influence coefficients for the concentration and density are related to the interaction parameters by
\begin{eqnarray}
    c_x(\rho) &= \frac{1}{2}\ell^2\rho^2\omega \label{Eq_cx} \\ 
    c_\rho(x) &= \frac{1}{2}\ell^2\left[\omega_{11}x^2 + 2\omega_{12}x(1-x) + \omega_{22}(1-x)^2\right] \label{Eq_crho}\\ 
    c_{\rho,x}(\rho,x) &= \frac{1}{2}\ell^2\rho\left[\omega_{11}x + \omega_{12}(1-2x)-\omega_{22}(1-x)\right] \label{Eq_crhox}
\end{eqnarray}
where $\omega = \omega_{11}+\omega_{22}-2\omega_{12}$ and $\ell$ is the length of a lattice cell (see SM Sec. S6 for the full derivation). Note that, in the language of Ref.~\cite{Feeney_Theoretical_2003}, Eqs.~(\ref{Eq_cx}) to (\ref{Eq_crhox}) define the influence coefficients \textit{a priori} with the assumption that the gradients contribute to the local energy density, but not to the entropy density.

To determine the interfacial profiles, we adopt a variational approach, based on a family of anzatz functions, choosing the optimal one by  minimizing the interfacial tension, given by Eq.~(\ref{Eq_sigma}). We have also obtained exact numerical solutions for the profiles by solving the equilibrium condition for the surface tension, which is obtained from Eq.~(\ref{Eq_sigma}),\cite{Poser_Surface_1979,Poser_Interfacial_1981,Kahl_Interfacial_2002}
\begin{equation}\label{Eq_Interface_EqCondition}
     \Delta\Omega(x,\rho, T) = \frac{1}{2}c_x(\rho)|\nabla x|^2 + \frac{1}{2}c_\rho(x)|\nabla \rho|^2 + c_{\rho,x}(\rho,x)\nabla\rho\cdot\nabla x
\end{equation}
as explained in Sec. S6 in the SM. We find that the variational approach is enough to capture the anomalous behavior of the interfacial properties with sufficient accuracy (see SM Sec. S6 for details). Throughout the main text, we report the variational results, based on the Fisher-Wortis profile, which accounts for the thermodynamic asymmetry between the two coexisting phases \cite{Fisher_Profile_1984,Anisimov_Mesothermo_2010}. A comparison with an alternative symmetric ansatz is also discussed in SM Sec. S6. The Fisher-Wortis ansatz is given in normalized form as a combination of symmetric and asymmetric components for both the density and concentration (molecular fraction) profiles as
\begin{align}
     \hat{\rho}(\hat{z}) = \frac{\rho(\hat{z})-\rho_\alpha}{\rho_\beta - \rho_\alpha} =& \hat{\rho}_\mathrm{sym}(\hat{z}) + \Delta\hat{\rho}_\mathrm{d}\hat{\rho}_\mathrm{asym}(\hat{z})  \label{Eq_rho_Profile}\\
    \hat{x}(\hat{z}) = \frac{x(\hat{z})-x_\alpha}{x_\beta - x_\alpha} =& \hat{x}_\mathrm{sym}(\hat{z}) + \Delta\hat{x}_\mathrm{d}\hat{x}_\mathrm{asym}(\hat{z}) \label{Eq_x_Profile}
\end{align}
where the symmetric contributions to the profiles are given by
\begin{align}
     \hat{\rho}_\mathrm{sym}(\hat{z}) =& \frac{1}{2}\left[\tanh{\left(\frac{\hat{z}}{\hat{\zeta}}\right)}+1\right]\\
     \hat{x}_\mathrm{sym}(\hat{z}) =& \frac{1}{2}\left[\tanh{\left(\frac{\hat{z}+\hat{\delta}}{\hat{\zeta}}\right)}+1\right]
\end{align}
and where the asymmetric contributions to the profiles are given by
\begin{align}
     \hat{\rho}_\mathrm{asym}(\hat{z}) =& \tanh^2{\left(\frac{\hat{z}}{\hat{\zeta}}\right)} + \frac{\ln\left[{\cosh{\left(\frac{\hat{z}}{\hat{\zeta}}\right)}}\right]}{\cosh^2\left(\frac{\hat{z}}{\hat{\zeta}}\right)}-(\rho_\beta-\rho_\alpha)\\
     \hat{x}_\mathrm{asym}(\hat{z}) =& \tanh^2{\left(\frac{\hat{z}+\hat{\delta}}{\hat{\zeta}}\right)} + \frac{\ln\left[{\cosh{\left(\frac{\hat{z}+\hat{\delta}}{\hat{\zeta}}\right)}}\right]}{\cosh^2\left(\frac{\hat{z}+\hat{\delta}}{\hat{\zeta}}\right)}-(x_\beta-x_\alpha)
\end{align}
in which $\hat{z}=z/\ell$ is the normalized coordinate perpendicular to the planar interface, the subscripts ``$\alpha$'' and ``$\beta$'' indicate the coexisting phases, $\hat{\zeta}=\zeta/\ell$ is the normalized interfacial thickness, and $\hat{\delta}=\delta/\ell$ is the normalized shift between the concentration and density profiles. The coefficient of the asymmetric terms in Eqs.~(\ref{Eq_rho_Profile}) and (\ref{Eq_x_Profile}) is the reduced diameter for the density and concentration, given by
\begin{align}
    \Delta\hat{\rho}_\text{d} &= \frac{\rho_\beta + \rho_\alpha}{2\rho_\text{c}} - 1\\
    \Delta\hat{x}_\text{d} &= \frac{x_\beta + x_\alpha}{2x_\text{c}} - 1
\end{align}
where $\rho_\text{c}$ and $x_\text{c}$ are the critical points determined from the non-reacting blinking-checkers model, referred to as virtual critical points, see Sec. \ref{Sec_Virtual_CPs}. The diameters with respect to the virtual critical points are provided in SM Sec. S7 for each system investigated. Relative to the liquid-vapor coexistence, for the liquid-liquid coexistence, the diameters are small, such that the asymmetric contribution to the liquid-liquid interfacial profiles is also minimal. 

Due to the lack of a theory to account for the interfacial profile asymmetry in compressible binary fluids, we choose the Fisher-Wortis ansatz, even though it was originally developed for a single component substance\cite{Fisher_Profile_1984,Anisimov_Mesothermo_2010}. This ansatz contains only two free parameters, $\hat{\zeta}$ and $\hat{\delta}$, less than the symmetric ansatz, and it partially reproduces the asymmetry of the exact solution (see details in SM Sec. S6).

%Moreover, it accounts for the asymmetry in the density profile quantitatively like the exact solution. The asymmetric shape of the exact solution justifies our choice of the Fisher-Wortis ansatz rather than the symmetric ansatz. However, when applied to the concentration profile, the Fisher-Wortis ansatz exaggerates the asymmetry in comparison to the exact solution. 

\section{Results and Discussion}
In this section, we demonstrate the anomalous behavior of the interfacial properties in the blinking-checkers lattice model. We also discuss the conditions for observing either an inflection point or extrema in the liquid-vapor interfacial tensions.

\subsection{Liquid-Vapor Interfacial tensions}
\begin{figure}[ht!]
    \centering
    \includegraphics[width=0.49\linewidth]{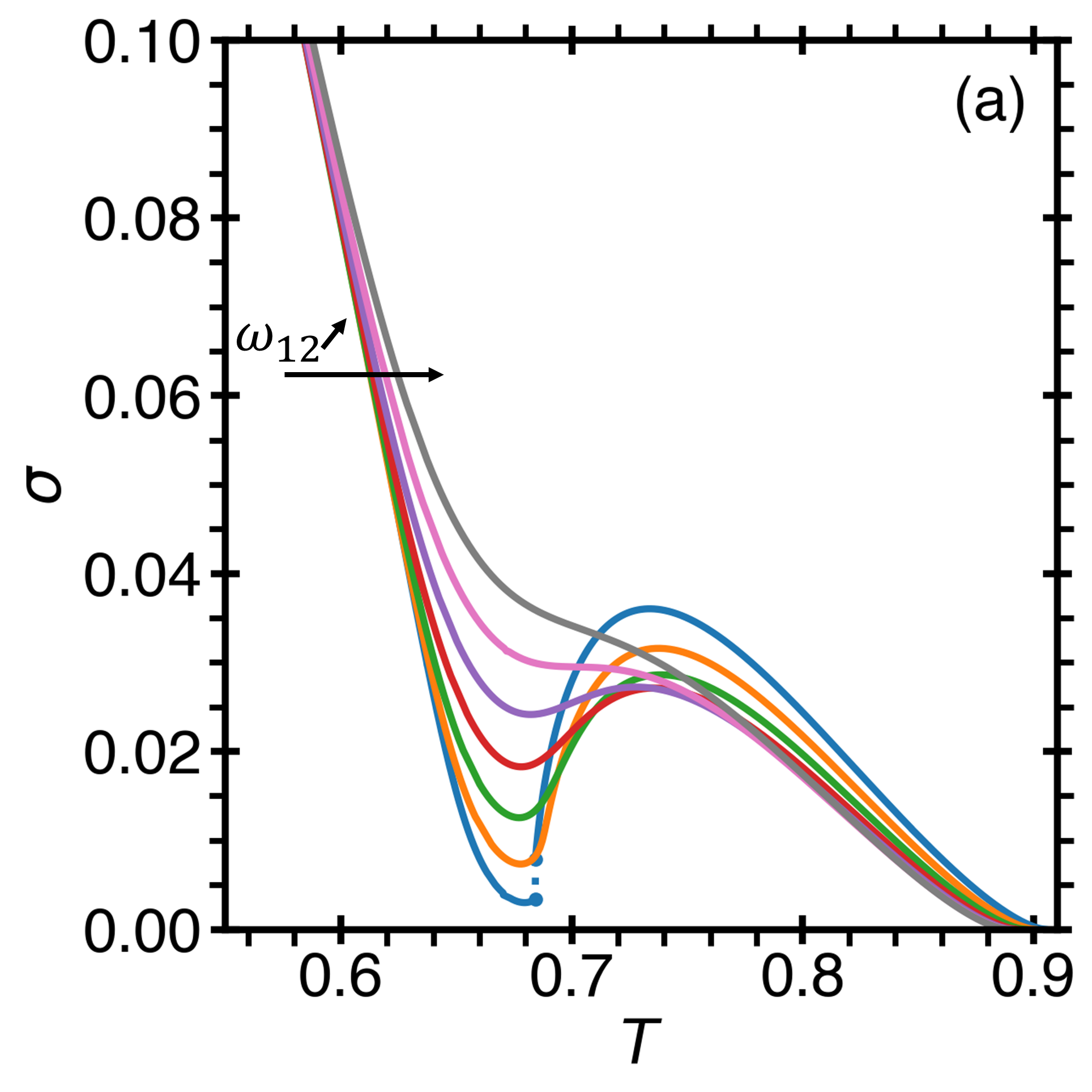}
    \includegraphics[width=0.49\linewidth]{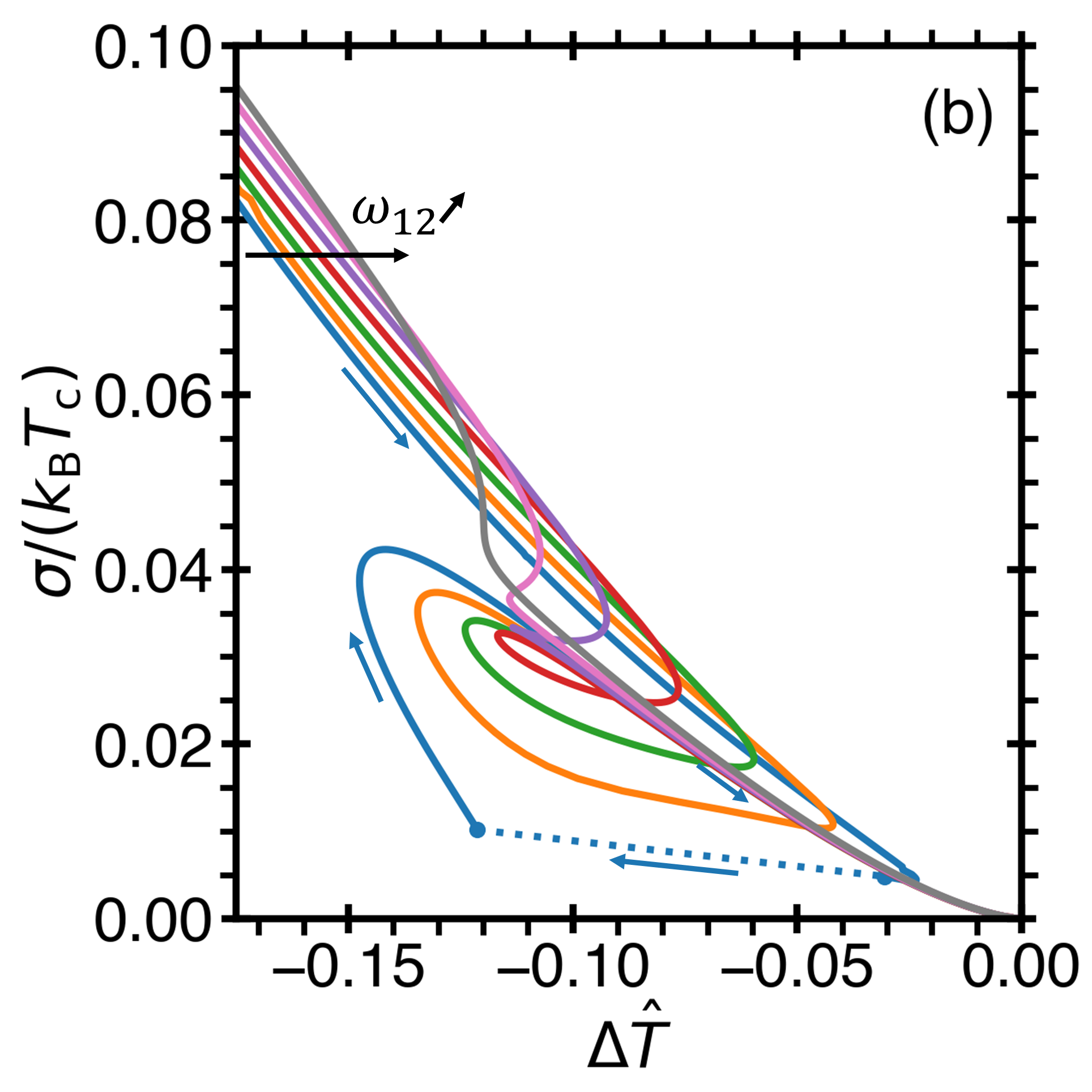}
    \includegraphics[width=0.49\linewidth]{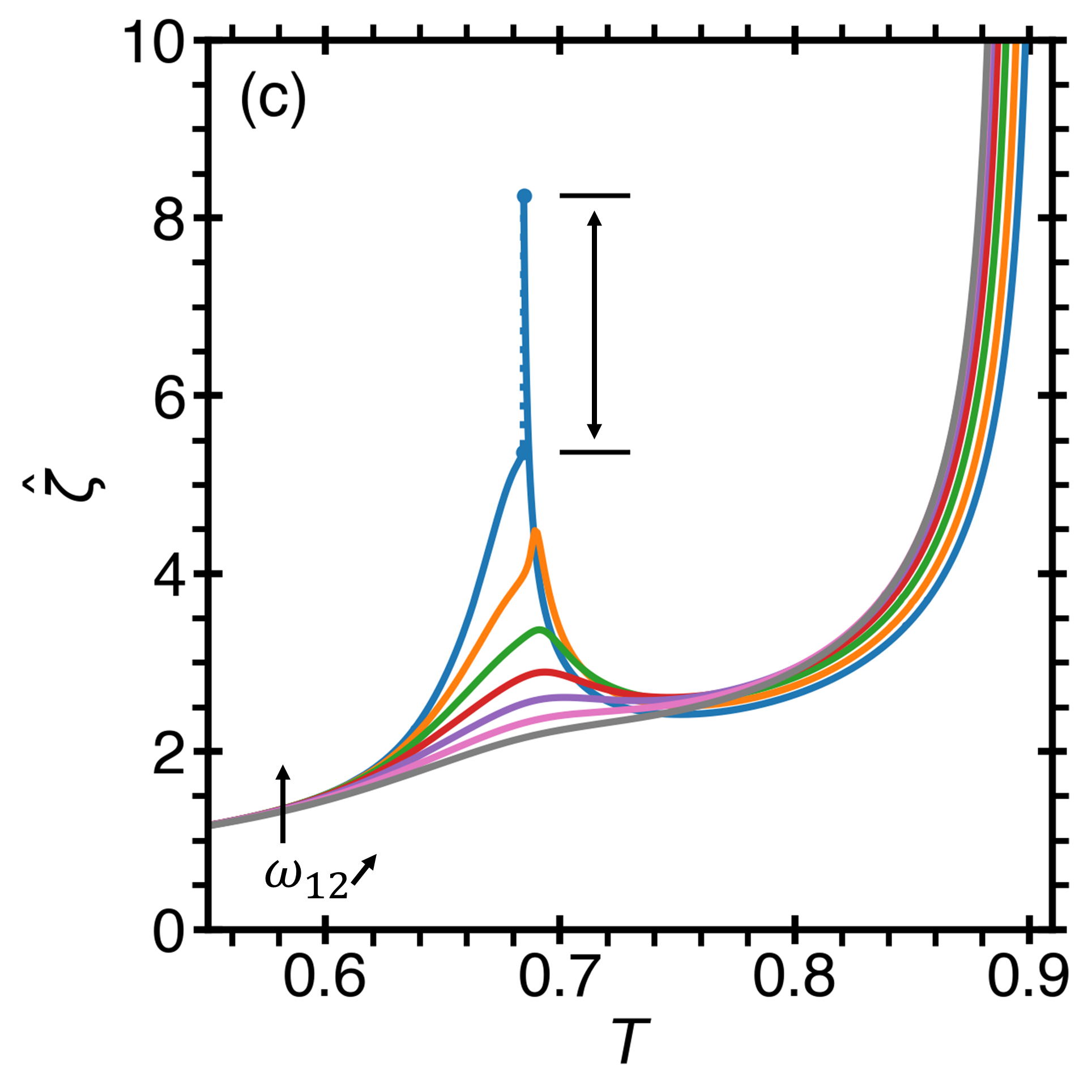}
    \includegraphics[width=0.49\linewidth]{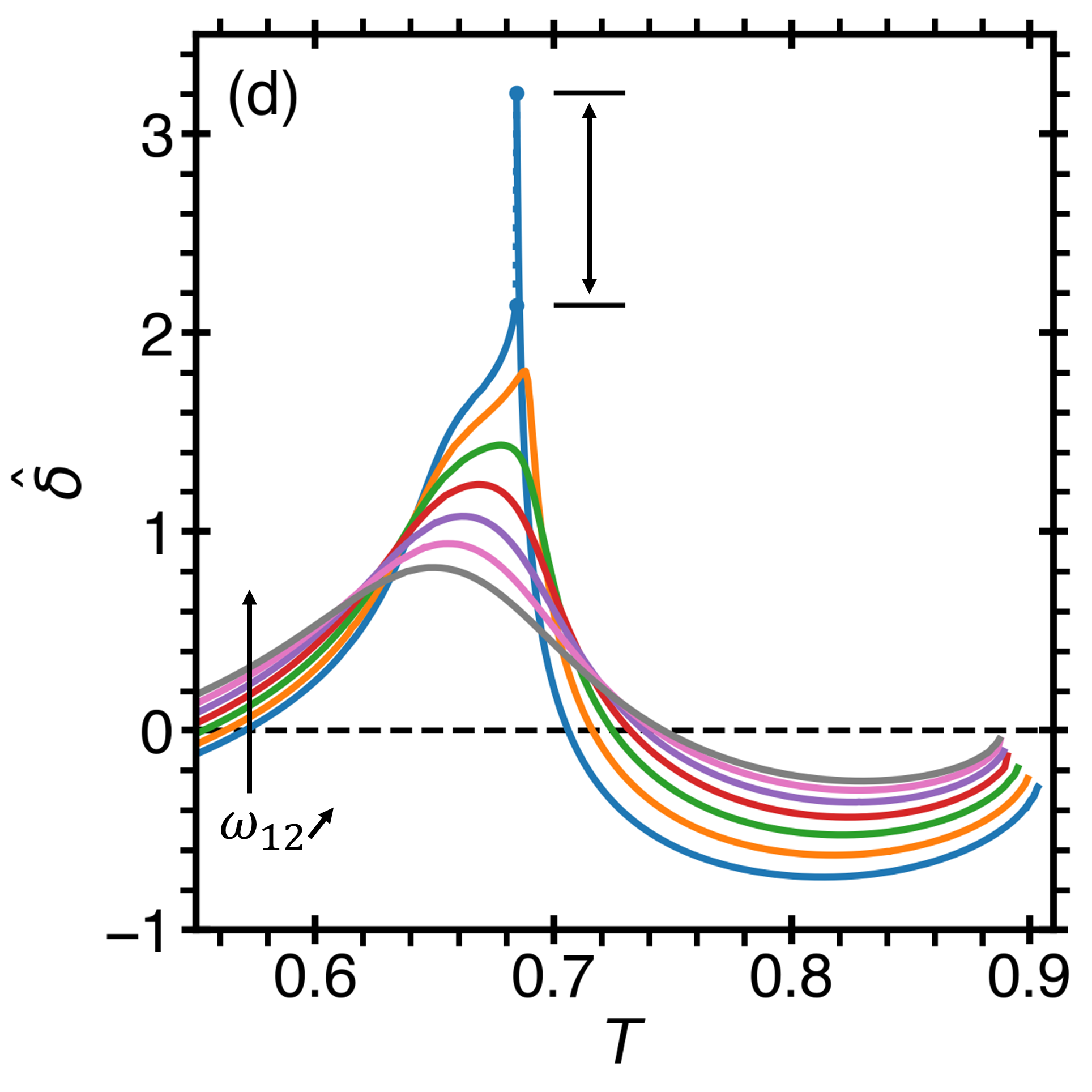}
    \caption{The liquid-vapor interfacial tension as a function of temperature (a), and also presented in reduced units (b), for the system with $\omega_{12}=1.00$ (blue), $\omega_{12}=1.04$ (orange), $\omega_{12}=1.08$ (green), $\omega_{12}=1.12$ (red), $\omega_{12}=1.16$ (purple), $\omega_{12}=1.20$ (pink), and $\omega_{12}=1.24$ (gray). In (b), the critical temperature is given by the ``virtual'' critical point (for the non-reactive binary mixture) for each concentration along the thermodynamic path selected by the interconversion reaction. The blue arrows indicate the direction of warming. (c) The reduced interfacial thickness, $\hat{\zeta}=\zeta/\ell$, and (d) the reduced relative distance between the concentration and density profiles, $\hat{\delta}=\delta/\ell$. In (a-d), the black arrow indicates the direction of increasing $\omega_{12}$, and the dotted lines indicate the discontinuity of the interfacial properties for the system with $\omega_{12}=1.00$ at the triple point, shown by the vertical bars in (c, d).}
    \label{Fig_sigma}
\end{figure}

Using the Fisher-Wortis ansatz, the liquid-vapor interfacial tension along the thermodynamic path (selected by interconversion) as a function of temperature is presented in Fig.~\ref{Fig_sigma}(a) for seven systems. We find that all scenarios exhibit either an inflection point or two extrema. Of the two scenarios that exhibited an inflection point, but not extrema, ($\omega_{12}=1.20$ and $\omega_{12}=1.24$) both were ``singularity free scenarios'' (exhibiting no liquid-liquid phase transition)\cite{Caupin_Polyamorphism_2021} whose thermodynamic path was relatively far away from the liquid-vapor critical line, see Fig.~\ref{Fig_Thermo_Path}. Each of the remaining scenarios exhibit a maximum and minimum depending on the proximity of the selected thermodynamic path to the liquid-vapor critical line, including the singularity-free system with $\omega_{12}=1.16$. During the review of the present work, we became aware	of a recent phenomenological density functional study of another water-like model~\cite{Singh_MaxTension_Water_2022} also reporting two extrema in the temperature dependence of the liquid-vapor interfacial tension. The scenarios for which the liquid-vapor coexistence was interrupted by the triple point ($\omega_{12}=1.00$) exhibit a discontinuity of the liquid-vapor surface tension at this point. 

The reduced interfacial tension, expressed through the distance to the virtual LV critical temperature $\Delta\hat{T}=1-T/[T_\text{c}(x)]$, is illustrated in Fig.~\ref{Fig_sigma}(b). Systems exhibiting two extrema in their interfacial tension demonstrate a ``looping'' pattern as the thermodynamic path approaches and then deviates from the virtual LVCL. As the interconverting systems approach their actual LV critical points, the surface tension asymptotically follows the meanfield power law $\sigma\sim |\Delta\hat{T}|^{3/2}$ (see SM Sec. S9 for details).

In the region where the surface tension reaches a minimum, the interfacial thickness, presented in Fig.~\ref{Fig_sigma}(c), correspondingly reaches a maximum. This phenomenon occurs since the thermodynamic path approaches the virtual LVCL. A DGT treatment of the liquid-vapor interface of real water\cite{Caupin_Cavitation_2005} reported the possibility of a minimum in the temperature dependence of the interfacial thickness (as observed in several cases here), depending on the equation of state used to describe metastable water. We also note that, in particular, for the system with $\omega_{12}=1.00$, the interfacial thickness exhibits a discontinuity at the triple point temperature (see Table~\ref{Table_TP_Params}). We also estimate that the DGT approximation breaks down when the interface becomes sharp. We estimate that a sharp liquid-vapor interface has an interfacial tension of $\sigma_\text{shp} \approx \omega_{11}/8 = 0.2$, which is reached around T = 0.5, where the interfacial thickness becomes proportional to the size of the lattice cell, $\hat{\zeta}=1$ (see SM Sec. S10 for details). For each system in the vicinity of the liquid-vapor critical point, we found that the interfacial thickness followed an asymptotic power law of the from $\hat{\zeta}\sim |\Delta\hat{T}|^{-0.38}$, which deviates from the van der Waals meanfield asymptotic power law\cite{Rowlinson_Capillarity_1982}, $\hat{\zeta}\sim |\Delta\hat{T}|^{-0.5}$, for the thickness of the order-parameter interface (see SM Sec. S9 for details). As predicted by the complete scaling theory \cite{Anisimov_General_1995,Wang_Asymmetry_2006,Wang_Asymmetry_2007}, the order parameter for the compressible binary mixture is a nonlinear combination of $\rho$ and $x$. Thus, the discrepancy in the asymptotic behavior of the interfacial thickness may be attributed to the assumption that the thickness for the density and concentration profiles is the same as for the order parameter in the Fisher-Wortis ansatz, see Eqs.~(\ref{Eq_rho_Profile}) and (\ref{Eq_x_Profile}).

The inflection points of the concentration and density profiles are related through the shift $\delta$, which was included in the concentration profile ansatz, Eq.~(\ref{Eq_x_Profile}). In the first-order approximation, $\delta$ can be separated into symmetric and asymmetric contributions as, $\delta = \delta_\text{sym}+\delta_\text{asym}$. The symmetric contribution is proportional to the difference in the centers of each profile, $\delta_\text{sym}\sim x[\hat{z}=0]-\rho[\hat{z}=0]$, while the asymmetric contribution is proportional to the difference in diameters, $\delta\sim\Delta\hat{x}_\text{d}-\Delta\hat{\rho}_\text{d}$ (see SM Sec. S11 for details). The effects of asymmetry on near-critical interfacial profiles in the scaling theory of inhomogeneous fluids was considered in Ref.\cite{Bertrand_Scaling_2010} In the region of the anomalous behavior of the surface tension, this shift reaches a maximum as illustrated in Fig.~\ref{Fig_sigma}(d). Similarly to the interfacial thickness, the shift is also discontinuous at the triple point temperature (see Table~\ref{Table_TP_Params}). Meanwhile, at the actual LV critical temperature $\delta$ approaches a finite value,  as both the density and concentration profiles become infinitely smooth and the numerical calculation become uncertain due to the large fluctuations (see SM Sec. S8 for details). 

\subsection{Liquid-Liquid Interfacial Tensions}
\begin{figure}[t!]
    \centering
    \includegraphics[width=0.99\linewidth]{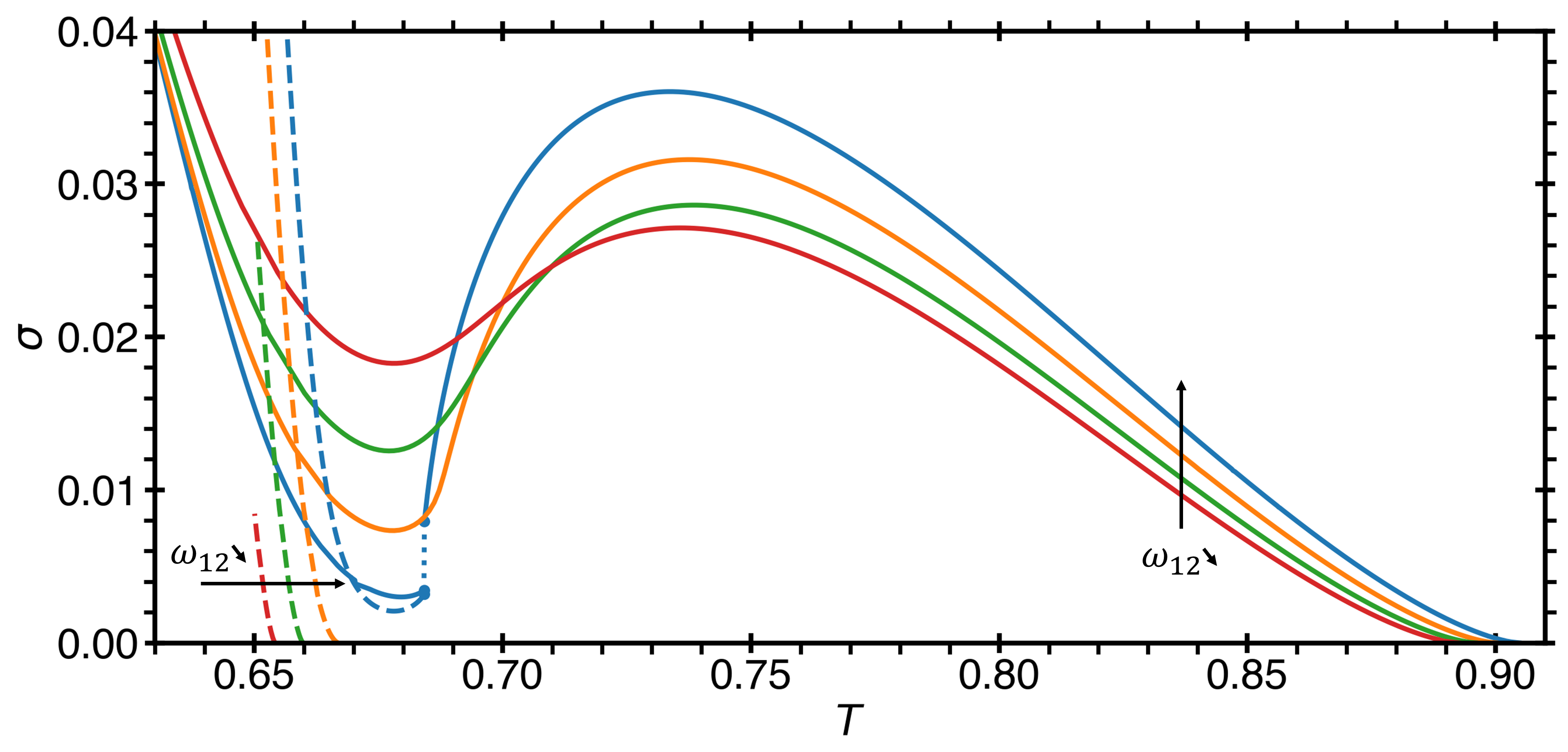}
    \caption{Comparison between the liquid-liquid (dashed curves) and liquid-vapor (solid curves) interfacial tensions as a function of temperature for the system with $\omega_{12}=1.00$ (blue), $\omega_{12}=1.04$ (orange), $\omega_{12}=1.08$ (green), $\omega_{12}=1.12$ (red). The dotted blue line indicates the discontinuity in the liquid-vapor interfacial tension, while the black arrows indicate the direction in which $\omega_{12}$ is decreasing.}
    \label{Fig_LL_n_LV_Comp}
\end{figure}

The liquid-liquid interfacial tensions were calculated for the four systems exhibiting liquid polyamorphism ($\omega_{12}=1.00$, $\omega_{12}=1.04$, $\omega_{12}=1.08$, and $\omega_{12}=1.12$) with use of the Fisher-Wortis ansatzes for the density and concentration profiles, Eqs.~(\ref{Eq_rho_Profile}) and (\ref{Eq_x_Profile}), and are illustrated in comparison with the liquid-vapor interfacial tension in Fig.~\ref{Fig_LL_n_LV_Comp} (more details in SM Sec. S8). We find that for three systems ($\omega_{12}=1.00$, $\omega_{12}=1.04$, and  $\omega_{12}=1.08$), the liquid-liquid interfacial tension crosses that of the liquid-vapor, being larger for lower temperatures \cite{Feeney_Theoretical_2003}. This is different from the conclusion of Feeney and Debenedetti\cite{Feeney_Theoretical_2003} that $\sigma_\mathrm{LL}$ is fundamentally lower than  $\sigma_\mathrm{LV}$ at the same temperature. Indeed, this behavior is observed when the bottleneck in the liquid-vapor coexistence is absent, as in the case of Ref.\cite{Feeney_Theoretical_2003}, or not very deep, as in our model for the system with $\omega_{12}=1.12$. Depending on the choice of parameters in the blinking-checkers model,  $\sigma_\mathrm{LL}$ may be large away from the LLCP; however, as the LLCP is approached, the ratio $\sigma_\mathrm{LL}/\sigma_\mathrm{LV}$ must vanish. 

We note that for the liquid-liquid interfacial tension, the DGT is a good approximation everywhere as the liquid-liquid coexistence approaches infinite pressures before forming a sharp interface between the two liquid phases. We estimate that a sharp interface forms where $\sigma_\text{shp}\approx \omega/8$, which goes from $\sigma_\text{shp} \approx 0.2$ for the system with $\omega_{12}=1.00$ to $\sigma_\text{shp} \approx 0.17$ for the system with $\omega_{12} = 1.12$, which is larger than any of the liquid-liquid interfacial tensions observed in the model (see details in SM Sec. S10). For the system with $\omega_{12} = 1.00$, the liquid-liquid interfacial tension is smaller than both the liquid 1 - vapor or the liquid 2 - vapor interfacial tensions (see Table~\ref{Table_TP_Params} for details). The reduced interfacial thicknesses, $\hat{\zeta}$, and the reduced shifts between the concentration and density profiles, $\hat{\delta}$, are provided in SM Sec. S8. In particular, we note that for the systems that reach a liquid-liquid critical point ($\omega_{12}=1.04$, $\omega_{12}=1.08$, and $\omega_{12}=1.12$), the liquid-liquid interfacial tension does not demonstrate any anomalous behavior. Furthermore, the interfacial tensions and interfacial thicknesses follow the predicted meanfield asymptotic power laws (see SM Sec. S9 for details). Moreover, in the system with $\omega_{12}=1.00$, the liquid-liquid interfacial tension exhibits a minimum prior to the triple point temperature.

\subsection{Interfacial Profiles}
\begin{figure}[tp!]
    \centering
    \includegraphics[width=0.49\linewidth]{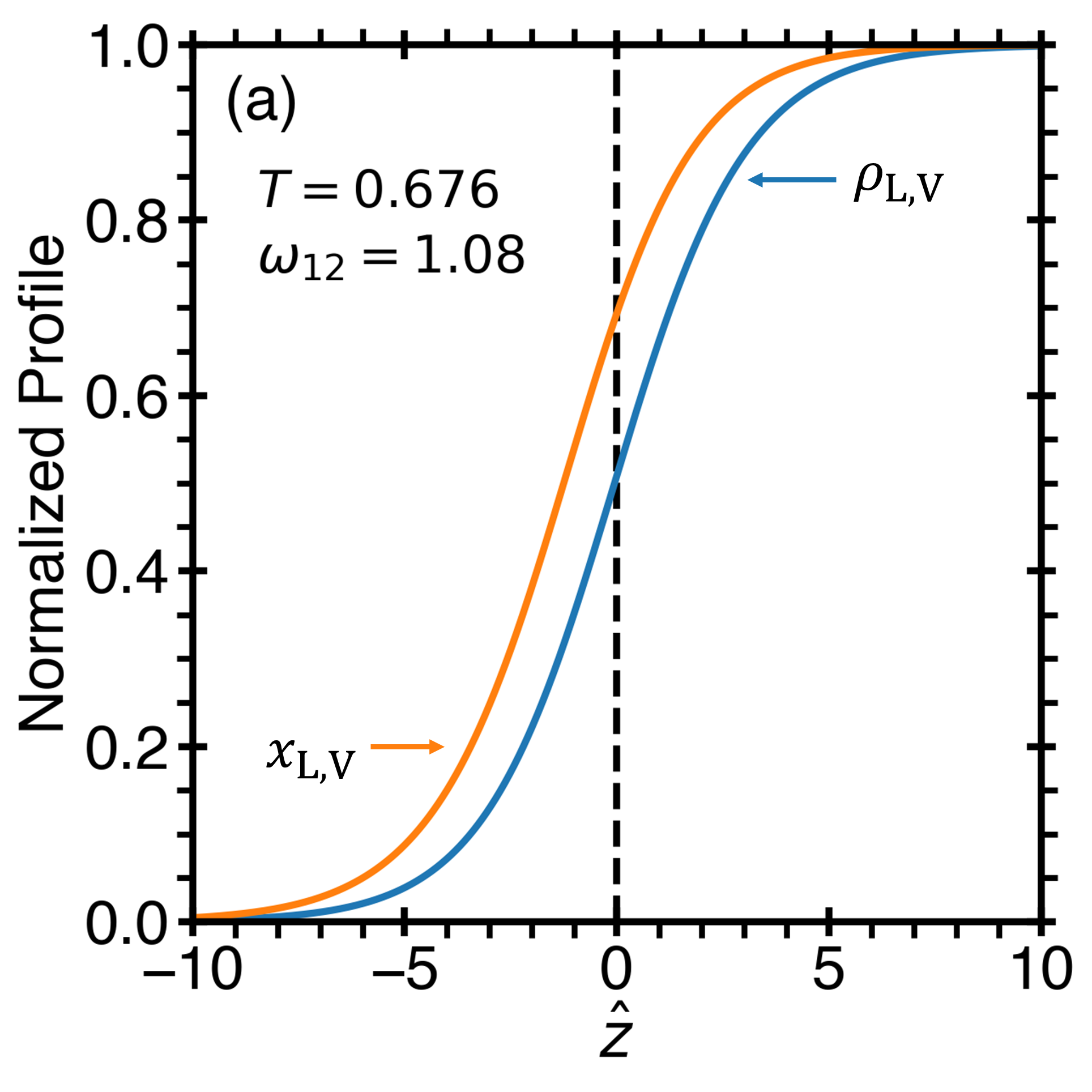}
    \includegraphics[width=0.49\linewidth]{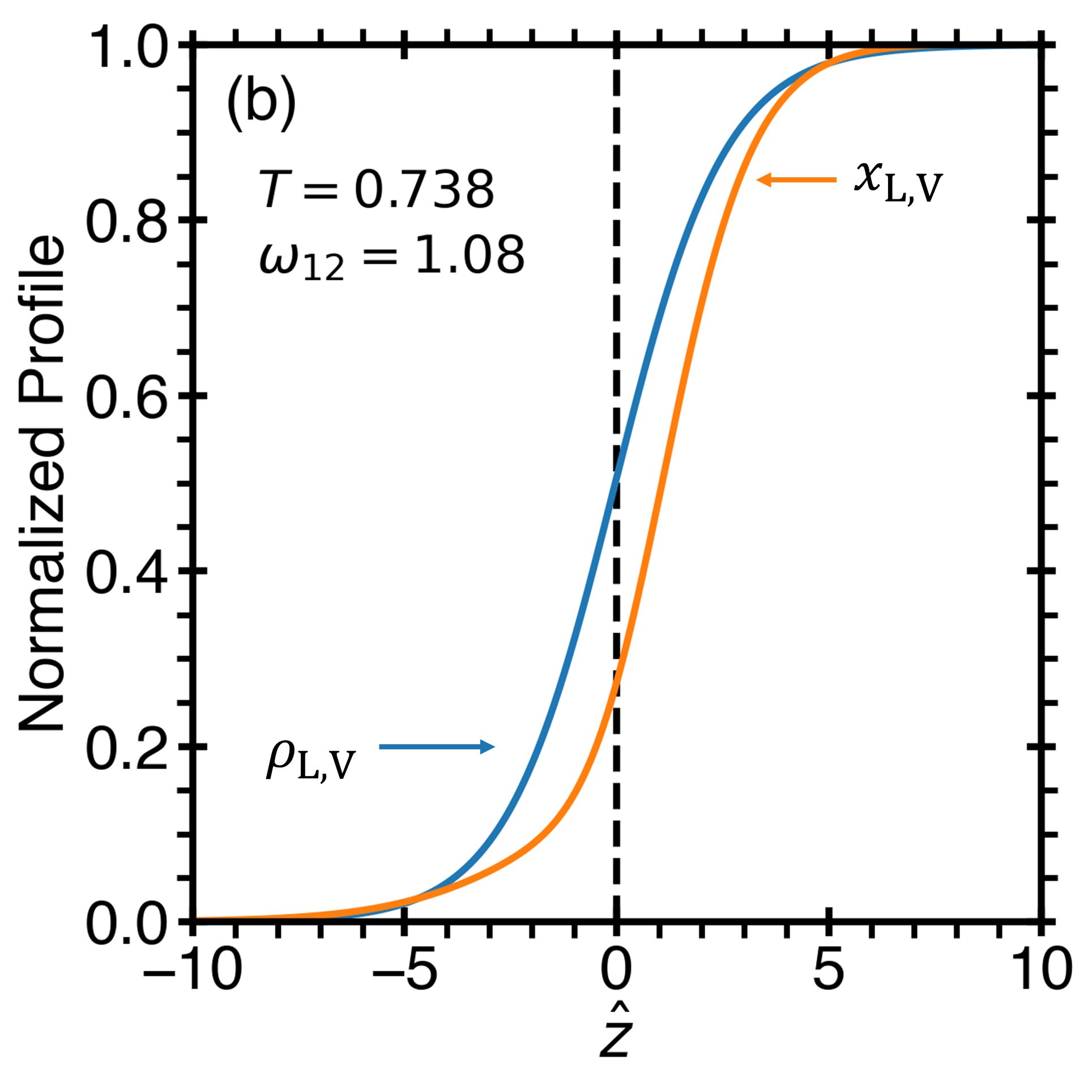}
    \includegraphics[width=0.49\linewidth]{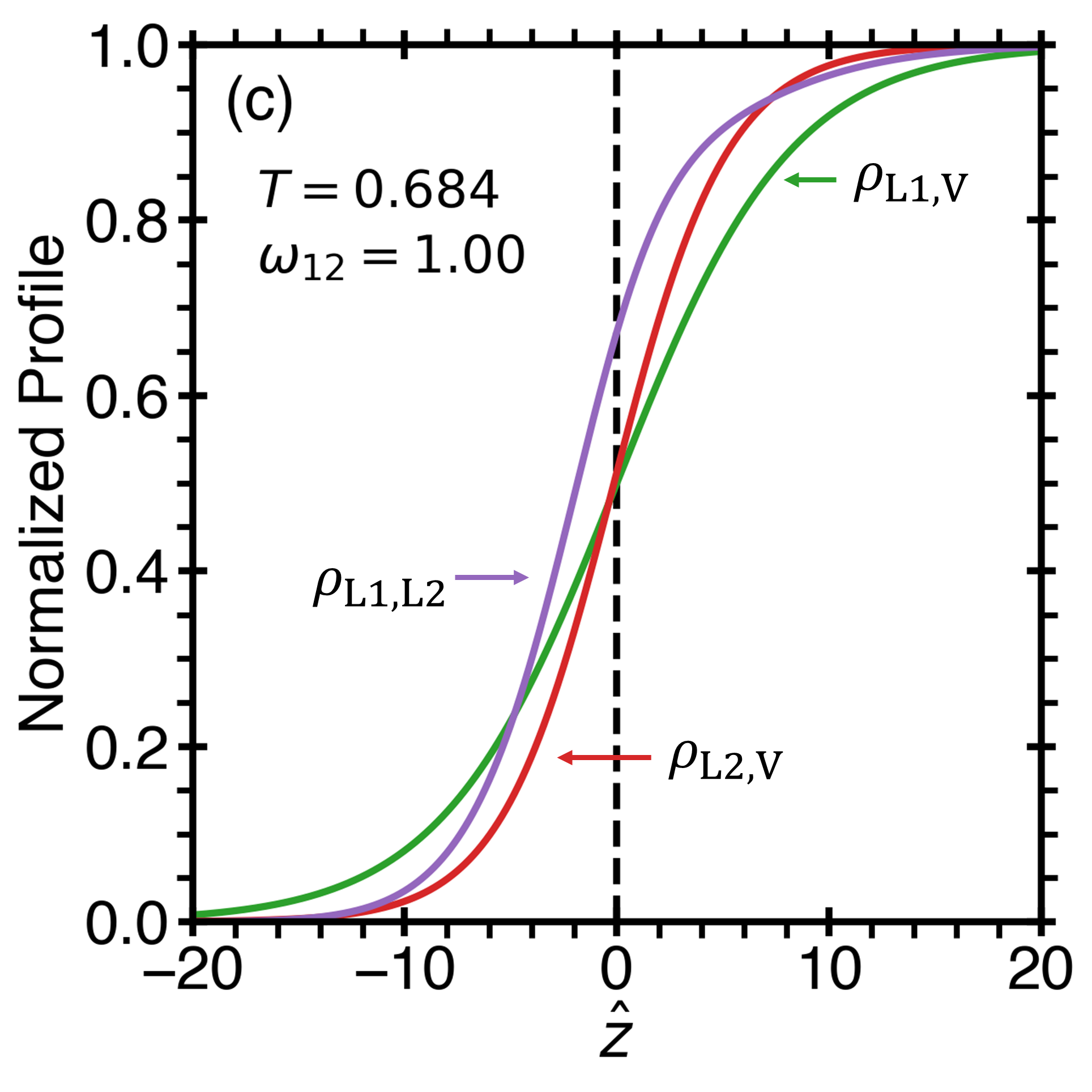}
    \includegraphics[width=0.49\linewidth]{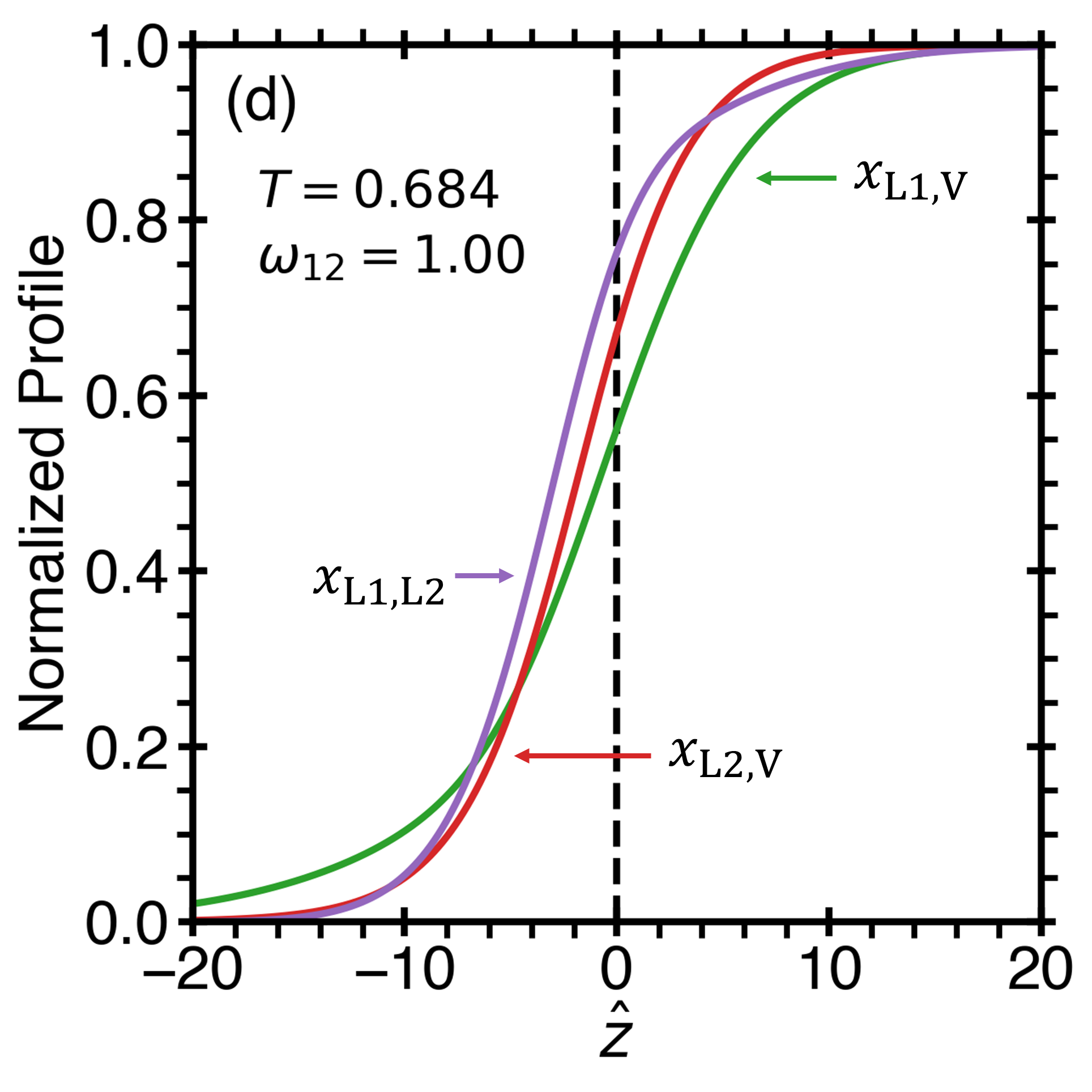}
    \caption{Normalized density and concentration liquid-vapor profiles as a function of the coordinate perpendicular to the planar interface, $\hat{z}=z/\ell$, given by Eqs.~(\ref{Eq_rho_Profile}) and (\ref{Eq_x_Profile}) for the system with $\omega_{11}=1.6$, $\omega_{22}=2.0$, $\omega_{12}=1.08$, $e=3$, and $s=4$ at the two temperatures (a, b) that correspond to the two extrema of the liquid-vapor interfacial tension (shown in Fig.~\ref{Fig_sigma}). Normalized (c) density and (d) concentration profiles for three-phase coexistence at the triple point, $T_\text{TP}=0.6843$, for the system with $\omega_{11}=1.6$, $\omega_{22}=2.0$, and $\omega_{12}=1.00$.}
    \label{Fig_TP_Profile}
\end{figure}

We now investigate the interfacial profiles for density and concentration. Figure~\ref{Fig_TP_Profile}(a) and (b) show the interfacial profiles predicted from the Fisher-Wortis ansatzes, Eqs.~(\ref{Eq_rho_Profile}) and (\ref{Eq_x_Profile}), for the system with $\omega_{12}=1.08$ at the temperatures that correspond to the maximum and minimum of the LV interfacial tension. We find that at the minimum, the interfacial profiles are relatively symmetric, while at the maximum, the concentration profile contains a large asymmetric contribution. The large asymmetry predicted by the Fisher-Wortis concentration ansatz occurs since the diameter of the concentration, $\Delta\hat{x}_\text{d}$, reaches a maximum at this temperature (see more details in SM Sec. S7). 

\begin{table}[htb]
	\caption{The surface tension $\sigma$, normalized interfacial thickness $\hat{\zeta}=\zeta/\ell$, and normalized shift $\hat{\delta}=\delta/\ell$ for the three coexisting phases (liquid 1, liquid 2, and vapor) for the system with $\omega_{12}=1.00$ at the triple point temperature ($T=0.6843$).}
	\begin{tabular}{lccc}
		\toprule
		& $\sigma$ & $\hat{\zeta}$ & $\hat{\delta}$ \\ \midrule
		L1-V   & 0.00793  & 8.242         & 3.203          \\
		L2-V   & 0.00334  & 5.372         & 2.145          \\
		L1-L2  & 0.00321  & 6.040         & 0.983       \\ \bottomrule 
	\end{tabular}
	\label{Table_TP_Params}
\end{table}

For the system with $\omega_{12}=1.00$, at the triple point temperature, $T_\text{TP}=0.6843$, all of the interfacial properties exhibit a discontinuity (see Table~\ref{Table_TP_Params}). Since the interfacial tension of the L1-V (low-density-liquid - vapor) interface is much larger than the other two interfacial tensions, then in accordance with Antonov's rule\cite{Rowlinson_Capillarity_1982}, $\sigma_\text{L2,V}+\sigma_\text{L1,L2} < \sigma_\text{L1,V}$, we predict that the L1-V interface will be enriched (wetted) by the L2 (high-density-liquid) phase to reduce the energetically unfavorable L1-V interface. This indicates that the non-monotonic behavior of the liquid-vapor interfacial tension may be caused by the surface enrichment of the L2-V coexistence by species 1. This behavior was confirmed by the interfacial profiles obtained in MC simulations of the blinking-checkers model near the minimum of the interfacial tension (see SM Sec. S12 for details). Notwithstanding this complete wetting phenomenon, we display in Fig.~\ref{Fig_TP_Profile}(c) and (d) the interfacial profiles for the density and concentration of the three coexisting phases at the triple point.

\newpage
\subsection{Conditions for Anomalous Interfacial Behavior}
We note that, based on the findings presented in this work, no general conclusion about the absence of a liquid-liquid transition can be drawn from the existence of an inflection point in the liquid-vapor interfacial tension. For instance, on one hand, in the present work, the inflection point is observed only for singularity free scenarios; on the other hand, the TIP4P/2005 model of water exhibits an inflection point\cite{Malek_Surface_2019,Wang_Second_2019,Gorfer_Highdensity_2022}, while it is thought to possess a liquid-liquid transition terminated by a critical point~\cite{Abascal_Widom_2010,Biddle_Twostructure_2017,Debenedetti_Second_2020}. We emphasize that the anomaly in the temperature dependence of the interfacial tension is linked to the anomaly in the liquid-vapor coexistence along the thermodynamic path selected by interconversion, and originates in the region where the equilibrium fraction of species, $x_{\text{e}}$, most dramatically changes, a concept that was first suggested by Hruby and Holten\cite{Hruby_Twostructure_2004}. It follows from our results that the shape of the LV coexistence is affected by two factors: the proximity to the virtual LVCL and the existence of the liquid-liquid phase transition. We note that this result is elucidated through the simplicity of the blinking-checkers lattice model, and may be observed in more complex (microscopic) models of fluids exhibiting polyamorphism and water-like anomalies. For instance, the anomalies in supercooled water can be interpreted as the results of the interconversion of two supramolecular structures. This interconversion occurs only at low temperatures, extremely far away from the liquid-vapor critical point. Thus, there is a maximum (at \SI{4}{\celsius}) and a minimum (cut-off by the limit of spontaneous ice formation) only in the liquid branch of the LV coexistence in water\cite{Gallo_Water_2016}. Consequently, the LV surface tension may exhibit a maximum or an inflection point depending on the depth of the minimum~\cite{Feeney_Theoretical_2003}.

%\section{Complete Scaling of Fluid Mixtures}

\section{Conclusion}
We investigated the interfacial properties of fluids exhibiting liquid polyamorphism and/or water-like anomalies modeled through a compressible binary lattice with molecular interconversion of species. We demonstrated that the change in the equilibrium fraction of the interconverting species as a function of temperature is the origin of various thermodynamic anomalies, \textit{e.g.} in density and in surface tension. We found that due to the proximity of the thermodynamic path, selected by the interconversion of species, to the liquid-vapor critical line of the non-reacting binary mixture, the liquid-vapor interfacial tension demonstrates an anomalous temperature dependence, exhibiting either two extrema or an inflection point. In the anomalous region, where the liquid-vapor interfacial tension exhibits a minimum, the interfacial thickness and the relative distance between the density and concentration interfacial profiles exhibit a maximum. Moreover, in the scenario where the fluid possesses a triple point between the three coexisting fluid phases, we predict a discontinuity in the interfacial properties as well as
complete wetting of the low-density-liquid and vapor interface by the high-density-liquid phase.

\section*{Supporting Information}
Further description and calculations of the blinking-checkers model, a comparison of the exact solution of the interfacial properties with phenomenological asymmetric ansatzes, and results of Monte Carlo simulations.

\begin{acknowledgement}
This work was inspired by the studies of the interfacial properties of polyamorphic liquids by Pablo Debenedetti and his coworkers. The authors thank Nikolay Shumovskyi for useful discussions on the Monte Carlo simulations of the blinking-checkers model. T.J.L. acknowledges the support of the Chateaubriand Fellowship of the Office for Science \& Technology of the Embassy of France in the United States. The research of T.J.L. and M.A.A. was partially supported by NSF award no. 1856479. The research of S.V.B. was supported by NSF Award No. 1856496, and partially supported through the Bernard W. Gamson Computational Science Center at Yeshiva College. F.C. acknowledges the support of the Agence Nationale de la Recherche, Grant No. ANR-19-CE30-0035-01. 
\end{acknowledgement}

\bibliography{ref}

\end{document}

\newpage
% Print the table of contents on a separate page
\tableofcontents
\newpage

%If included this into the main text, use the environment:
%\begin{suppinfo}
%\end{suppinfo}

\section{Equilibrium Interconversion Condition for the Blinking-Checkers Model}

The Helmholtz free energy of the non-reacting binary lattice model, $f$ (Eq.~(1) in the main text), may be expressed through the partial densities, $\rho_1 =\rho x$ and $\rho_2 = \rho (1-x)$, as given by
\begin{equation}\label{SM_Eq_GenFree}
    \begin{split}
        f(T,\rho_1,\rho_2) =& \varphi_1^\circ(T)\rho_1 + \varphi_2^\circ(T)\rho_2 - (\rho_1 + \rho_2)(\omega_{11} \rho_1 + \omega_{22} \rho_2) + \omega \rho_1 \rho_2 \\ &+ T [\rho_1 \ln \rho_1 + \rho_2 \ln \rho_2 + (1 - \rho_1 - \rho_2)\ln(1 - \rho_1 - \rho_2)]
    \end{split}
\end{equation}
where $\varphi_1^\circ(T)$ and $\varphi_2^\circ(T)$ are two temperature functions, which depend on the arbitrary choices of zero energy and zero entropy\cite{Anisimov_Polyamorphism_2018}, while $\omega = \omega_{11}+\omega_{22}-2\omega_{12}$. The chemical potentials of each species in solution is given by
\begin{align}
    \mu_1 = \frac{\partial f}{\partial \rho_1}\bigg|_{\rho_2,T} &= \varphi_1^\circ(T) - 2\omega_{11}\rho_1 - (\omega_{11}+\omega_{22}-\omega)\rho_2 + T\ln\left(\frac{\rho_1}{1-\rho_1-\rho_2}\right) \label{SM_Eqn_mu1} \\
    \mu_2 = \frac{\partial f}{\partial \rho_2}\bigg|_{\rho_1,T} &= \varphi_2^\circ(T) - (\omega_{11}+\omega_{22}-\omega)\rho_1 - 2\omega_{22}\rho_2 + T\ln\left(\frac{\rho_2}{1-\rho_1-\rho_2}\right) \label{SM_Eqn_mu2}
\end{align}
The difference between the chemical potentials, $\mu_{12}= \mu_1-\mu_2$, for the non-reacting binary lattice model in solution is given by
\begin{equation}
    \mu_{12} = \varphi_{12}^\circ - 2(\omega_{11} -\omega_{12})\rho_1 + 2(\omega_{22}-\omega_{12})\rho_2 +T\ln\left(\frac{\rho_1}{\rho_2}\right)
\end{equation}
where $\varphi_{12}^\circ = \varphi_1^\circ-\varphi_2^\circ$. The chemical-reaction equilibrium condition requires (as described in Sec. 2.1 of the main text), $\partial f/\partial x|_{T,\rho}=0$, or equivalently, $\mu_{12}=0$. Thus, in the interconverting blinking-checkers lattice model, $\varphi_{12}^\circ$ is defined as
\begin{equation}
    \varphi_{12}^\circ = -(e-Ts)
\end{equation}
where $e$ and $s$ are the energy and entropy change of the reaction respectively. %We note that in this model, there is no volume change of the reaction or any other higher ordered terms (like the isobaric expansivity change, isothermal compressibility change, or heat capacity change of reaction)\cite{Anisimov_Polyamorphism_2018}. 

In this work, all units are expressed in a dimensionless form relative to the species 2 - species 2 energy of interactions, $\epsilon_{22}$. Thus, the units of energy, temperature, pressure, and surface tension are, respectively, $-z\epsilon_{22}/4$, $-z\epsilon_{22}/(4k_\text{B})$, $-z\epsilon_{22}/(4\ell^3)$, and $-z\epsilon_{22}/(4\ell^2)$, in which $z$ is the number of nearest neighbors, $\ell$ is the length of a lattice cell, and we adopt Boltzmann's constant, $k_\text{B}$, as $k_\text{B} = 1$. The number density, $\rho$, and fraction of particles of species 1, $x$, are expressed through non-dimensional quantities of the form $\rho = (N_1 + N_2)/N$ and $x = N_1/(N_1 + N_2)$, where $N$, $N_1$, and $N_2$, are the number of lattice sites, particles of type 1, and particles of type 2, respectively. All distances are reduced by the length of a lattice cell ($\ell$), and to indicate this, the corresponding quantities are expressed with a ``$\wedge$''.

\clearpage
\newpage

\section{Liquid-Vapor and Liquid-Liquid Coexistence}

\begin{figure}[h!]
    \centering
    \includegraphics[width=0.49\linewidth]{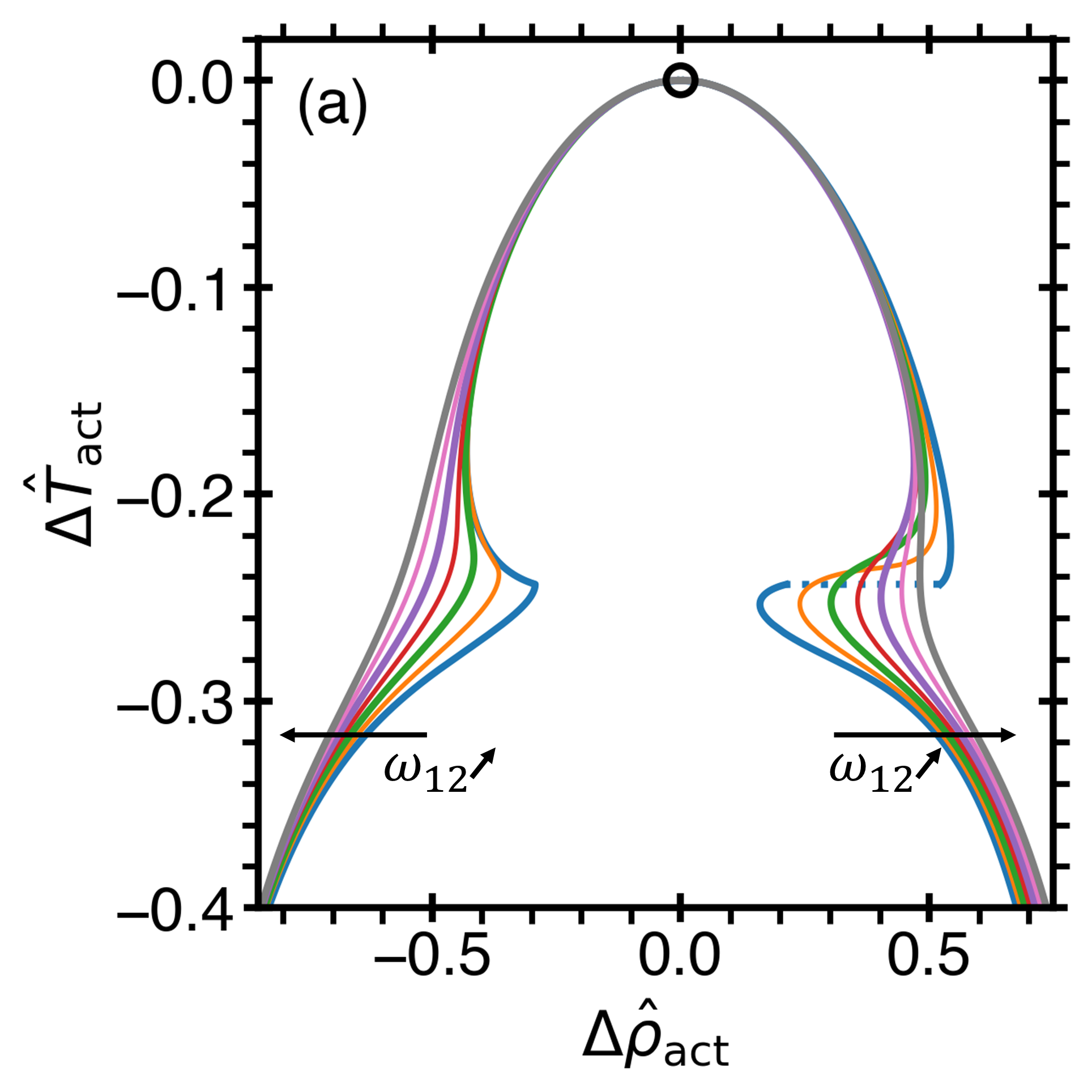}
    \includegraphics[width=0.49\linewidth]{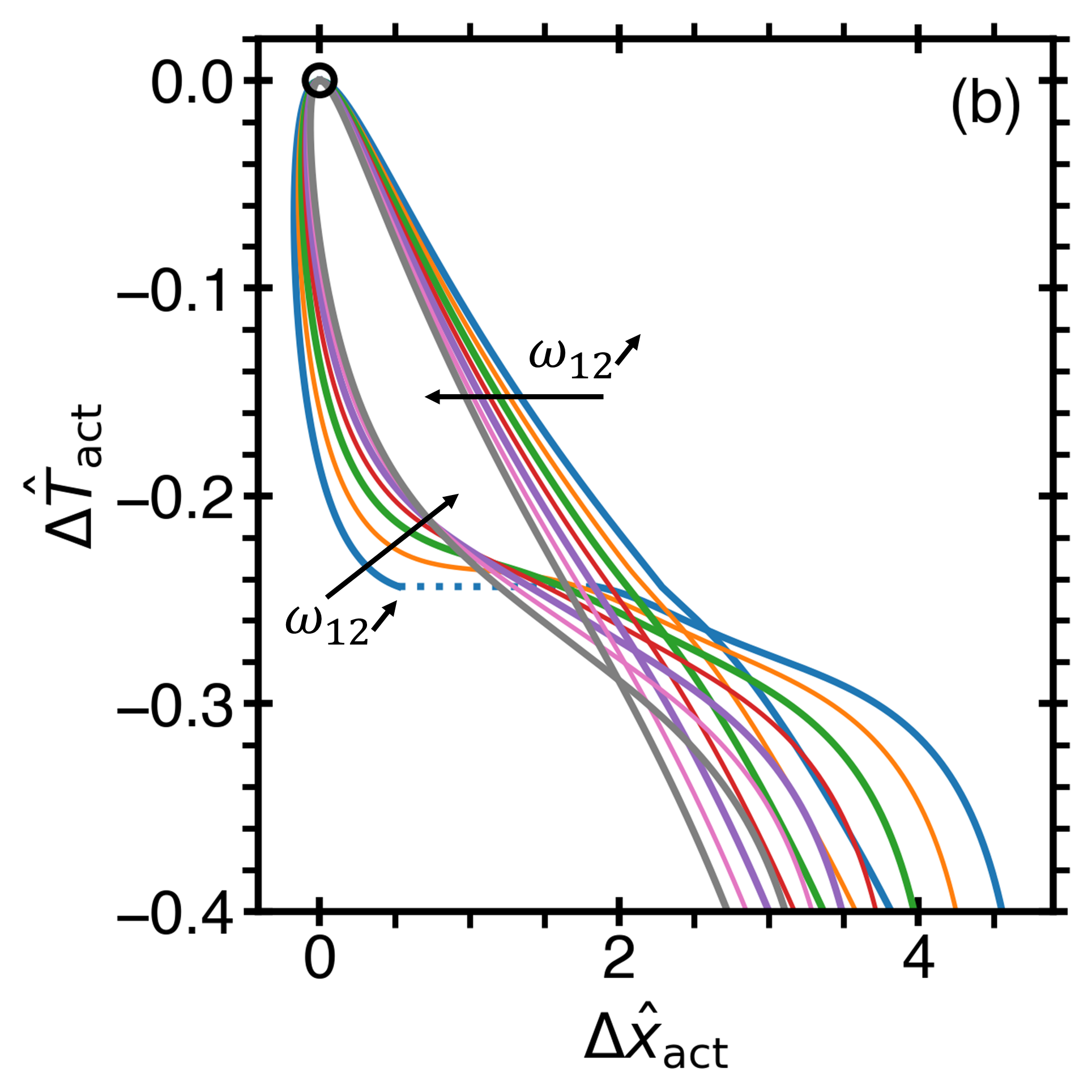}
    \includegraphics[width=0.49\linewidth]{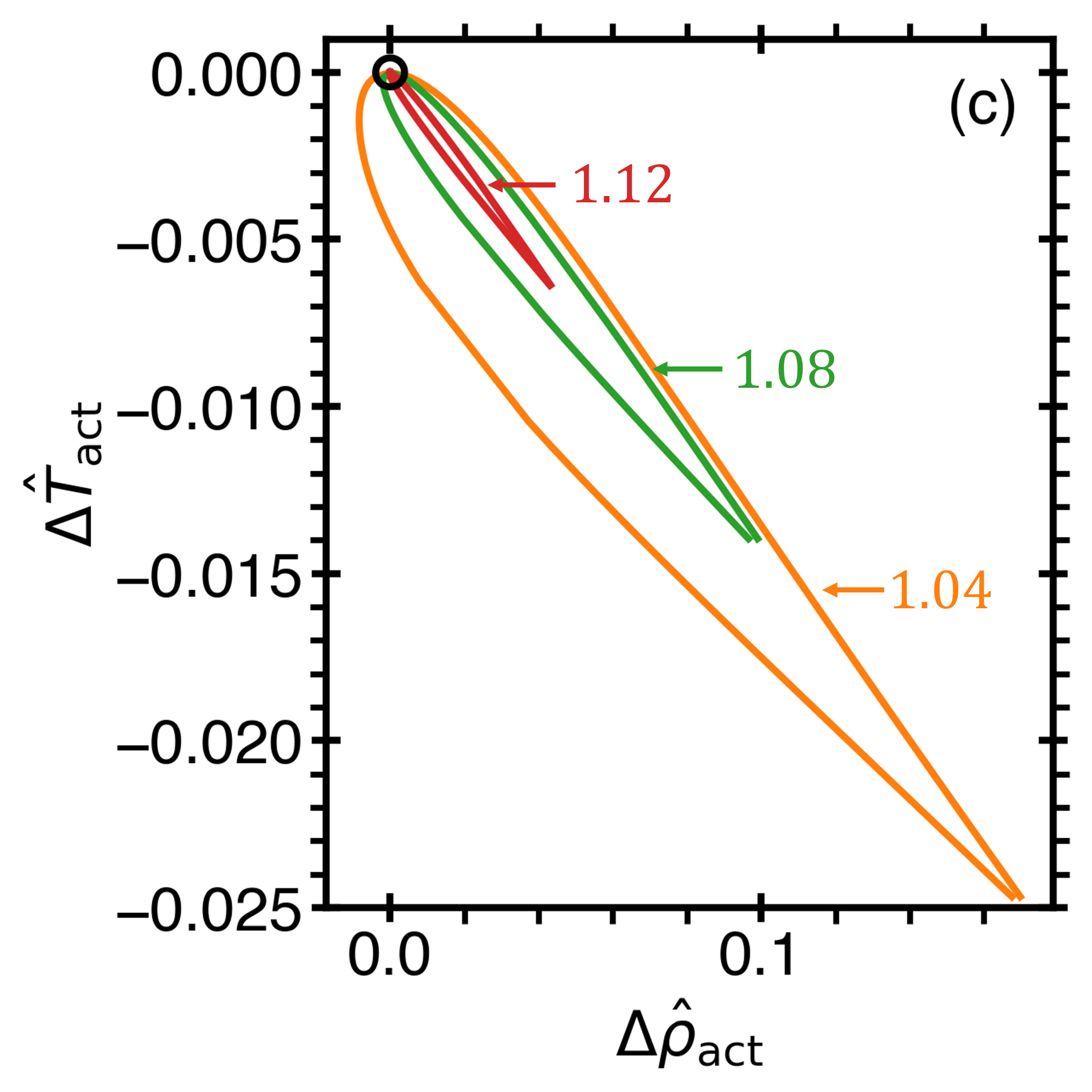}
    \includegraphics[width=0.49\linewidth]{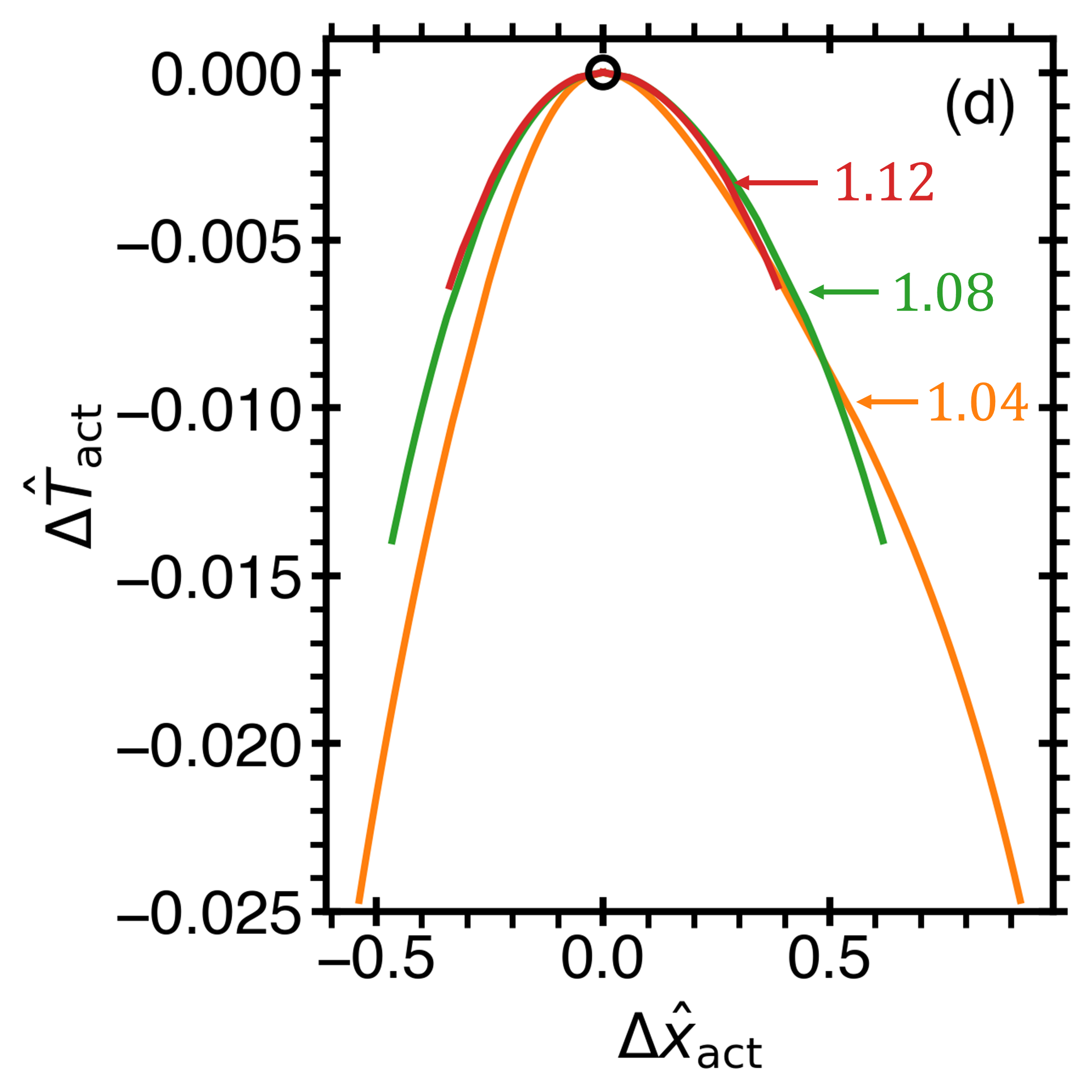}
    \caption{Liquid-vapor (a,b) and liquid-liquid (c,d) coexistence curves for the seven systems with $\omega_{11}=1.6$, $\omega_{22}=2.0$, $e=3$, $s=4$, and with various values of $\omega_{12}$: $\omega_{12}=1.00$ (blue), $\omega_{12}=1.04$ (orange), $\omega_{12}=1.08$ (green), $\omega_{12}=1.12$ (red), $\omega_{12}=1.16$ (purple), $\omega_{12}=1.20$ (pink), and $\omega_{12}=1.24$ (gray) as considered in the main text. (a,c) The reduced temperature, $\Delta\hat{T}_\mathrm{act}=1-T/T_\mathrm{c}^\mathrm{act}$, vs reduced density, $\Delta\hat{\rho}_\mathrm{act}=1-\rho/\rho_\mathrm{c}^\mathrm{act}$, and (b,d) the reduced temperature vs reduced concentration, $\Delta\hat{x}_\mathrm{act}=1-x/x_\mathrm{c}^\mathrm{act}$, where $T_\mathrm{c}^\mathrm{act}$, $\rho_\mathrm{c}^\mathrm{act}$, and $x_\mathrm{c}^\mathrm{act}$ are the ``actual'', as a result of interconversion, LVCP (or LLCP), provided in Tables~\ref{Table_SM_LVCPs} and \ref{Table_SM_LLCPs}. In (a,b), the arrows indicate the direction of increasing $\omega_{12}$.}
    \label{Fig_SM_LVcoex}
\end{figure}

\clearpage

%\begin{figure}[h!]
%    \centering
%    \includegraphics[width=0.49\linewidth]{SM_Figures/LL_T_rho_coex.png}
%    \includegraphics[width=0.49\linewidth]{SM_Figures/LL_T_x_coex.png}
%    \caption{Liquid-liquid temperature-density (a) and temperature-concentration (b) coexistence curves for the four systems (with $\omega_{11}=1.6$, $\omega_{22}=2.0$, $e=3$ and $s=4$) exhibiting liquid polyamorphism: $\omega_{12}=1.00$ (blue), $\omega_{12}=1.04$ (orange), $\omega_{12}=1.08$ (green), $\omega_{12}=1.12$ (red). The liquid-liquid critical points are given in Table~\ref{Table_SM_LLCPs}.}
%    \label{Fig_SM_LLcoex}
%\end{figure}

\begin{table}[h!]
\begin{tabular}{lcccc}
\toprule
$\omega_{12}$ & $T_\mathrm{c}^\mathrm{act}$ & $\rho_\mathrm{c}^\mathrm{act}$ & $x_\mathrm{c}^\mathrm{act}$ & $P_\mathrm{c}^\mathrm{act}$ \\ \midrule
1.00          & 0.905          & 0.548             & 0.174          & 0.208          \\
1.04          & 0.900          & 0.546             & 0.183          & 0.204          \\
1.08          & 0.896          & 0.544             & 0.193          & 0.201          \\
1.12          & 0.893          & 0.541             & 0.202          & 0.198          \\
1.16          & 0.890          & 0.538             & 0.211          & 0.195          \\
1.20          & 0.889          & 0.534             & 0.219          & 0.192          \\
1.24          & 0.888          & 0.531             & 0.227          & 0.189      \\
\bottomrule
\end{tabular}
\caption{Actual liquid-vapor critical points of interconverting systems, referred to as ``actual'' critical points in the main text, for the seven systems considered in this work (with $\omega_{11}=1.6$, $\omega_{22}=2.0$, $e=3$, and $s=4$).}
\label{Table_SM_LVCPs}
\end{table}

\begin{table}[h!]
\begin{tabular}{lcccc}
\toprule
$\omega_{12}$ & $T_\mathrm{c}^\mathrm{act}$ & $\rho_\mathrm{c}^\mathrm{act}$ & $x_\mathrm{c}^\mathrm{act}$ & $P_\mathrm{c}^\mathrm{act}$ \\ \midrule
1.04          & 0.667          & 0.851             & 0.419          & 0.211          \\
1.08         & 0.660          & 0.904             & 0.463          & 0.358          \\
1.12          & 0.654          & 0.958             & 0.488          & 0.726        \\ \bottomrule 
\end{tabular}
\caption{Liquid-liquid critical points (LLCP) for the three systems exhibiting liquid polyamorphism and a LLCP (with $\omega_{11}=1.6$, $\omega_{22}=2.0$, $e=3$, and $s=4$).}
\label{Table_SM_LLCPs}
\end{table}

We note the difference between the critical points selected by the thermodynamic path, Tables~\ref{Table_SM_LVCPs} and \ref{Table_SM_LLCPs}, which we refer to as ``actual'' critical points, and the critical points of the non-reacting binary mixture, each of which is connected to a corresponding point along the thermodynamic path by a critical isochore at fixed composition, which we refer to as ``virtual'' critical points (see sec 2.2 in the main text). The virtual critical points are defined for the binary mixture at each overall concentration, $x$, and overall density, $\rho$, through the conservation equation
\begin{equation}\label{SM_Eq_Overallx}
    \rho_\mathrm{V} x_\mathrm{V}\upsilon_\mathrm{V} + \rho_\mathrm{L} x_\mathrm{L} \upsilon_\mathrm{L} = \rho x
\end{equation}
where $\rho_\mathrm{V}$, $x_\mathrm{V}$, $\rho_\mathrm{L}$, and $x_\mathrm{L}$ are the liquid and vapor coexisting values of the density and concentration, while $\upsilon_\mathrm{V}$ and $\upsilon_\mathrm{L}$ are the volumes of the liquid (L) or vapor (V) phases. Eliminating the volumes via $\upsilon_\mathrm{V} + \upsilon_\mathrm{L} = 1$ and $\rho_\mathrm{V}\upsilon_\mathrm{V} + \rho_\mathrm{L}\upsilon_\mathrm{L} = \rho$ in Eq.~(\ref{SM_Eq_Overallx}), gives the overall concentration and density in terms of just the coexisting parameters as
\begin{equation}\label{SM_Eq_overallxSimple}
    x_\mathrm{V}\rho\rho_\mathrm{V} - x_\mathrm{L}\rho\rho_\mathrm{L} + x_\mathrm{L}\rho_\mathrm{L}\rho_\mathrm{V} -x_\mathrm{V}\rho_\mathrm{L}\rho_\mathrm{V} = x\rho(\rho_\mathrm{V}-\rho_\mathrm{L})
\end{equation}
Equation~(\ref{SM_Eq_overallxSimple}) defines the overall density and concentration for any set of coexisting densities and concentrations. For the non-reacting binary mixture each overall fraction corresponds to a plane of coexisting values. However, for the interconverting binary-mixture, the chemical-reaction equilibrium condition selects only a single value of the overall fraction for each point along the liquid-vapor coexistence. This means that the thermodynamic behavior of the reacting system may be reproduced in a series of non-reacting binary mixtures by carefully preparing systems for each overall concentration at the equilibrium conditions determined from the reacting system. Consequently, each of these non-reacting binary mixtures, prepared for each overall concentration, has its own critical point. Expressing the free energy through $\rho_1=\rho x$ and $\rho_2 = \rho(1-x)$, as in Eq.~(\ref{SM_Eq_GenFree}), and introducing the notation, $f_y = \partial f/\partial y$, the critical line is determined from the following thermodynamic stability conditions,
\begin{equation}
    f_{\rho_1\rho_1}f_{\rho_2\rho_2} - f^2_{\rho_1\rho_2} = 0
\end{equation}
\begin{equation}
    f_{\rho_1\rho_1\rho_1}\left(\frac{f_{\rho_1\rho_2}}{f_{\rho_1\rho_1}}\right)^3 -3f_{\rho_1\rho_1\rho_2}\left(\frac{f_{\rho_1\rho_2}}{f_{\rho_1\rho_1}}\right)^2 + 3f_{\rho_1\rho_2\rho_2}\left(\frac{f_{\rho_1\rho_2}}{f_{\rho_1\rho_1}}\right) - f_{\rho_2\rho_2\rho_2} = 0
\end{equation}
see SM of Ref.\cite{Caupin_Polyamorphism_2021} for more details. Collectively, these critical points, realized only for the non-reacting binary system, make up the virtual critical line in the reacting system. At each point along the thermodynamic path, the behavior of the thermodynamic properties, being state functions, are connected to each unique virtual critical point by critical isochores at constant composition.

\section{Monte Carlo Simulations of Blinking-Checkers Model: Preliminary Results}\label{SM_Sec_MC_Results}

We perform Monte-Carlo simulations of the blinking-checkers model on a cubic lattice with $L\times L\times L$ sites with periodic boundary conditions. In these preliminary calculations, we investigate a system of size $L=64$. Each lattice site, $i$, may be in one of three possible states: $s_i=0$ (empty), $s_i=1$ (particle of type 1), and $s_i=2$ (particle of type 2). We keep the number of empty sites, $N_0$, fixed, but we vary the number of particles of types 1 and 2 (given by $N_1$ and $N_2$, respectively) through an interconversion reaction, such that $N_0+N_1+N_2=N=L^3$. We define the overall density and mole fractions in accordance with the description presented in the Methods of the main text, as $\rho=(N_1+N_2)/N$, $x_1=N_1/(N_1+N_2)$, and $x_2=N_2/(N_1+N_2)$. Each lattice site, $i$, has $z=6$ neighbors, which can be in states, $s_{ij}$, where $j=1,2,...,z$. The potential energy of each site is computed as $u_i=\sum_j^z\epsilon(s_{i},s_{ij})$, where $\epsilon(k,l)=2\omega_{kl}/z$ is a symmetric matrix with $\omega_{0k}=\omega_{k0}=0$. In this example, we use $\omega_{11}=2.0$, $\omega_{12}=1.4$, and $\omega_{22}=2.5$. The total energy of the system is computed as $U=\sum_i^N u_i/2 +e N_2$, where, $e$ is the internal energy of a particle of type 2. We also assume that particles of type 2 have internal entropy, $s$, which can be regarded as the energy and the entropy of the reaction. We use $s=4$, $e=3$ in this example.

At each Monte Carlo step, we attempt to perform a Kawasaki move\cite{kawasaki_diffusion_1966}, simulating diffusion, and a Glauber flip\cite{glauber_timedependent_1963}, simulating an interconversion reaction\cite{Shum_Phase_2021}. In a Kawasaki move, we randomly select an empty site and an occupied site and attempt to swap their states, while in a Glauber flip, we randomly select an occupied site and attempt to change its state from $1$ to $2$, or vice versa. For either step, we compute $\Delta F = \Delta U -T\Delta S$, where $\Delta U$ is the change in the total energy caused by this attempt. In a Kawasaki move, $\Delta S=0$, while in a Glauber flip $\Delta S=\pm s$, where ``$+$'' indicates when a state of this site changes from 1 to 2, while ``$-$'' indicates the opposite scenario. In accordance with the Metropolis criterion\cite{metropolis_basic_1963}, we accept the new state with probability $p=\exp(-\Delta F/T)$ for $\Delta F>0$, while for $\Delta F<0$, the new state is always accepted.  

\begin{figure}[ht!]
    \centering
    \includegraphics[width=0.48\linewidth]{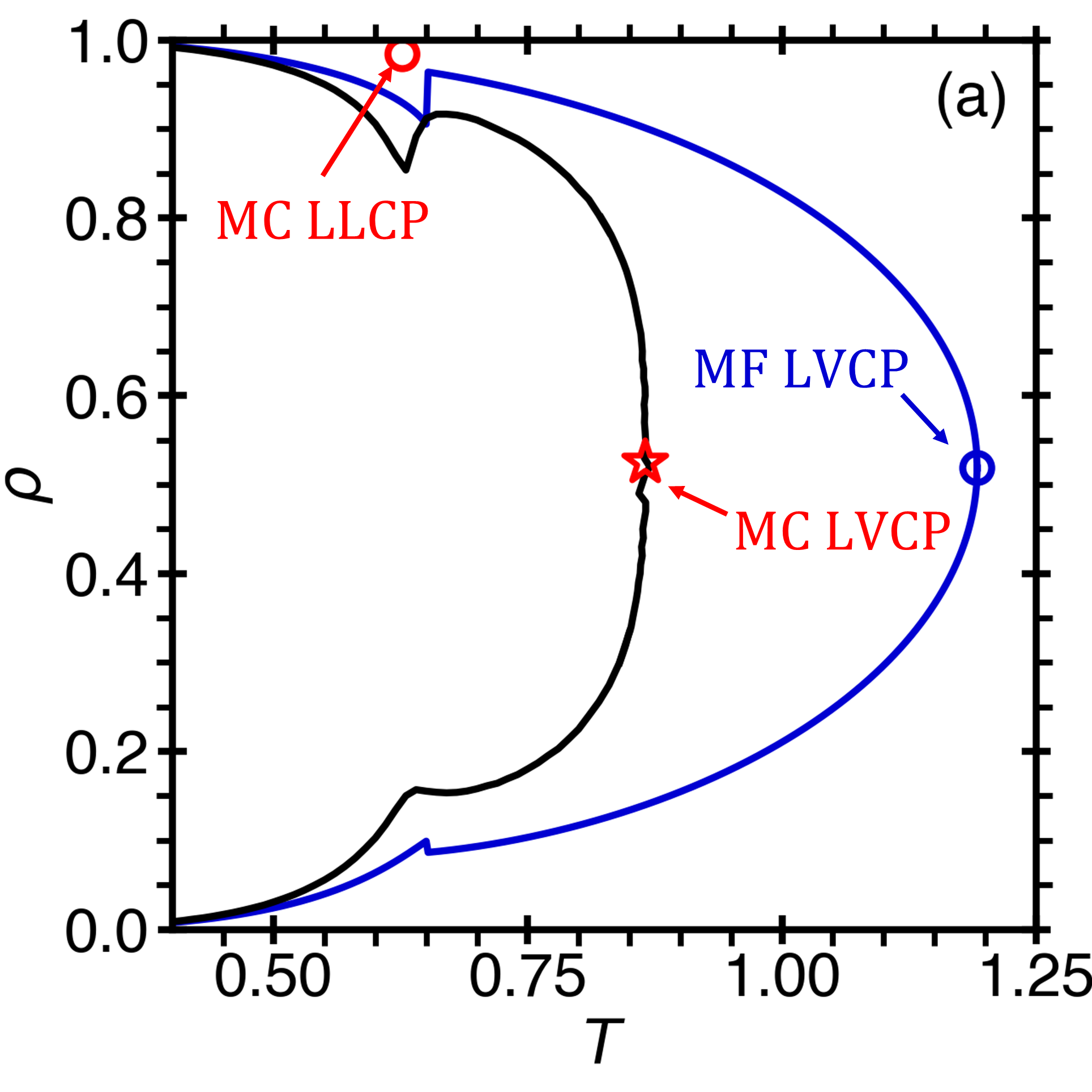}
    \includegraphics[width=0.48\linewidth]{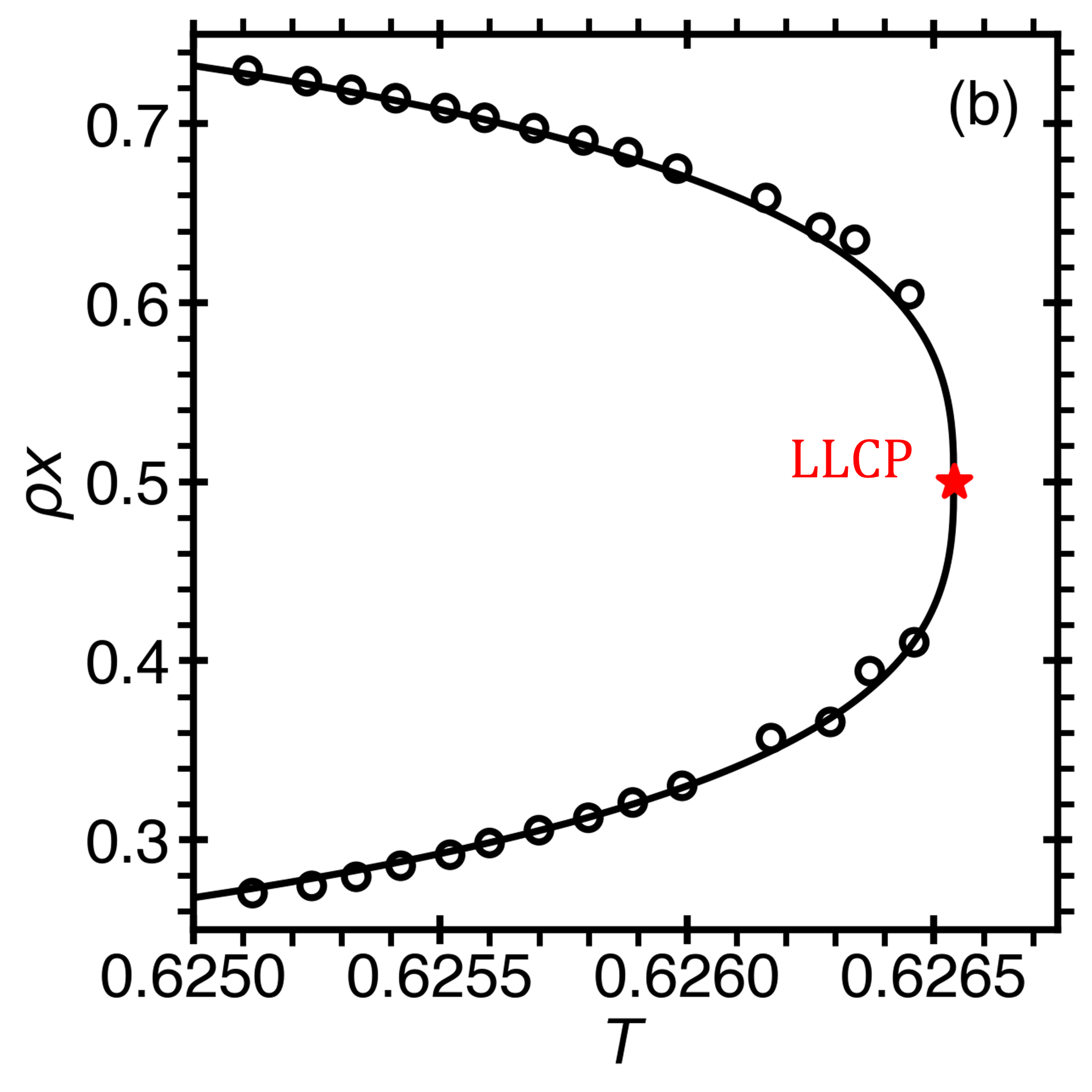}
    \includegraphics[width=0.48\linewidth]{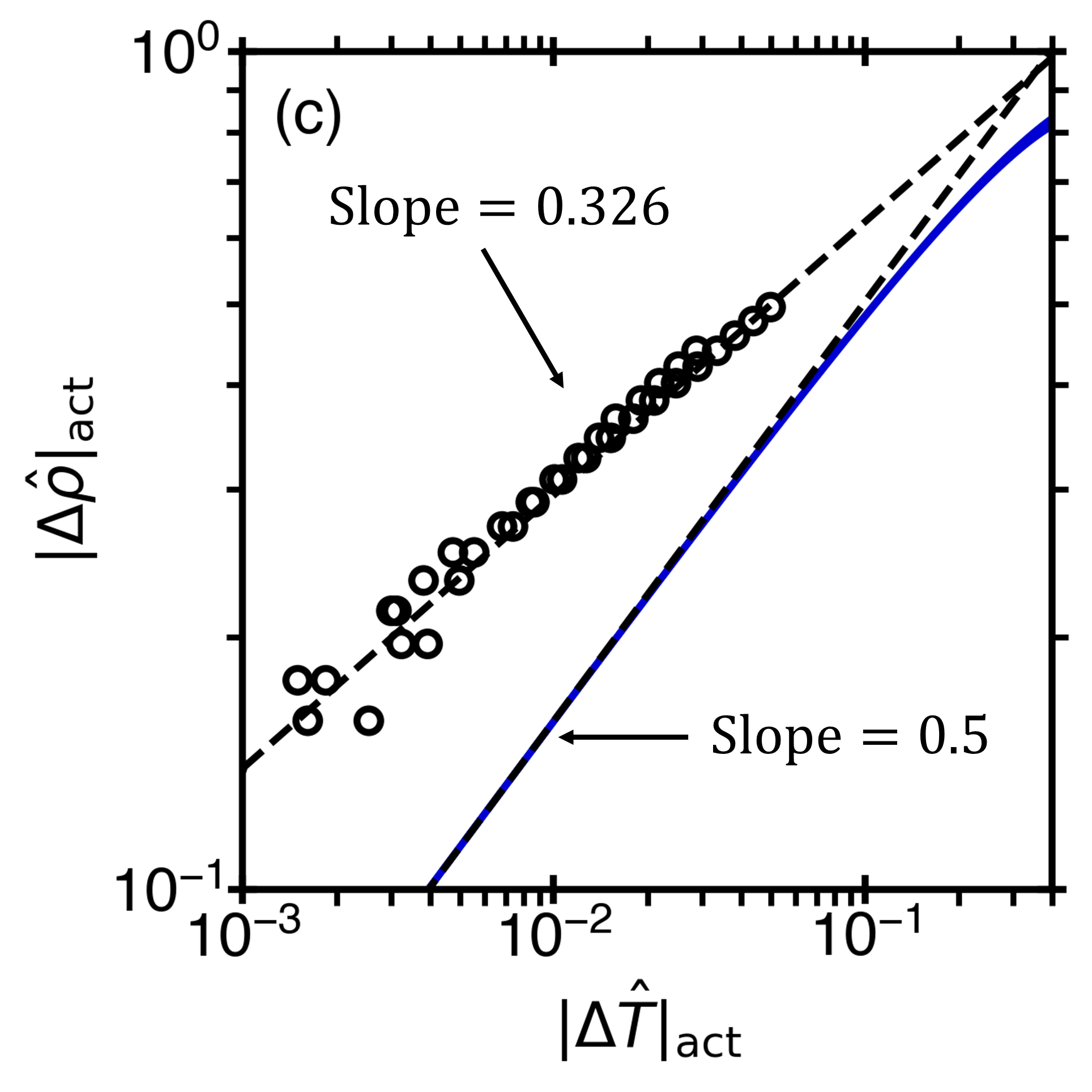}
    \includegraphics[width=0.48\linewidth]{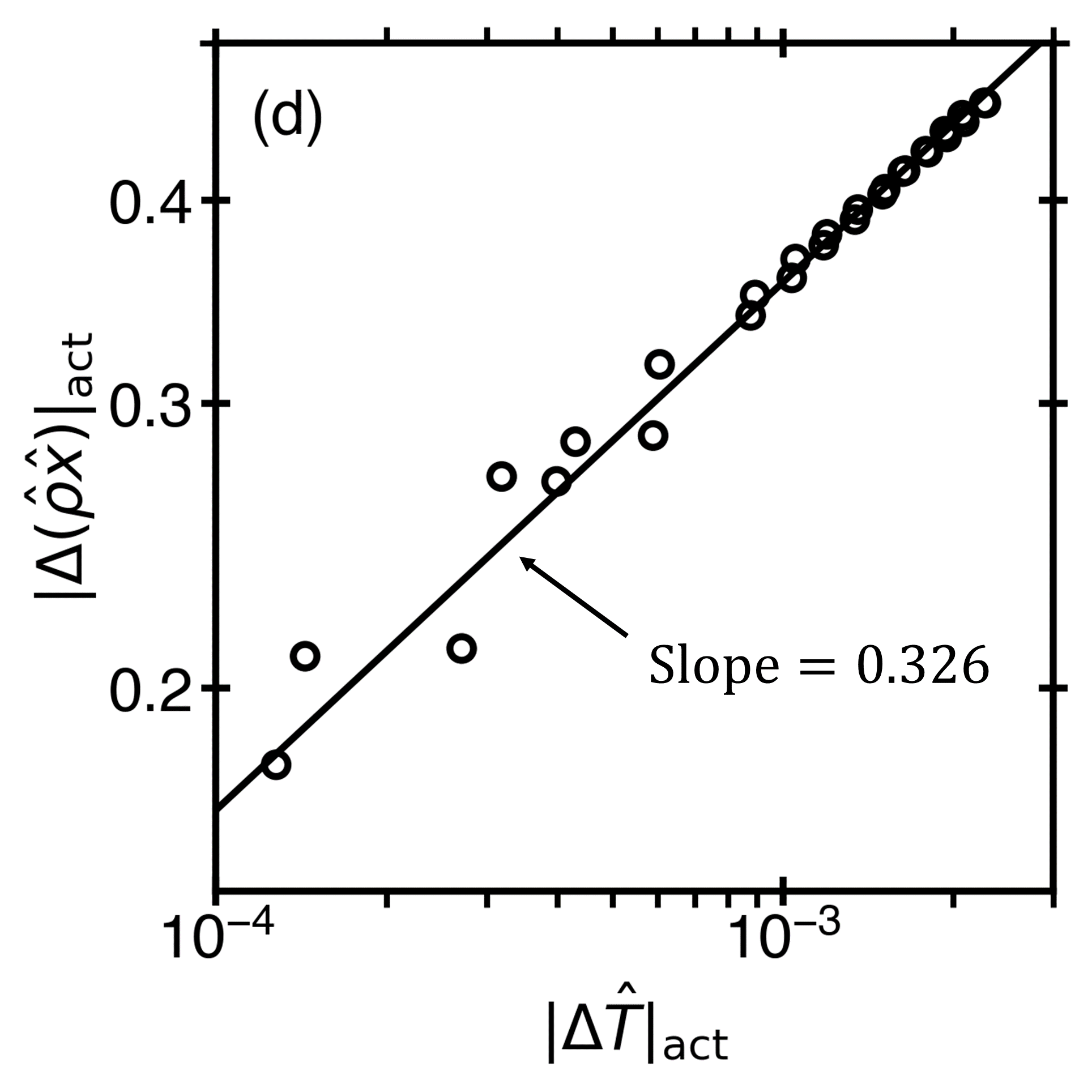}
    \caption{Comparison of the LV and LL critical behavior in the Monte Carlo (MC) and meanfield (MF) blinking-checkers lattice models for systems with $e=3$, $s=4$, $\omega_{11}=2.0$, $\omega_{22}=2.5$, and $\omega_{12}=1.4$. (a) liquid-vapor coexistence in the MC model (black) and MF model (blue). Liquid-vapor critical points are marked by the red star (MC LVCP) and blue circle (MF LVCP), while the liquid-liquid critical point is marked by the red circle (MC LLCP). (b) liquid-liquid coexistence in the MC model. The open circles represent simulation results, while the black curve is a fit based on the scaling theory that asymptotically close to the critical point for systems in the three-dimensional Ising universality class\cite{fisher_scaling_1983}, $\Delta(\hat{\rho}\hat{x})_\mathrm{act}\sim|\Delta\hat{T}|^{0.326}_\mathrm{act}$, where $\Delta(\hat{\rho}\hat{x})_\mathrm{act} = 1-\rho x/(\rho_\mathrm{c}^\mathrm{act}x_\mathrm{c}^\mathrm{act})$ and $\Delta\hat{T}_\mathrm{act} = 1-T /T_\mathrm{c}^\mathrm{act}$. This universality class is confirmed for both the LV and LL critical points in (c,d) where $\Delta\hat{\rho}_\mathrm{act} = 1-\rho/\rho_\mathrm{c}^\mathrm{act}$ . Note that in (c), asymptotically close to the LVCP, the MF power law, $\Delta\hat{\rho}\sim|\Delta\hat{T}|^{0.5}_\mathrm{act}$, is demonstrated.}
    \label{Fig_MC_Comparison}
\end{figure}

The entropy of the system can be obtained by thermodynamic integration, $S=\int  dU/T$, which does not explicitly depends on $s$. We compute the pressure of the system using the method described in Ref.\cite{Mauro_Pressure_2020} through $n_\lambda=100$ ghost sites, each for a specific $\lambda$ with $\Delta \lambda=0.01$. The ghost sites exchange states with randomly selected sites in the lattice. The simulations are performed for $2^{17}N$ Monte Carlo steps, which is sufficient for the system to reach equilibrium at all temperatures and densities.

Figure~\ref{Fig_MC_Comparison}(a,b) demonstrates the behavior of the MC model. We find an actual  LV critical point at $T_\mathrm{c}^\mathrm{LV} = 0.865$ and $\rho_\mathrm{c}^\mathrm{LV} = 0.535$, while we observe an actual LL critical point at $T_\mathrm{c}^\mathrm{LL} = 0.62654$ and $(x_\mathrm{c}\rho_\mathrm{c})^\mathrm{LL} = 0.5$. In the meanfield blinking-checkers model for the same interaction parameters, we find an actual LV critical point at $T_\mathrm{c}^\mathrm{LV} = 1.19$ and $\rho_\mathrm{c}^\mathrm{LV} = 0.519$, while no actual meanfield LLCP is observed, as it exists in the metastable region, below the LV coexistence. Therefore, we note that the effects of critical fluctuations cause a decrease in the critical temperature by $\sim 37\%$. Moreover, Figs.~\ref{Fig_MC_Comparison}(c,d) demonstrate that the MC version of the blinking-checkers model is classified in the three-dimensional Ising model universality class for both the LV and LL critical points. Interfacial profiles obtained in the MC approach, demonstrating the surface enrichment of species 1 near the triple point temperature (as discussed in Sec. 3.3 of the main text), are presented in Fig.~\ref{SM_Fig_MC_Enrich}.

\section{Excess Grand Thermodynamic Potential}
The excess grand thermodynamic potential may be expressed either for the non-reacting blinking-checkers model, as described by Eq.~(4) in the main text (in which $\varphi_{12}^\circ$ is unspecified), or for the interconverting blinking-checkers model (in which $\varphi_{12}^\circ$ is specified). In this section, we demonstrate that due to the chemical-reaction equilibrium condition, both of the expressions are identical.

In the non-reacting binary mixture, the excess grand thermodynamic potential, $\Delta\Omega$, is given by Eq.~(4), where the chemical potentials of species 1 and 2 in solution are given by Eq.~(\ref{SM_Eqn_mu1}) and (\ref{SM_Eqn_mu2}), see the SM of Ref.\cite{Caupin_Polyamorphism_2021} for more details. Evaluating $\mu_1$ and $\mu_2$ along coexistence, the excess grand potential is given by
\begin{equation}\label{Eq_SM_BinGrandPot}
    \begin{split}
        \Delta\Omega =& f(T,\rho,x) + P^{\mathrm{cxc}}+ 2\omega_{22}\rho^{\mathrm{cxc}}x^{\mathrm{cxc}}\rho x - (\omega-\omega_{11}-\omega_{22})\rho^{\mathrm{cxc}}(1-x^{\mathrm{cxc}})\rho x - T\rho x \ln\left[\frac{\rho^{\mathrm{cxc}}x^{\mathrm{cxc}}}{1-\rho^{\mathrm{cxc}}}\right] \\ &+2\omega_{11}\rho^{\mathrm{cxc}}\rho(1-x) - (\omega+\omega_{11}-\omega_{22})\rho^{\mathrm{cxc}}x^{\mathrm{cxc}}\rho(1-x) -T\rho(1-x)\ln\left[\frac{\rho^{\mathrm{cxc}}(1-x^{\mathrm{cxc}})}{1-\rho^{\mathrm{cxc}}}\right]
    \end{split}
\end{equation}
where $\omega = \omega_{11}+\omega_{22}-2\omega_{12}$ and the superscript ``cxc'' refers the quantity being evaluated along the fluid-phase coexistence. For the interconverting blinking-checkers model, the excess grand thermodynamic potential, $\Delta\Omega_\mathrm{rxn}$, is defined as
\begin{equation}\label{Eq_SM_RxnGrandPot}
    \Delta\Omega_\mathrm{rxn} = f_\mathrm{rxn}(T,\rho,x)-\rho\mu^\mathrm{cxc} + P^\mathrm{cxc}
\end{equation}
where $f_\mathrm{rxn}$ is the Helmholtz free energy per lattice site of the reacting mixture (in which $\varphi_{12}^\circ = -(e-Ts)$ is specified), $\mu^\mathrm{cxc} = \partial f_\mathrm{rxn}/\partial\rho|_{x,T}$ is the chemical potential per lattice site of the reacting mixture, and evaluated along coexistence; and $P^\mathrm{cxc}$ is the pressure per lattice site along coexistence. Evaluating all derivatives and subtracting $\Delta\Omega_\mathrm{rxn}$, given through Eq.~(\ref{Eq_SM_RxnGrandPot}), from $\Delta\Omega$, given by Eq.~(\ref{Eq_SM_BinGrandPot}), one obtains
\begin{equation}\label{Eq_SM_GrandDifference}
    \begin{split}
        \Delta\Omega - \Delta\Omega_\mathrm{rxn} =& \rho(x-x^\mathrm{cxc})(e-Ts)+\rho^\mathrm{cxc}\rho(x-x^\mathrm{cxc})(\omega_{11}-\omega_{22}) \\ &-\omega\rho^\mathrm{cxc}\rho(x-x^\mathrm{cxc})(1-2x^\mathrm{cxc})-T\rho(x-x^\mathrm{cxc})\ln\left(\frac{x^\mathrm{cxc}}{1-x^\mathrm{cxc}}\right)
    \end{split}
\end{equation}
To understand the difference in the grand thermodynamic potentials given by Eq.~(\ref{Eq_SM_GrandDifference}), we need to define the chemical-reaction equilibrium condition. This condition is given by $\partial f_\mathrm{rxn}/\partial x|_{T,\rho} =0$, and with $f_\mathrm{rxn}$, given in the discussion of Eq.~(1) in the main text, the equilibrium condition is
\begin{equation}\label{Eq_SM_RxnEq}
    e-Ts +\rho^\mathrm{cxc}[\omega_{11}-\omega_{22}-\omega(1-2x^\mathrm{cxc})]-T\ln\left(\frac{x^\mathrm{cxc}}{1-x^\mathrm{cxc}}\right) = 0
\end{equation}
(see more details in the SM of Ref.\cite{Caupin_Polyamorphism_2021}). After multiplying both sides of Eq.~(\ref{Eq_SM_RxnEq}) by $\rho(x-x^\mathrm{cxc})$, one observes that the right hand side of Eq.~(\ref{Eq_SM_GrandDifference}) is the equilibrium condition, which by definition is zero. Therefore, the difference between the excess grand thermodynamic potential expressed for the non-reacting mixture and with that for the interconverting mixture is, $\Delta\Omega-\Delta\Omega_\mathrm{cxc} = 0$. This result is unsurprising as $\Delta\Omega$ is a state function independent of the choice of thermodynamic path.

\section{Influence Parameters}
In this section, we derive the microscopic influence parameters for the blinking-checkers model to be used in the density gradient theory (DGT)\cite{Debye_Dissymmetry_1959,Yang_Molecular_1976,Rowlinson_Capillarity_1982,Lu_Density_1985,Kahl_Interfacial_2002,Stephan_Interfaces_2020} for the calculations of the interfacial tension. We follow the approach of Debye\cite{Debye_Dissymmetry_1959}, and define the influence coefficients \textit{a priori} with the assumption that the gradients contribute to the local energy density, but not to the entropy density\cite{Feeney_Theoretical_2003}.

We assume a local spatial-dependent density and concentration, $\rho=\rho(\vec{r}')$ and $x = x(\vec{r}')$, where $\vec{r}'$ is the spatial coordinate, and generate a Taylor series around a neighboring point $\vec{r}$ to second-order, which yields
\begin{align}
    \rho(\vec{r}') &= \rho(\vec{r}) + (\vec{r}'-\vec{r})\nabla\rho(\vec{r}) + \frac{1}{2}(\vec{r}'-\vec{r})^2\nabla^2\rho(\vec{r}) \label{Eq_SM_rhoTaylor}\\
    x(\vec{r}') &= x(\vec{r}) + (\vec{r}'-\vec{r})\nabla x(\vec{r}) + \frac{1}{2}(\vec{r}'-\vec{r})^2\nabla^2 x(\vec{r}) \label{Eq_SM_xTaylor}
\end{align}
We define the total interaction energy, $E$, through each of four possible species interactions by integrating over $\vec{r}'$. For the species 1 - species 1 (denoted as ``11'') interaction, assuming all interaction parameters are independent of position, this gives $E_{11} = -\omega_{11}\int \dd{\vec{r}}'[(\vec{r}')x(\vec{r}')\rho(\vec{r}')]$. With use of Eqs.~\ref{Eq_SM_rhoTaylor}) and (\ref{Eq_SM_xTaylor}), this gives the following expressions for the four interaction energies:
\begin{align}
    E^\mathrm{(2)}_{11} &= -\omega_{11}\ell^2\left(\nabla x\nabla\rho+\frac{1}{2}\left[x\nabla^2\rho + \rho\nabla^2 x\right]\right)\\
    E^\mathrm{(2)}_{12} &= -\omega_{12}\ell^2\left(-\nabla x\nabla\rho+\frac{1}{2}\left[(1-x)\nabla^2\rho - \rho\nabla^2 x\right]\right)\\
    E^\mathrm{(2)}_{21} &= -\omega_{21}\ell^2\left(\nabla x\nabla\rho+\frac{1}{2}\left[x\nabla^2\rho + \rho\nabla^2 x\right]\right)\\
    E^\mathrm{(2)}_{22} &= -\omega_{22}\ell^2\left(-\nabla x\nabla\rho+\frac{1}{2}\left[(1-x)\nabla^2\rho - \rho\nabla^2 x\right]\right)
\end{align}
where the superscript ``(2)'' indicates only the contribution to the interaction energy from the second-order gradient terms, the notation ``$(\vec{r})$'' has been dropped for simplicity, and $\ell = |\vec{r}'-\vec{r}|$ is the distance between two lattice cells. Note that all odd terms go to zero upon integration over $\vec{r}'$ due to the symmetry of the lattice.

We assume that the gradient terms only contribute to the local internal energy per lattice site, $u=U/N$, and do not effect the local entropy. The internal energy per lattice site is given by
\begin{equation}\label{Eq_SM_internalContribution}
    u = \frac{1}{2}\int\dd{\vec{r}}\left[\rho x(E_{11}+E_{12}) + \rho (1-x)(E_{21}+E_{22})\right]
\end{equation}
Simplifying Eq.~(\ref{Eq_SM_internalContribution}) with use of Green's first identity\cite{Schwartz_Vector_1960} gives the contribution from the interactions of the surface to the excess free energy. Thus, we obtain
\begin{equation}\label{Eq_SM_sigma}
    \sigma = \int\left[\Delta\Omega(x,\rho, T) + \frac{1}{2}c_x(\rho)|\nabla x|^2 + \frac{1}{2}c_\rho(x)|\nabla \rho|^2 + c_{\rho,x}(\rho,x)\nabla\rho\cdot\nabla x\right]\dd{\vec{r}}
\end{equation}
where the three influence parameters are given by
\begin{align}
    c_x(\rho) &= \frac{1}{2}\ell^2\rho^2\omega \label{Eq_cx} \\ 
    c_\rho(x) &= \frac{1}{2}\ell^2\left[\omega_{11}x^2 + 2\omega_{12}x(1-x) + \omega_{22}(1-x)^2\right] \label{Eq_crho}\\ 
    c_{\rho,x}(\rho,x) &= \frac{1}{2}\ell^2\rho\left[\omega_{11}x + \omega_{12}(1-2x)-\omega_{22}(1-x)\right] \label{Eq_crhox}
\end{align}
We note that there are three limits that may be observed for Eq.~(\ref{Eq_SM_sigma}), in which the system reverts to either the lattice-gas or binary-lattice models\cite{Anisimov_Polyamorphism_2018}. For these three cases, the integrand of Eq.~(\ref{Eq_SM_sigma}) becomes:
\begin{itemize}
    \item The limit of pure species 1 ($x = 1$),
    \begin{equation}\label{Eq_SM_sigmaXeq1}
        \Delta\Omega(x=1,\rho, T) + \frac{1}{4}\ell^2\omega_{11}|\nabla\rho|^2
    \end{equation}
    \item The limit of pure species 2 ($x=0$),
    \begin{equation}
        \Delta\Omega(x=0,\rho, T) + \frac{1}{4}\ell^2\omega_{22}|\nabla\rho|^2
    \end{equation}
    \item The limit of very high density ($\rho = 1$),
    \begin{equation}\label{Eq_SM_sigmaRhoeq1}
        \Delta\Omega(x,\rho=1, T) + \frac{1}{4}\ell^2\omega|\nabla x|^2
    \end{equation}
\end{itemize}

Lastly, the equilibrium condition is found by the minimization of Eq.~(\ref{Eq_SM_sigma}) with respect to density or concentration via the Euler-Lagrange derivative\cite{Poser_Surface_1979,Poser_Interfacial_1981,Rowlinson_Capillarity_1982}, which yields two expressions,
\begin{align}
    \frac{\partial(\Delta\Omega)}{\partial \rho} + \frac{1}{2}\frac{\partial c_x}{\partial \rho}|\nabla x|^2 + \frac{1}{2}\frac{\partial c_\rho}{\partial x}\nabla x\nabla\rho &= c_\rho \nabla^2\rho + c_{x,\rho}\nabla^2 x \label{SM_Eq_Euler1}\\
    \frac{\partial(\Delta\Omega)}{\partial x} + \frac{1}{2}\frac{\partial c_\rho}{\partial x}|\nabla\rho|^2 + \frac{1}{2}\frac{\partial c_x}{\partial \rho}\nabla x\nabla\rho &= c_{x,\rho}\nabla^2 \rho + c_x \nabla^2 x \label{SM_Eq_Euler2}
\end{align}
where we have used the fact that $\partial c_{x,\rho}/\partial \rho = (1/2)\partial c_\rho/\partial x$ and $\partial c_{x,\rho}/\partial x = (1/2)\partial c_x/\partial \rho$. Upon integration, these equations give the equilibrium condition for the interfacial tension\cite{Poser_Surface_1979,Poser_Interfacial_1981,Kahl_Interfacial_2002}
\begin{equation}\label{SM_Eq_EquiCond}
    \Delta\Omega(x,\rho, T) = \frac{1}{2}c_x(\rho)|\nabla x|^2 + \frac{1}{2}c_\rho(x)|\nabla \rho|^2 + c_{\rho,x}(\rho,x)\nabla\rho\cdot\nabla x
\end{equation}

\section{Comparison of Exact Solution with Phenomenological Asymmetric Ansatzes}
We considered a variety of different ansatzes that would minimize the interfacial tension, Eq.~(3) in the main text. We found that there were no one free-parameter ansatzes that were able to minimize the interfacial tension with sufficient accuracy, while several ansatzes with three free parameters were sufficient. In the main text, we consider the Fisher-Wortis ansatz, which has two free parameters ($\hat{\delta}$ and $\hat{\zeta}$). In this section, we compare the two-parameter Fisher-Wortis ansatz to an alternative symmetric three-parameter ansatz and the exact solution of the interfacial tension equilibrium condition, Eq.~(\ref{SM_Eq_EquiCond}). 

\begin{figure}[ht!]
    \centering
    \includegraphics[width=0.49\linewidth]{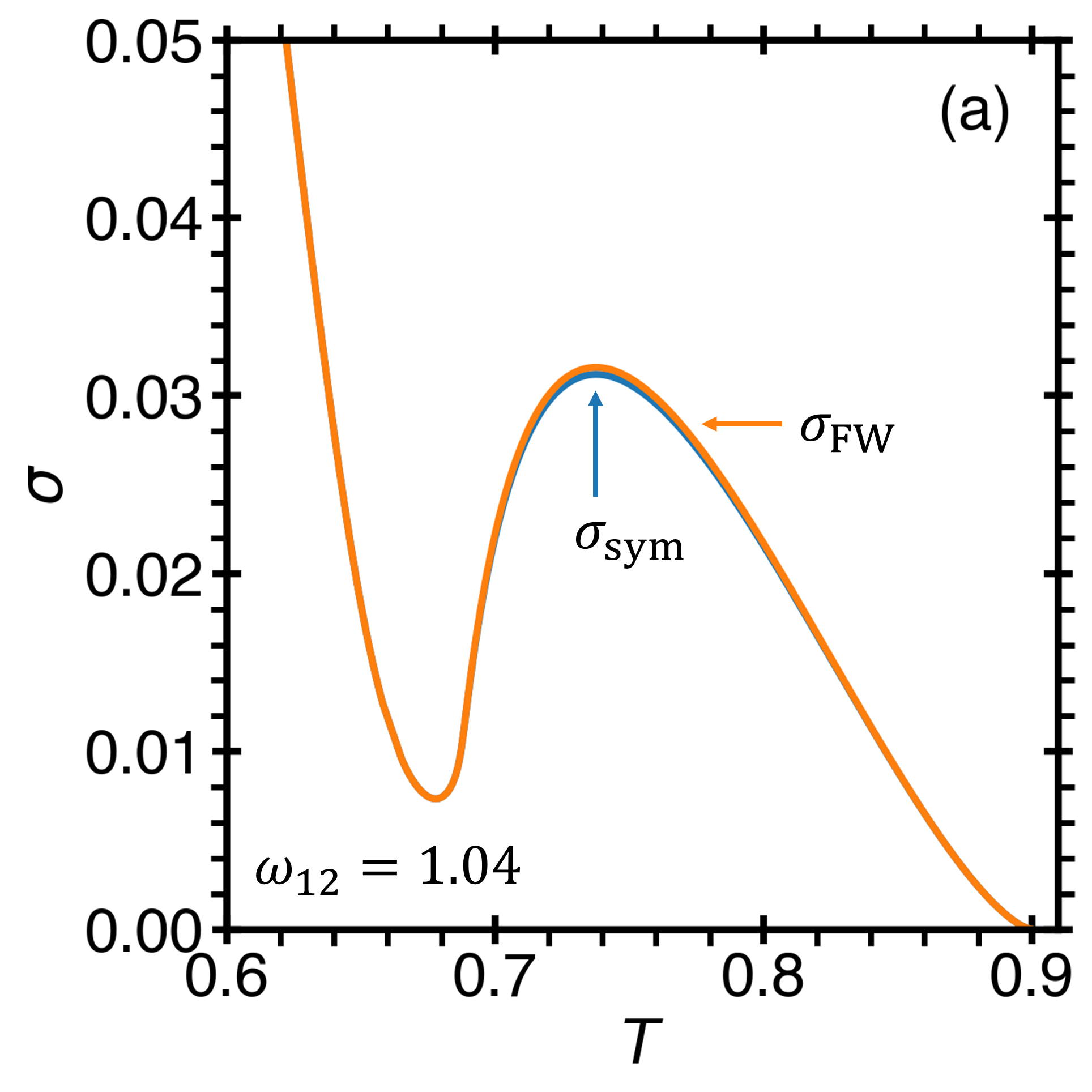}
    \includegraphics[width=0.49\linewidth]{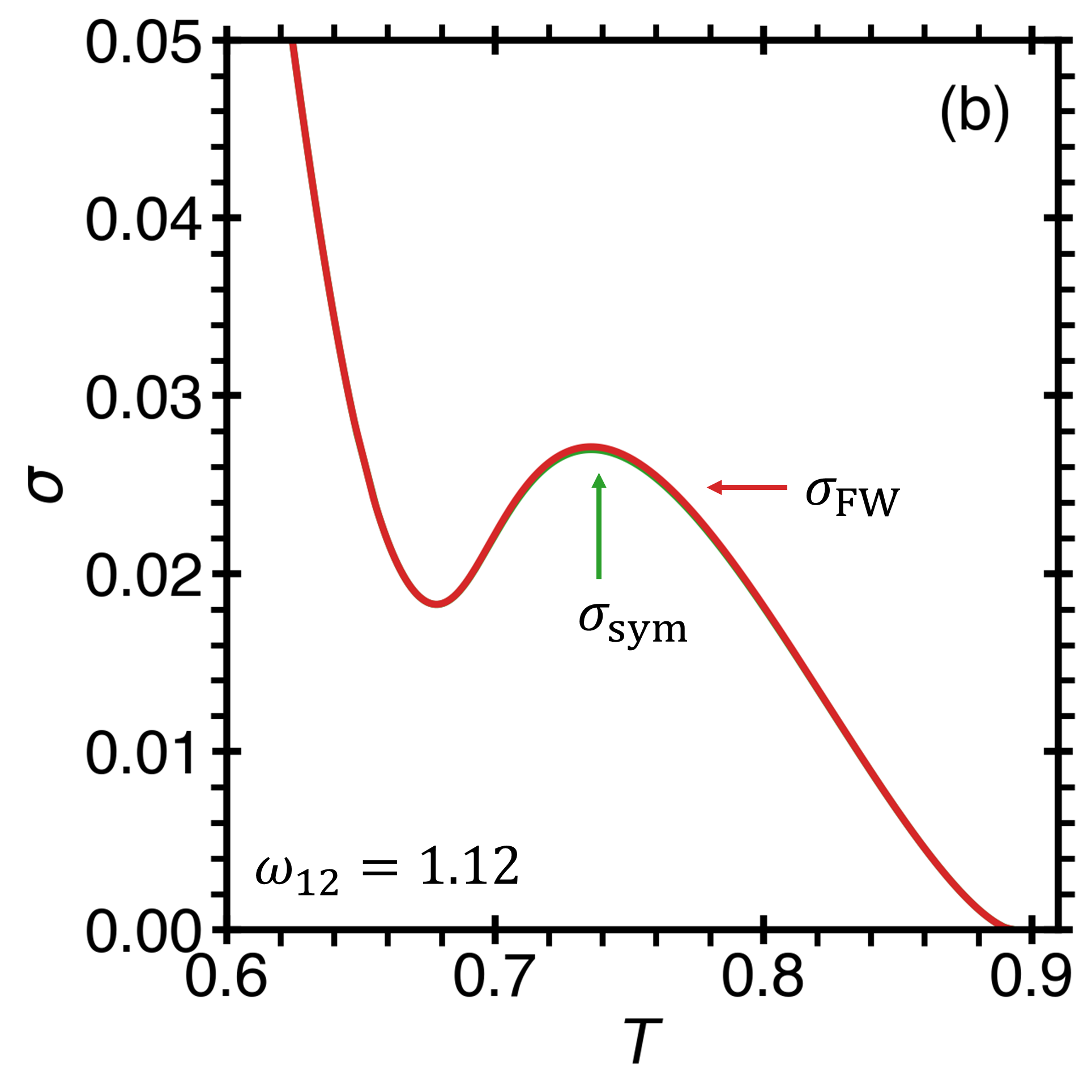}
    \includegraphics[width=0.6\linewidth]{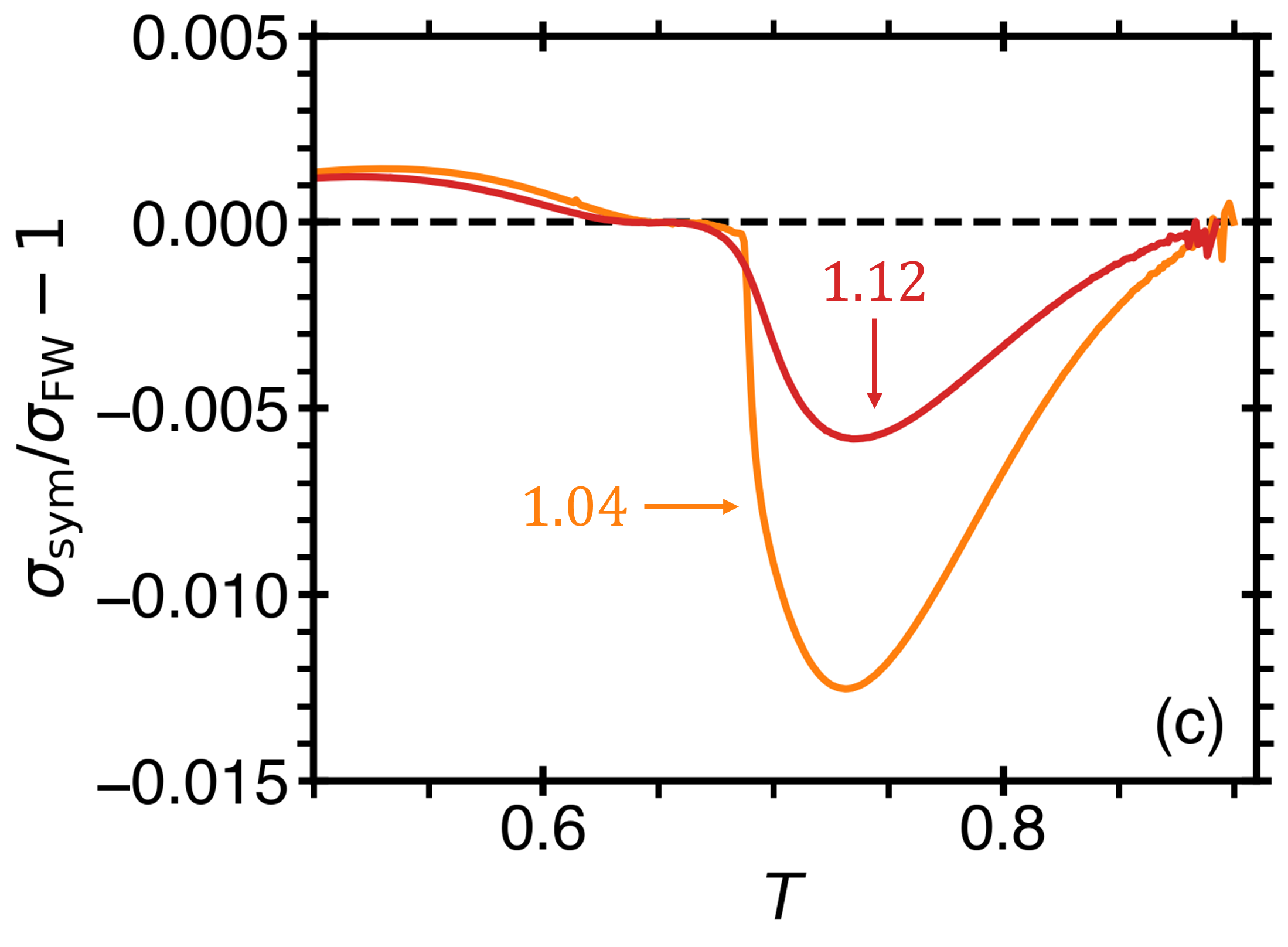}
    \caption{Comparison between the Fisher-Wortis (FW) and an alternative symmetric ansatz (sym). obtained for the liquid-vapor coexistence, for two systems with $\omega_{12}=1.04$ (a) and $\omega_{12}=1.12$ (b) with $\omega_{11}=1.6$, $\omega_{22}=2.0$, $e=3$, and $s=4$. (c) The relative deviation between the symmetric and FW ansatzes.}
    \label{SM_Fig_Ansatz_Comp}
\end{figure}

The symmetric three-parameter ansatz is given in normalized form by
\begin{align}
    \hat{\rho}(\hat{z}) &= \frac{\rho(\hat{z})-\rho_\mathrm{V}}{\rho_\mathrm{L}-\rho_\mathrm{V}} = \frac{1}{2}\left[\tanh\left(\frac{\hat{z}}{\zeta_\rho}\right)-1\right]\label{SM_Eq_rhoSymm}\\
    \hat{x}(\hat{z}) &= \frac{x(\hat{z})-x_\mathrm{V}}{x_\mathrm{L}-x_\mathrm{V}} = \frac{1}{2}\left[\tanh\left(\frac{\hat{z}+\hat{\delta}}{\zeta_x}\right)-1\right]\label{SM_Eq_xSymm}
\end{align}
in which the three parameters are the shift between the concentration and density profiles, $\hat{\delta}$, and the interfacial thicknesses of the density, $\hat{\zeta}_\rho$, and concentration, $\hat{\zeta}_x$. A comparison of the liquid-vapor interfacial tension between the Fisher-Wortis (FW) and the symmetric ansatzes of Eqs.~(\ref{SM_Eq_rhoSymm}) and (\ref{SM_Eq_xSymm}) is presented in Fig.~\ref{SM_Fig_Ansatz_Comp}. We find that the FW ansatz describes the LV interfacial tension with sufficient accuracy everywhere except near the maximum of $\sigma_\mathrm{LV}$. However, in this region, the symmetric ansatz differs from the FW ansatz by $\simeq 1\%$ or less. Therefore, we deemed that the FW ansatz is the most sufficient way to describe the interfacial properties, since it has only two free parameters. 

\begin{figure}[t]
    \centering
    \includegraphics[width=0.6\linewidth]{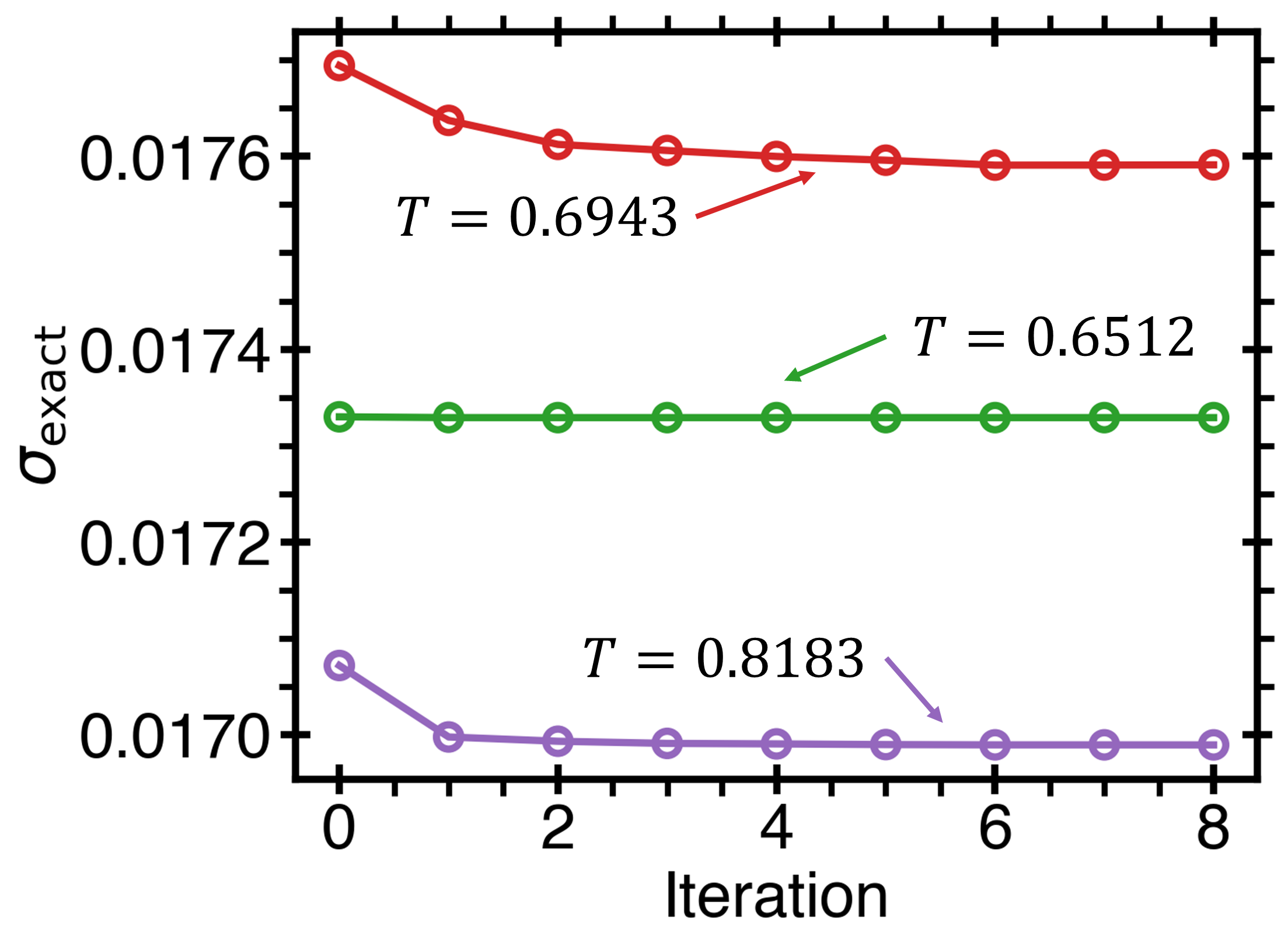}
    \caption{Numerical calculations of the liquid-vapor surface tension for three temperatures in the system with $\omega_{12}=1.04$, $\omega_{11}=1.00$, $\omega_{22}=2.00$, $e=3$, and $s=4$. The FW ansatz is given at iteration 0. The temperatures were chosen based on surface tensions with similar values, as predicted by the FW ansatz.}
    \label{SM_Fig_Exact_Iterations}
\end{figure}

\begin{figure}[t!]
    \centering
    \includegraphics[width=0.49\linewidth]{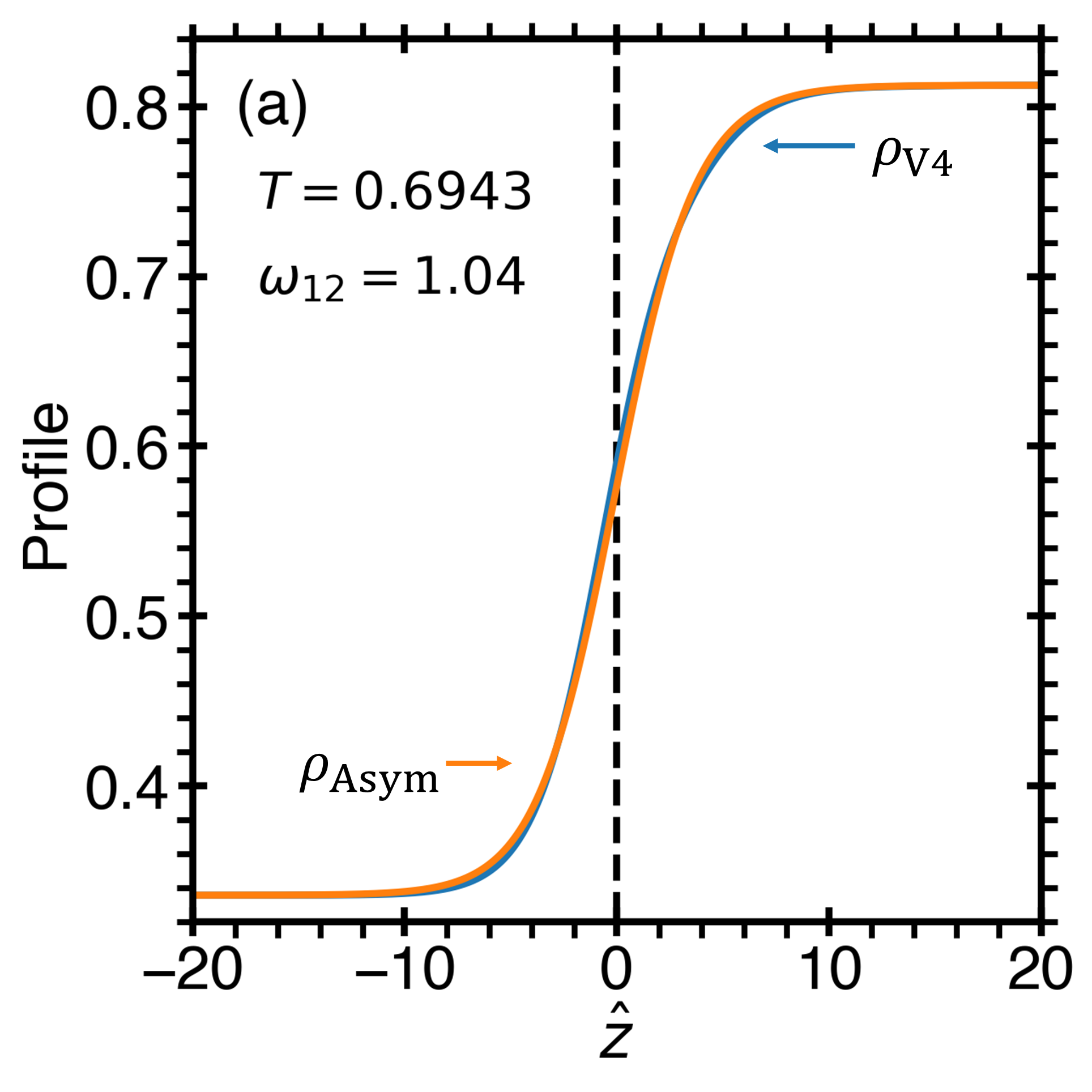}
    \includegraphics[width=0.49\linewidth]{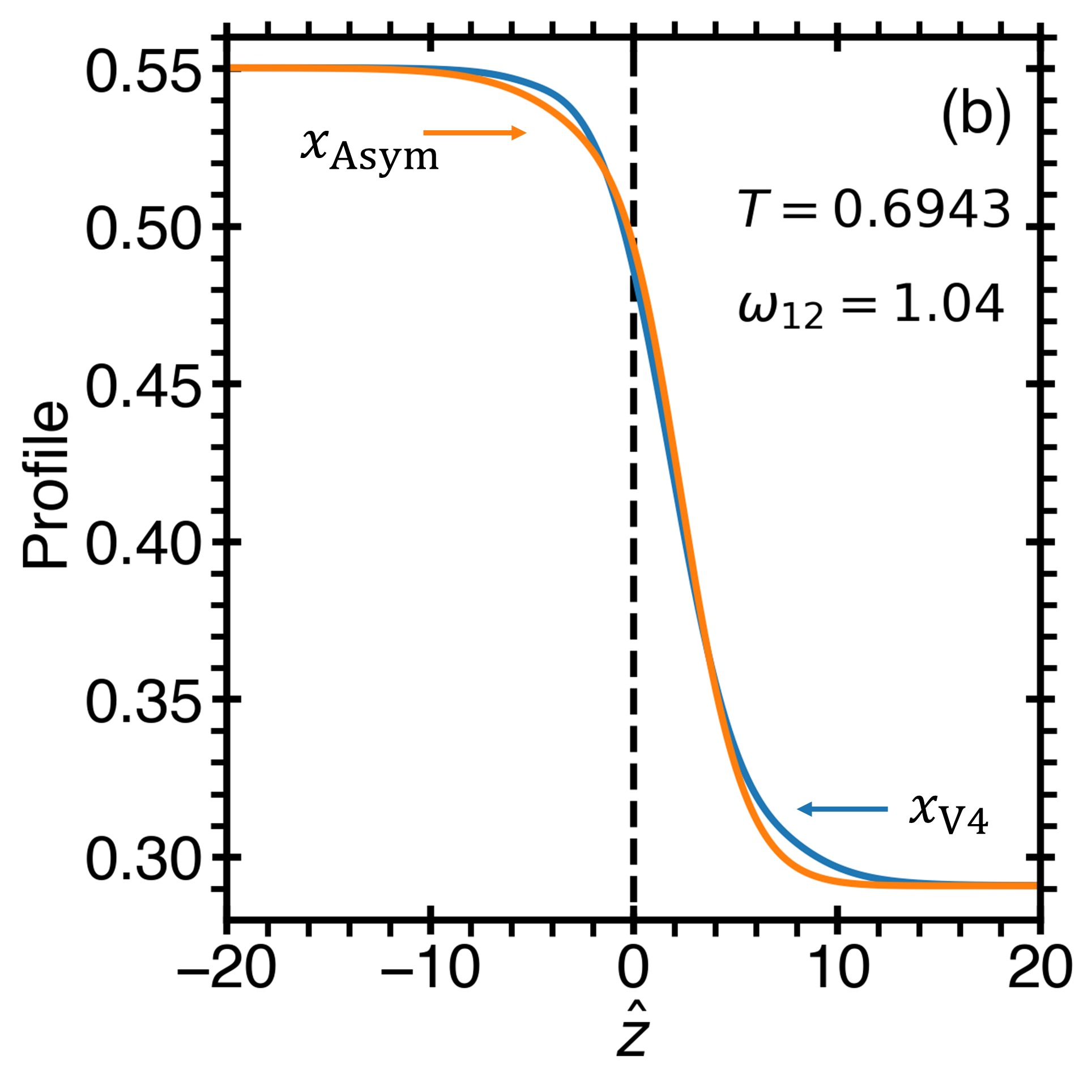}
    \includegraphics[width=0.49\linewidth]{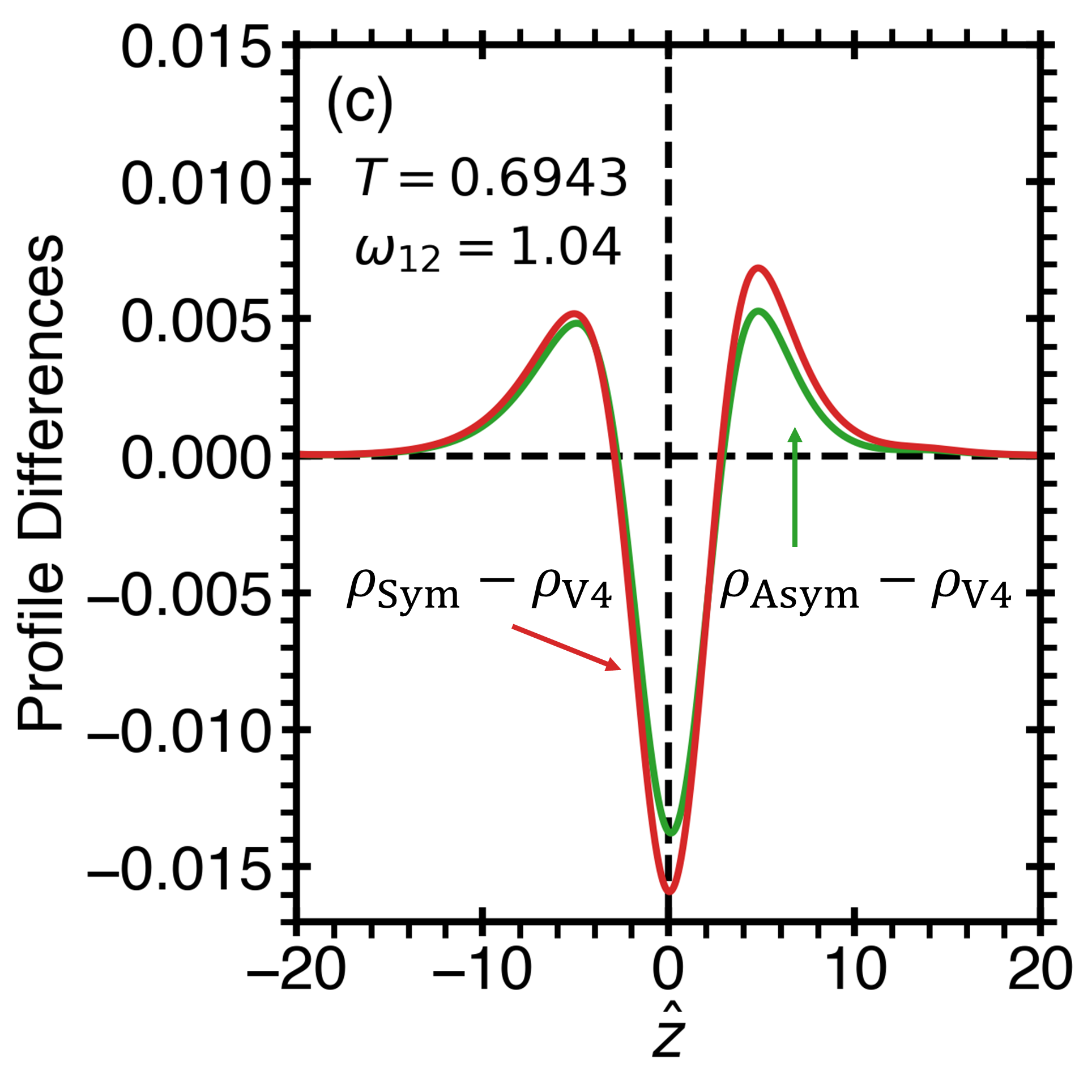}   
    \includegraphics[width=0.49\linewidth]{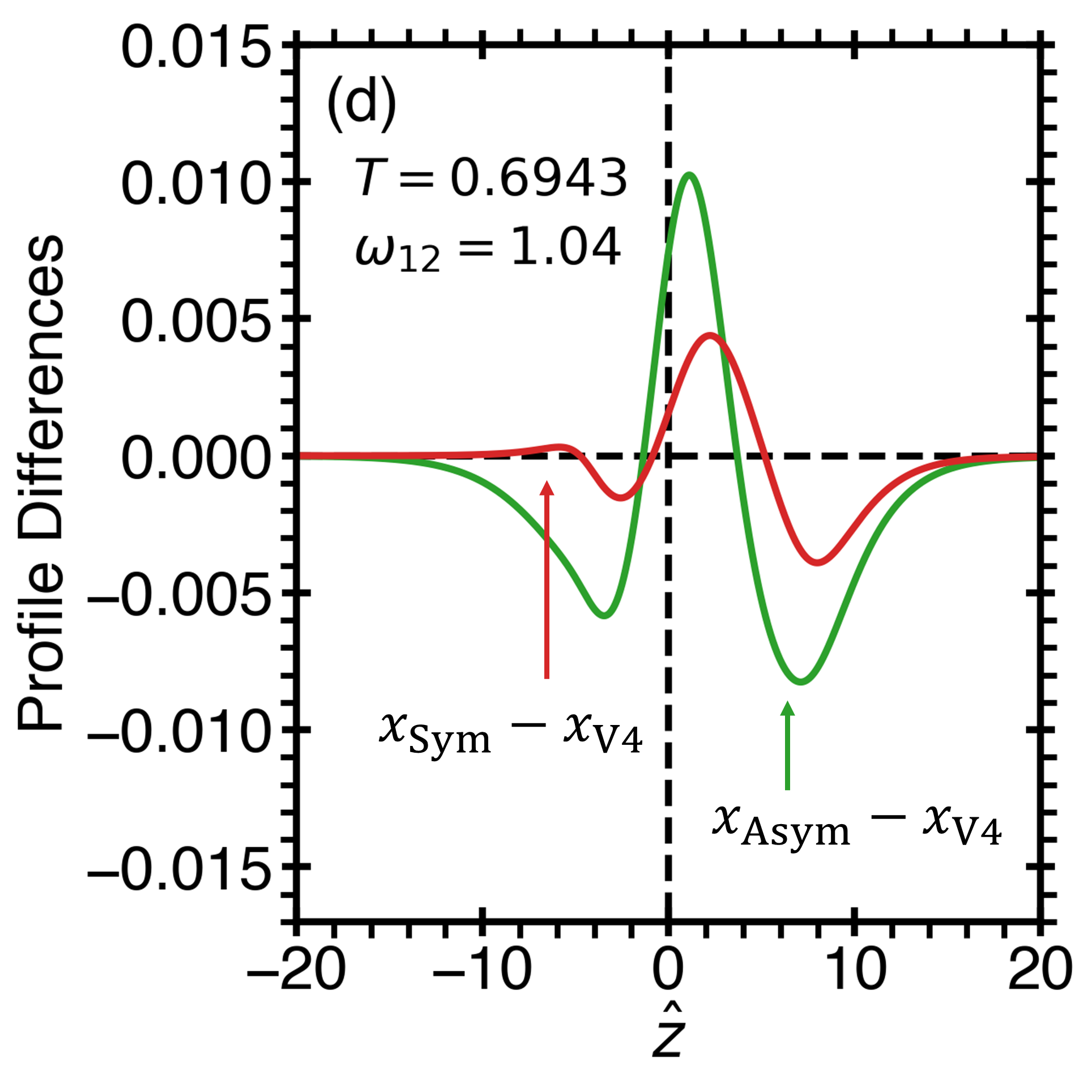} 
    \caption{Comparison of the liquid-vapor density (a) and concentration (b) interfacial profiles, as obtained from numerical calculations after eight iterations, $\rho_\mathrm{V4}$ and $x_\mathrm{V4}$, and for the FW ansatz, $\rho_\mathrm{V0}$ and $x_\mathrm{V0}$, for the system with $\omega_{12}=1.04$, $\omega_{11}=1.00$, $\omega_{22}=2.00$, $e=3$, and $s=4$ at temperature, $T=0.6943$ (red curve in Fig.~\ref{SM_Fig_Exact_Iterations}). (c,d) The difference between the exact solution for the density and concentration profiles and the FW and symmetric ansatzes.
    In (a-d), the different profiles were aligned along their Gibbs dividing surface, such that the excess density is zero for all profiles.}
    \label{SM_Fig_Exact_Profiles}
\end{figure}

In addition to the ansatz method, we also calculated the interfacial properties numerically by solving the interfacial tension equilibrium condition. We found that the Euler-Lagrange derivatives, Eqs.~(\ref{SM_Eq_Euler1}) and (\ref{SM_Eq_Euler2}), were highly unstable, as one would need to know both a coordinate position of both the density and concentration profiles as well as their derivative at this position. Thus, to solve these two equations for the density and concentration profiles, would require the adjustment of four unknown variables. Instead, we numerically solved Eq.~(\ref{SM_Eq_EquiCond}), the integral of Eqs.~(\ref{SM_Eq_Euler1}) and (\ref{SM_Eq_Euler2}), which requires two functions, $\rho(z)$ and $x(z)$, and two coordinate positions. Using one of the ansatzes for one of the unknown functions, we alternated between solving Eq.~(\ref{SM_Eq_EquiCond}) for $\rho(z)$ and $x(z)$ in an iterative process to determine the exact solution. Figure~\ref{SM_Fig_Exact_Iterations} illustrates the iterative solution method for three temperatures in the system with $\omega_{12}=1.04$. The solutions for the profiles using the iterative method are presented in comparison to the FW ansatz in Fig.~\ref{SM_Fig_Exact_Profiles}. We note the asymmetric shape of the exact solution also serves to justify our choice of the FW ansatz, which more closely matches the shape of the exact solution for the profiles than the symmetric ansatz. To discuss the amount of asymmetry in each profile, let $\phi(z)$ represent either the density or concentration profiles, and $z_{1/2}$ be the $z$-value where $\phi(z)$ reaches its midpoint, $\phi(z) = [\phi(+\infty)+\phi(-\infty)]/2$. We define the degree of asymmetry $D_\mathrm{Asym}$ via
\begin{equation}\label{SM_Eq_DegAsym}
    D_\mathrm{Asym} = D_0\sqrt{\dfrac{\int_{0}^{+\infty} \left\{\left[\phi(+\infty)-\phi(z_{1/2} + z')\right] + \left[\phi(-\infty) -\phi(z_{1/2} - z')\right]\right\}^2\mathrm{d}z'}{\int_{0}^{+\infty} \left\{\left[\phi(+\infty)-\phi(z_{1/2} + z')\right]^2 + \left[\phi(-\infty) -\phi(z_{1/2} - z')\right]^2\right\}\mathrm{d}z'}}
\end{equation}
where the prefactor, $D_0$, may be $-1$ or $1$, as given by
\begin{equation}\label{SM_Eq_DegAsym_Amp}
    D_0 = \mathrm{Sign}\left[\int_{0}^{+\infty} \left[\phi(+\infty)-\phi(z_{1/2} + z')\right] + \left[\phi(-\infty) -\phi(z_{1/2} - z')\right]\mathrm{d}z' \right]
\end{equation}
$D_\mathrm{Asym}=0$, if and only if the profile is symmetric; otherwise, $D_\mathrm{Asym}$ adopts values from $-1$ to $1$ depending on whether the profile spreads more towards the low or high values of $\phi$, respectively. A summary of the asymmetry for the system with $\omega_{12}=1.04$ at $T=0.6943$ is provided in Table~\ref{SM_Table_DegAsym} below. Thus, we find that the exact solution is indeed asymmetric. Ultimately, since the exact solution differs from the FW ansatz by a fraction of a percent (see Fig.~\ref{SM_Fig_Exact_Iterations}), we determined that the FW ansatz was the most efficient way to describe the interfacial properties with sufficient accuracy.

\begin{table}[t!]
\begin{tabular}{lcc}
\toprule
Ansatz & $\rho$-Profile & $x$-Profile \\ \midrule
Sym    & 0              & 0           \\
FW     & 0.016        & 0.103       \\
Exact  &  0.104        & -0.060   \\ \bottomrule 
\end{tabular}
\caption{The degree of asymmetry for each profile of the system with $\omega_{12}=1.04$ at $T=0.6943$, determined from the symmetric ansatz (sym), the Fisher-Wortis ansatz (FW), or the exact solution as calculated from Eq.~(\ref{SM_Eq_DegAsym}).}
\label{SM_Table_DegAsym}
\end{table}

\clearpage
\newpage

\section{Liquid-Vapor Diameters of Density and Concentration}

\begin{figure}[h!]
    \centering
    \includegraphics[width=0.49\linewidth]{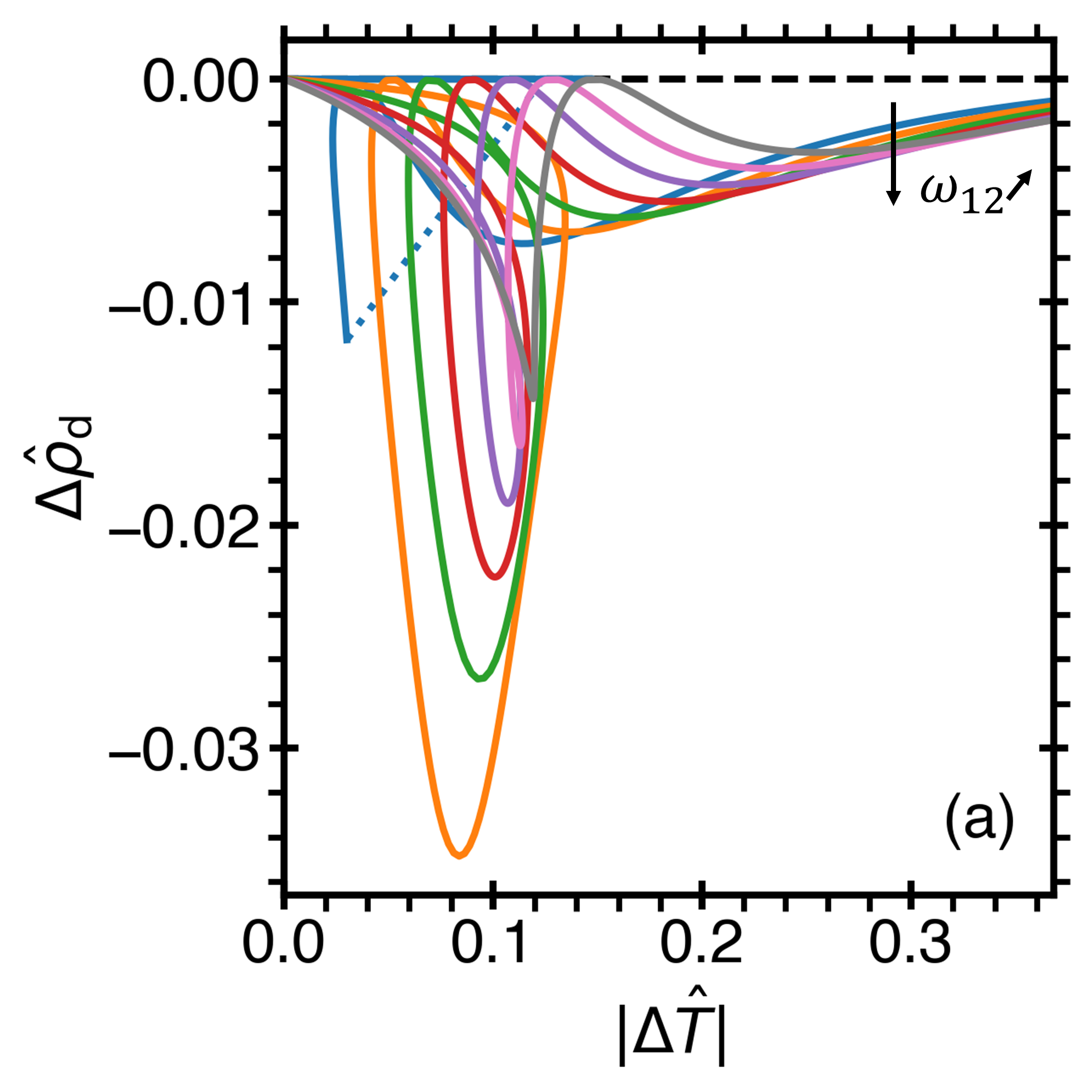}
    \includegraphics[width=0.49\linewidth]{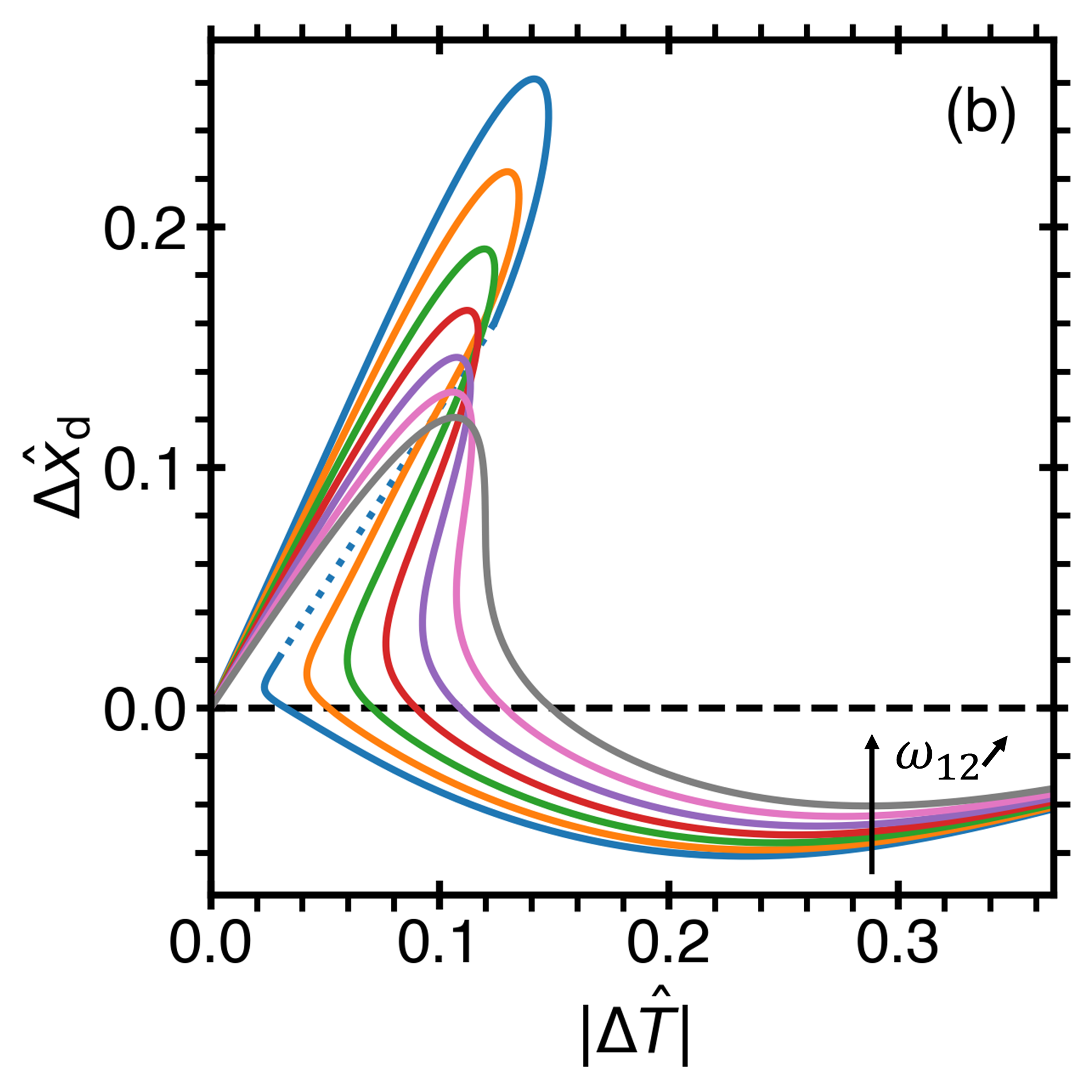}
    \caption{Liquid-vapor diameters of (a) the density, given by Eq.~(15) in the main text, and (b) the concentration, given by Eq.~(16) in the main text, as a function of the distance to the virtual LVCL along the thermodynamic path selected by interconversion for systems with $\omega_{11}=1.6$, $\omega_{22}=2.0$, $e=3$, $s=4$, and with various values of $\omega_{12}$: $\omega_{12}=1.00$ (blue), $\omega_{12}=1.04$ (orange), $\omega_{12}=1.08$ (green), $\omega_{12}=1.12$ (red), $\omega_{12}=1.16$ (purple), $\omega_{12}=1.20$ (pink), and $\omega_{12}=1.24$ (gray). In (a,b), the black arrow indicates the direction of increasing $\omega_{12}$, and the dotted blue lines indicate the discontinuity for the system with $\omega_{12}=1.00$ at the triple point.}
    \label{Fig_SM_RectiDiam}
\end{figure}

\newpage

\section{Liquid-Liquid Diameters and Interfacial Properties}

\begin{figure}[h!]
    \centering
    \includegraphics[width=0.49\linewidth]{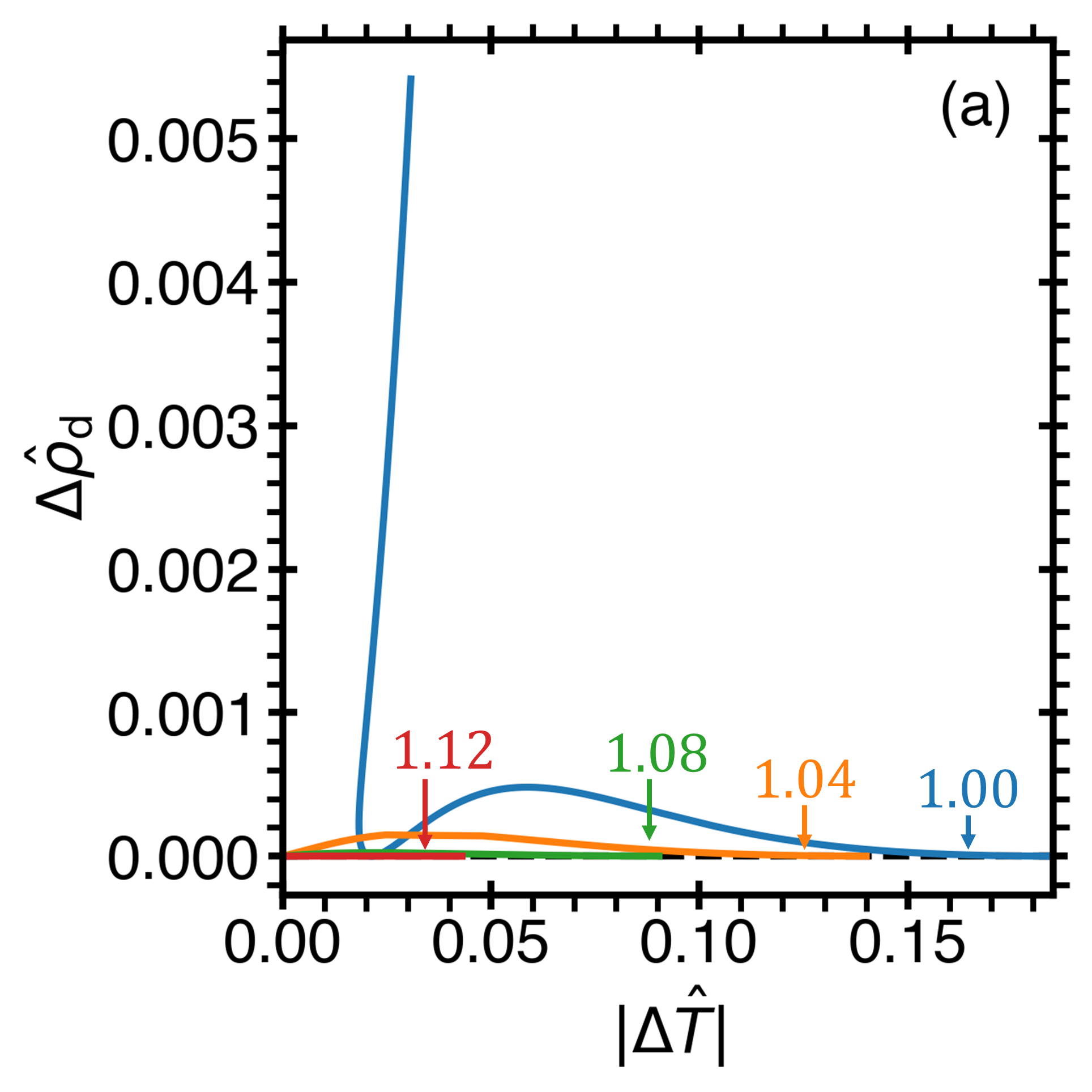}
    \includegraphics[width=0.49\linewidth]{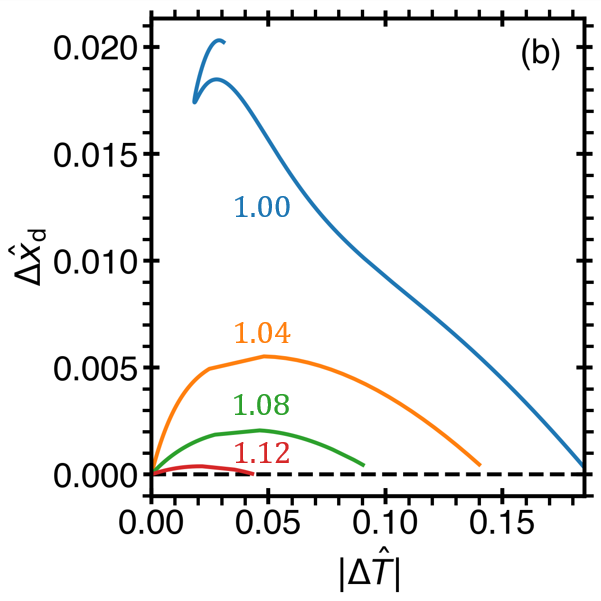}
    \caption{Liquid-liquid diameters of (a) the density, given by Eq.~(15) in the main text, and (b) the concentration, given by Eq.~(16) in the main text, as a function of the distance to the virtual liquid-liquid critical line (LLCL) along the thermodynamic path selected by interconversion for systems exhibiting liquid polyamorphism with $\omega_{11}=1.6$, $\omega_{22}=2.0$, $e=3$, $s=4$, and with various values of $\omega_{12}$: $\omega_{12}=1.00$ (blue), $\omega_{12}=1.04$ (orange), $\omega_{12}=1.08$ (green), $\omega_{12}=1.12$ (red).}
    \label{Fig_SM_LL_RectiDiam}
\end{figure}

\begin{figure}[th!]
    \centering
    \includegraphics[width=0.49\linewidth]{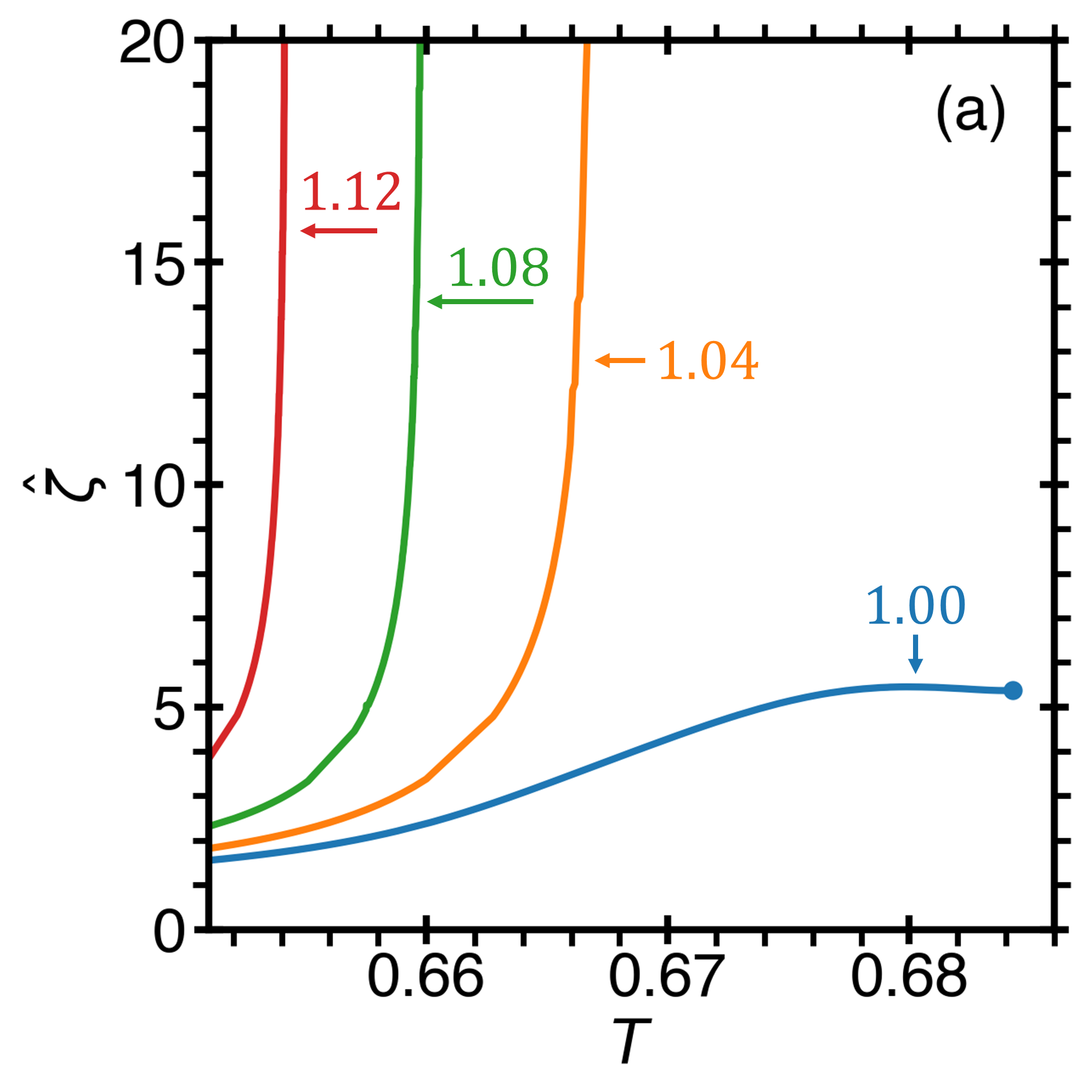}
    \includegraphics[width=0.49\linewidth]{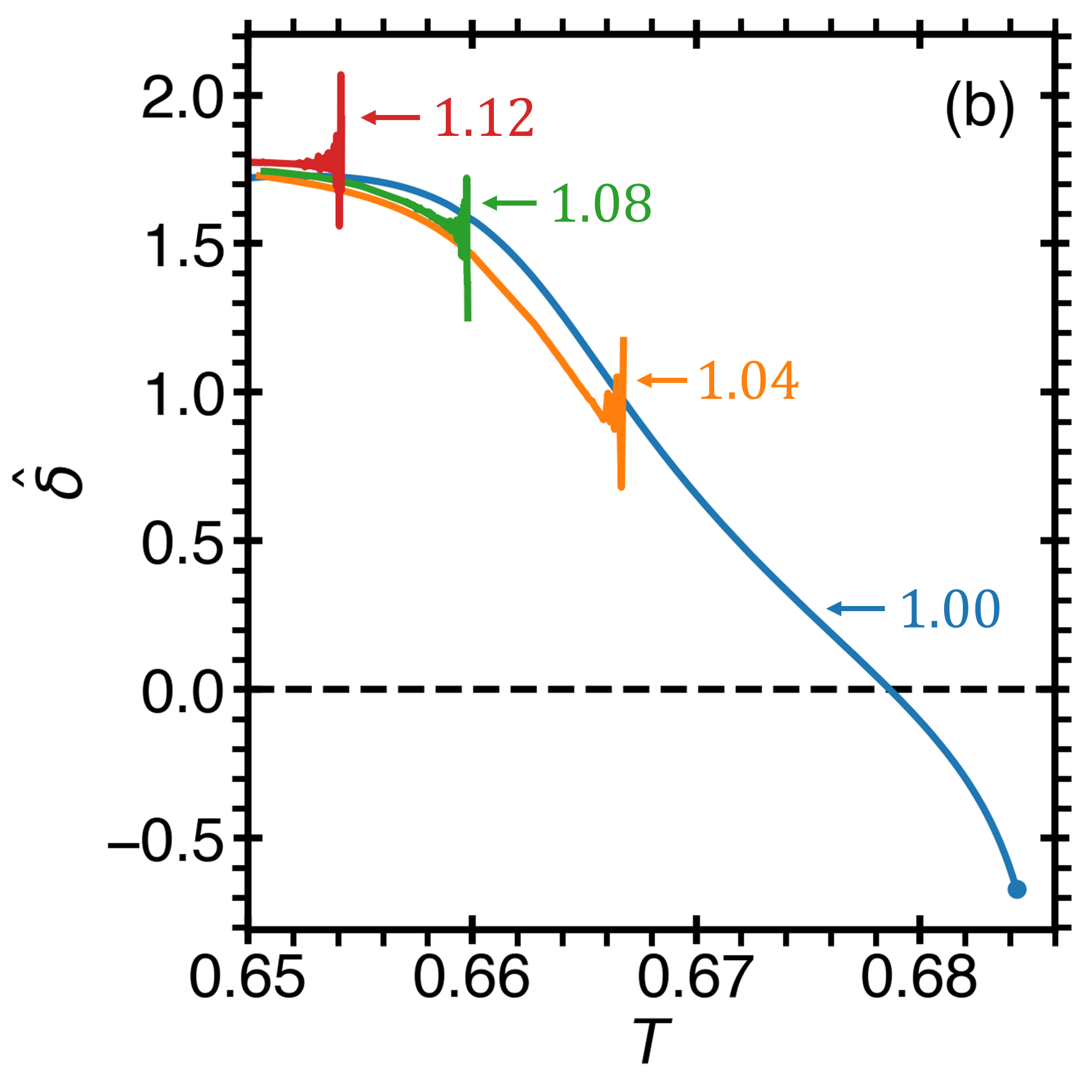}
    \caption{Liquid-liquid interfacial properties of the systems exhibiting liquid polyamorphism with $\omega_{11}=1.6$, $\omega_{22}=2.0$, $e=3$, $s=4$, and with various values of $\omega_{12}$: $\omega_{12}=1.00$ (blue), $\omega_{12}=1.04$ (orange), $\omega_{12}=1.08$ (green), $\omega_{12}=1.12$ (red). (a) the reduced thickness, $\hat{\zeta}=\zeta/\ell$ of the liquid-liquid interface, and (b) the reduced shift, $\hat{\delta}=\delta/\ell$, between the density and concentration liquid-liquid profiles. In (a,b) the thickness and shift reach a finite value (marked with a blue circle) at the triple point temperature.}
    \label{Fig_SM_LL_Props}
\end{figure}

\clearpage
\newpage

\section{Asymptotic Meanfield Behavior}

\begin{figure}[h!]
    \centering
    \includegraphics[width=0.49\linewidth]{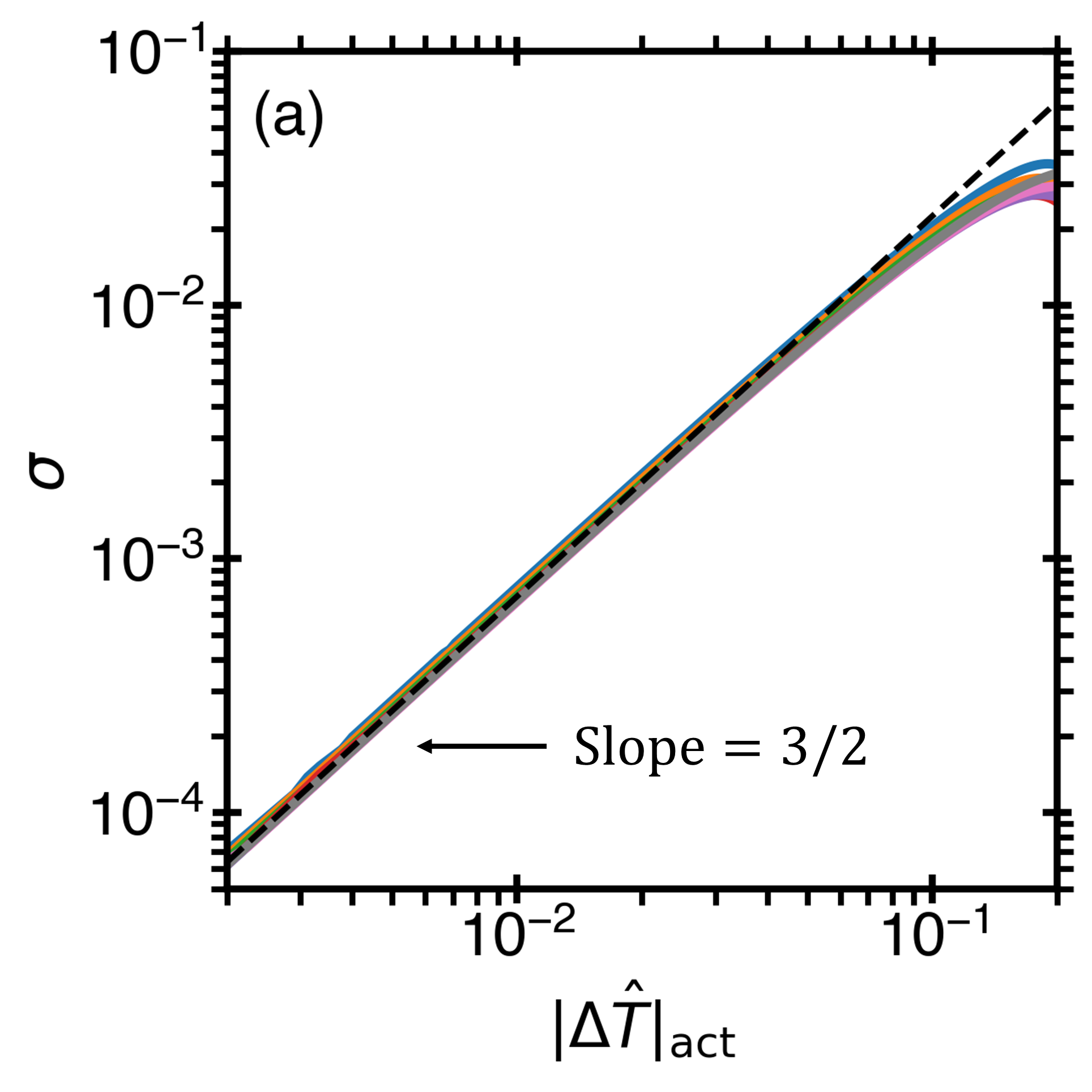}
    \includegraphics[width=0.49\linewidth]{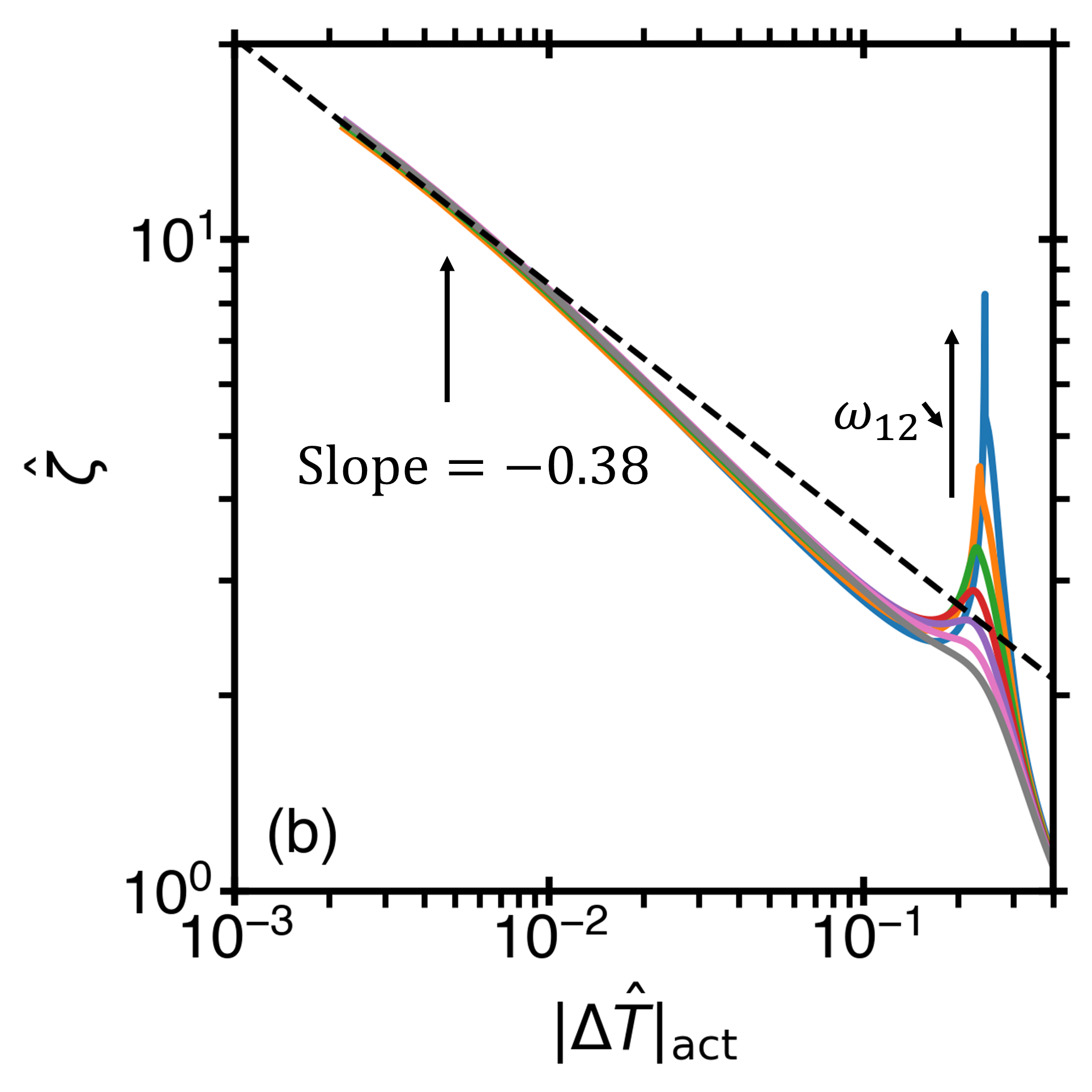}
    \caption{(a) The behavior of the liquid-vapor interfacial tension follows the power law, $\sigma = \sigma_0 |\Delta\hat{T}|^{3/2}$, where the amplitude was found to be $\sigma_0 = 0.71$, asymptotically close to the actual liquid-vapor critical temperature (see Table~\ref{Table_SM_LVCPs}). (b) The behavior of the reduced liquid-vapor interfacial thickness, $\hat{\zeta}=\zeta/\ell$, follows the power law, $\hat{\zeta}=\hat{\zeta}_0|\Delta\hat{T}|^{-0.38}$, where the amplitude was found to be $\hat{\zeta}_0 = 1.50$ asymptotically close to the actual critical point.}
    \label{Fig_SM_LVasympt}
\end{figure}

\begin{figure}[h!]
    \centering
    \includegraphics[width=0.49\linewidth]{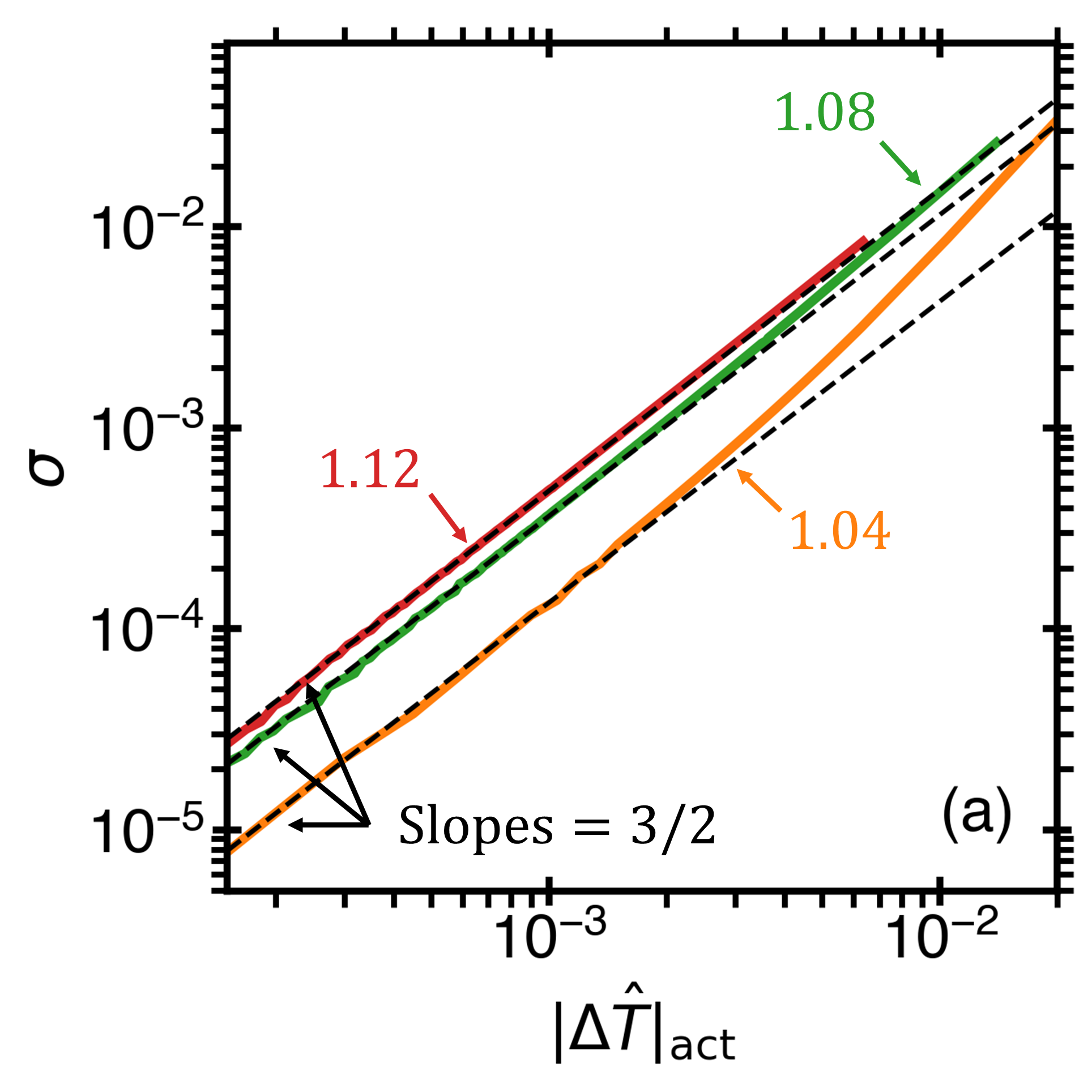}
    \includegraphics[width=0.49\linewidth]{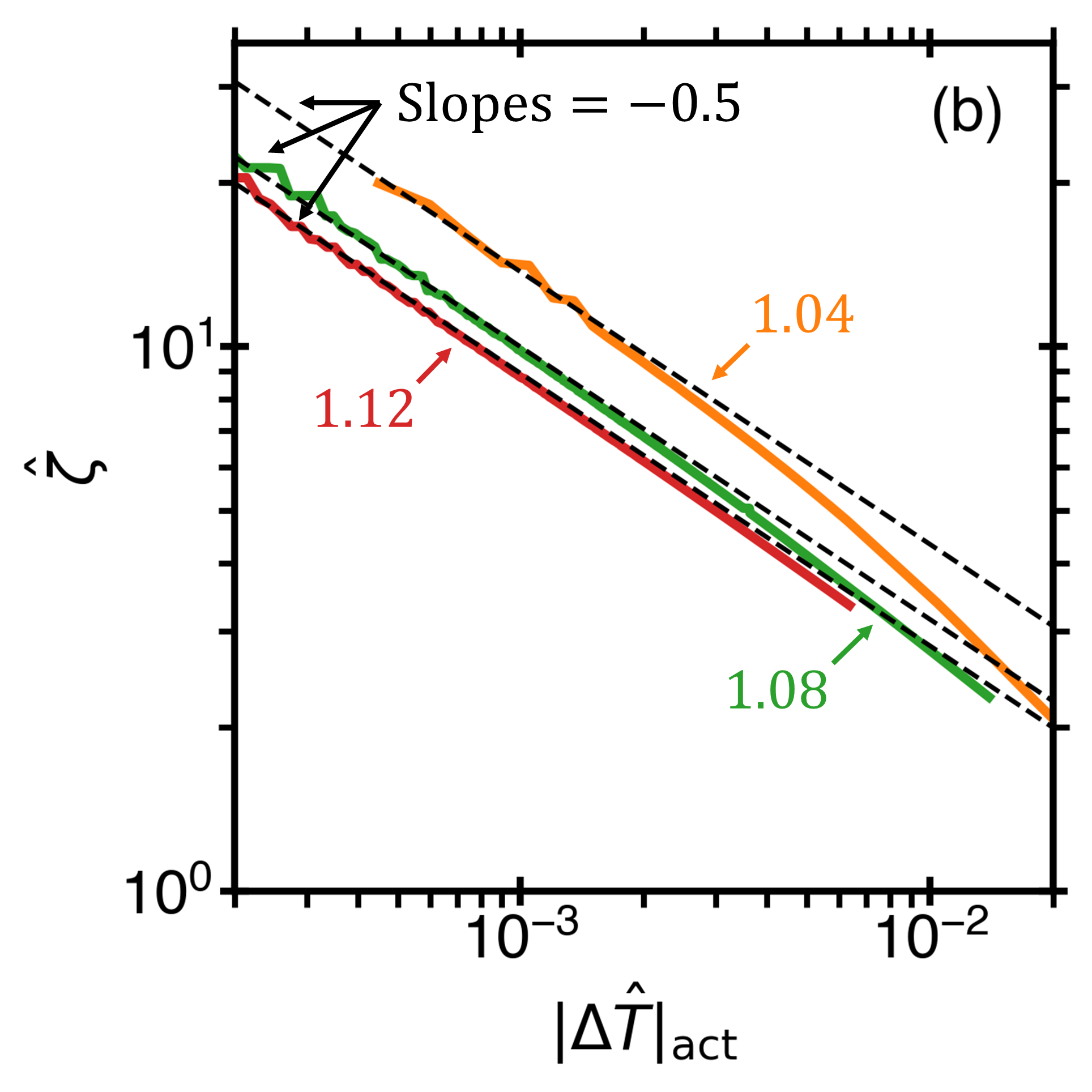}
    \caption{(a) The behavior of the liquid-liquid interfacial tension follows the meanfield power law, $\sigma = \sigma_0 |\Delta\hat{T}|^{3/2}$, (dashed lines) asymptotically close to the actual liquid-liquid critical temperature (see Table~\ref{Table_SM_LLCPs}). (b) The behavior of the reduced liquid-liquid interfacial thickness, $\hat{\zeta}=\zeta/\ell$, follows the meanfield power law, $\hat{\zeta}=\hat{\zeta}_0|\Delta\hat{T}|^{-1/2}$, (dashed lines). In (a,b) the systems exhibiting liquid polyamorphism and a liquid-liquid critical point are shown: $\omega_{12}=1.04$ (orange), $\omega_{12}=1.08$ (green), $\omega_{12}=1.12$ (red), and the amplitudes of the asymptotic meanfield power laws are provided in Table~\ref{Table_SM_Asymp_Amps}.}
    \label{Fig_SM_LLasympt}
\end{figure}

\begin{table}[h!]
\begin{tabular}{lcc}
\toprule
$\omega_{12}$ & $\sigma_0$ & $\hat{\zeta}_0$ \\ \midrule
1.04          & 4.26       & 0.433         \\
1.08          & 11.48      & 0.315         \\
1.12          & 15.35      & 0.282         \\ \bottomrule  
\end{tabular}
\caption{Asymptotic amplitudes of the liquid-liquid interfacial tension and liquid-liquid correlation length of concentration fluctuations for the three systems exhibiting liquid polyamorphism and a liquid-liquid critical point. The asymptotic meanfield behavior is illustrated in Fig.~\ref{Fig_SM_LLasympt}.}
\label{Table_SM_Asymp_Amps}
\end{table}

\begin{figure}[ht!]
    \centering
    \includegraphics[width=0.49\linewidth]{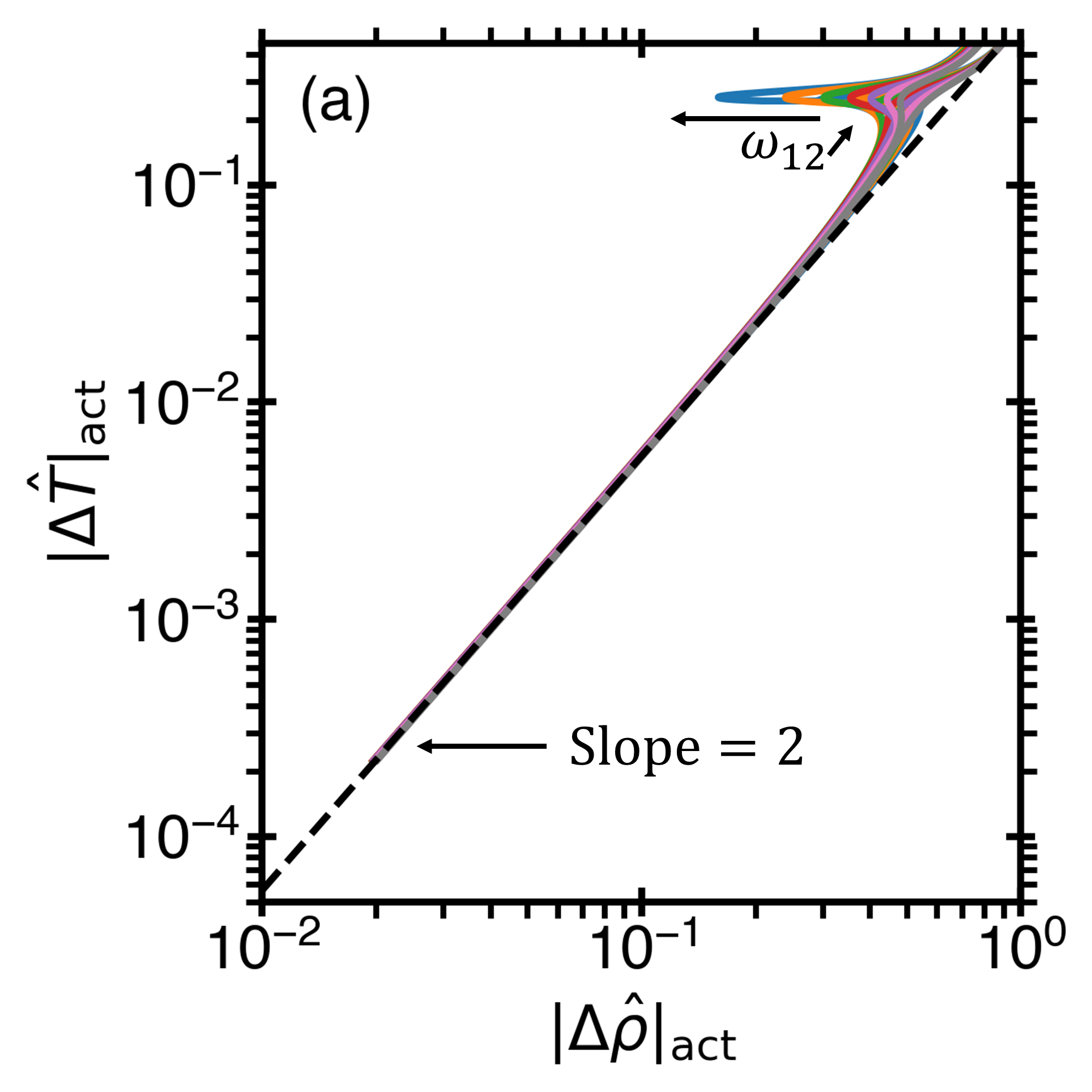}
    \includegraphics[width=0.49\linewidth]{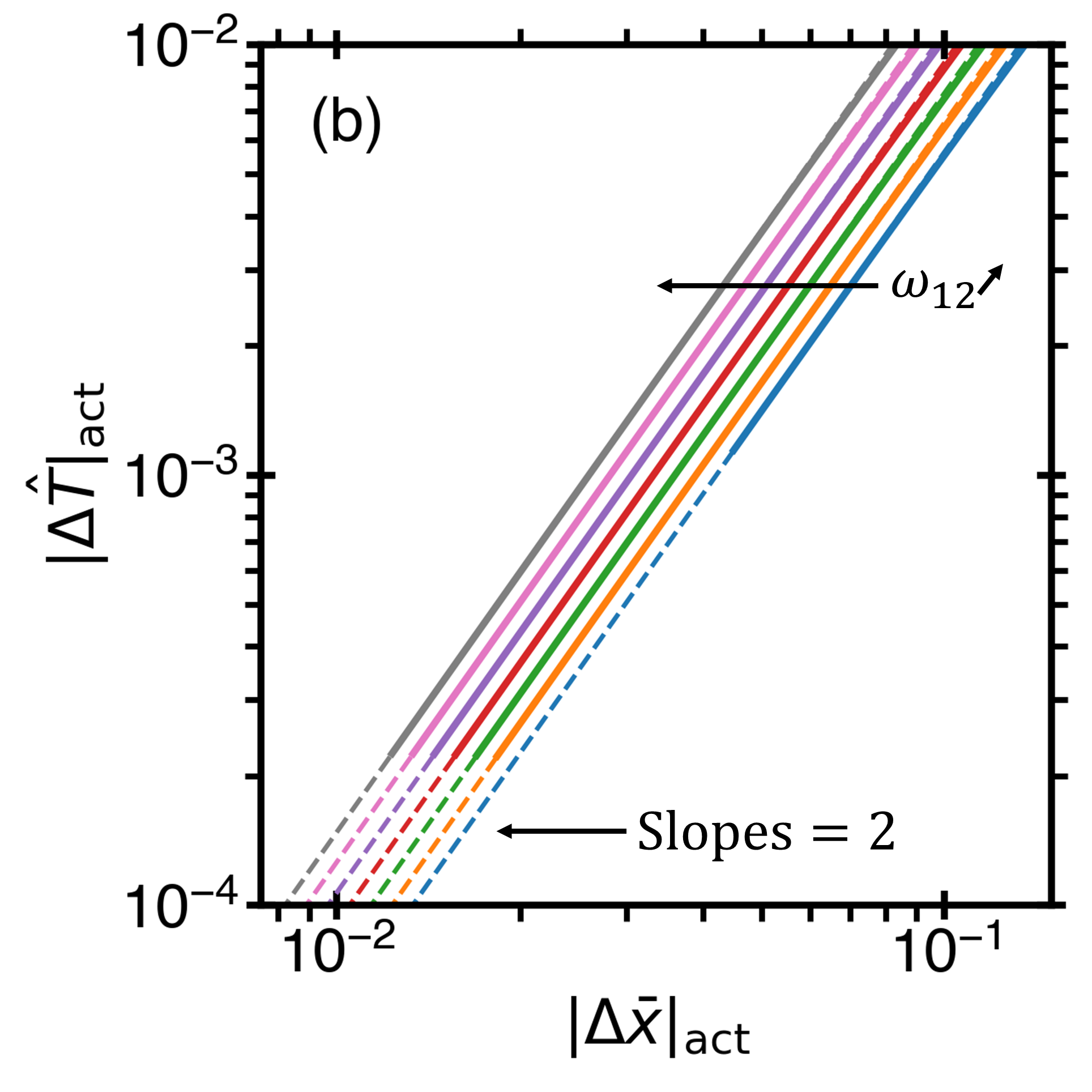}
    \includegraphics[width=0.49\linewidth]{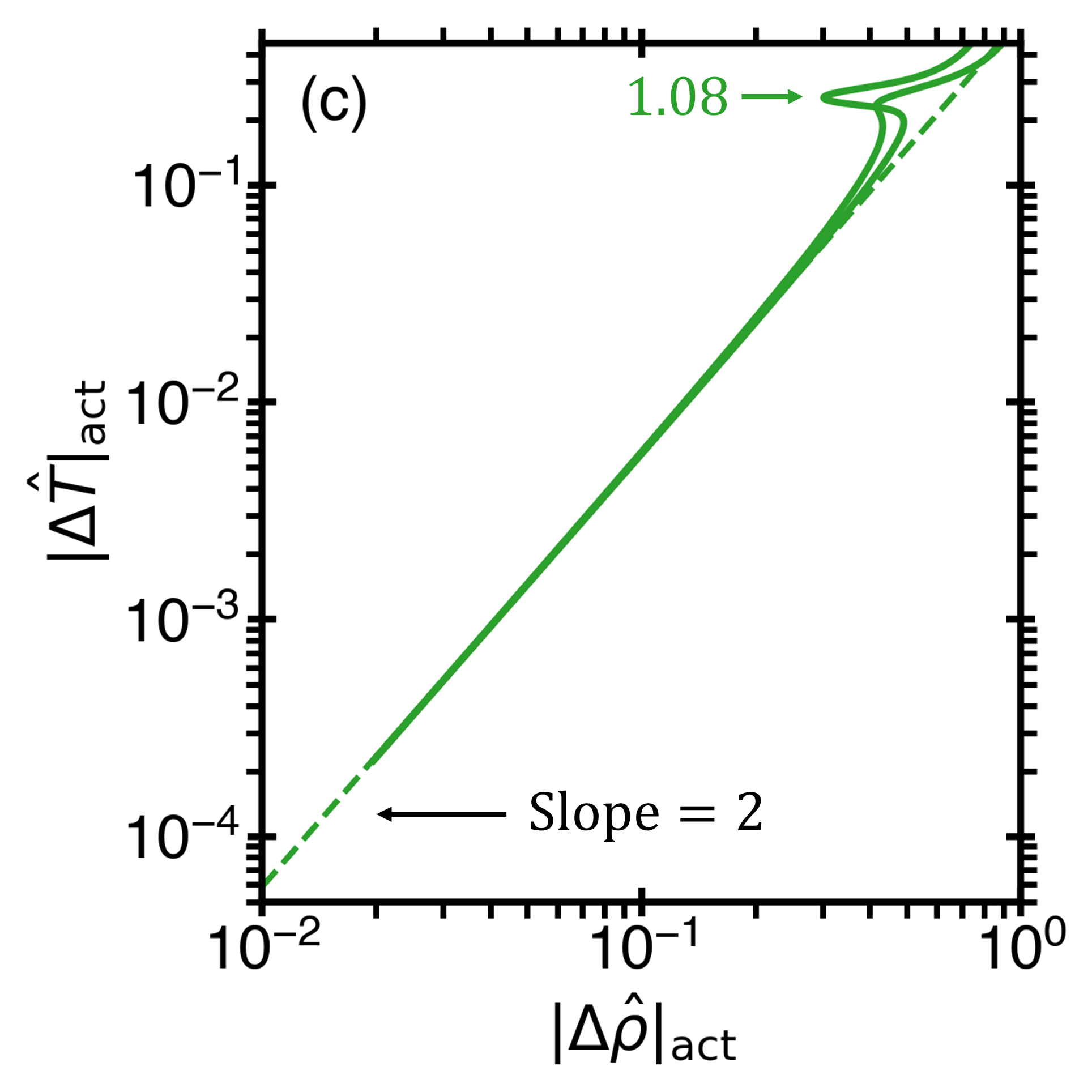}
    \includegraphics[width=0.49\linewidth]{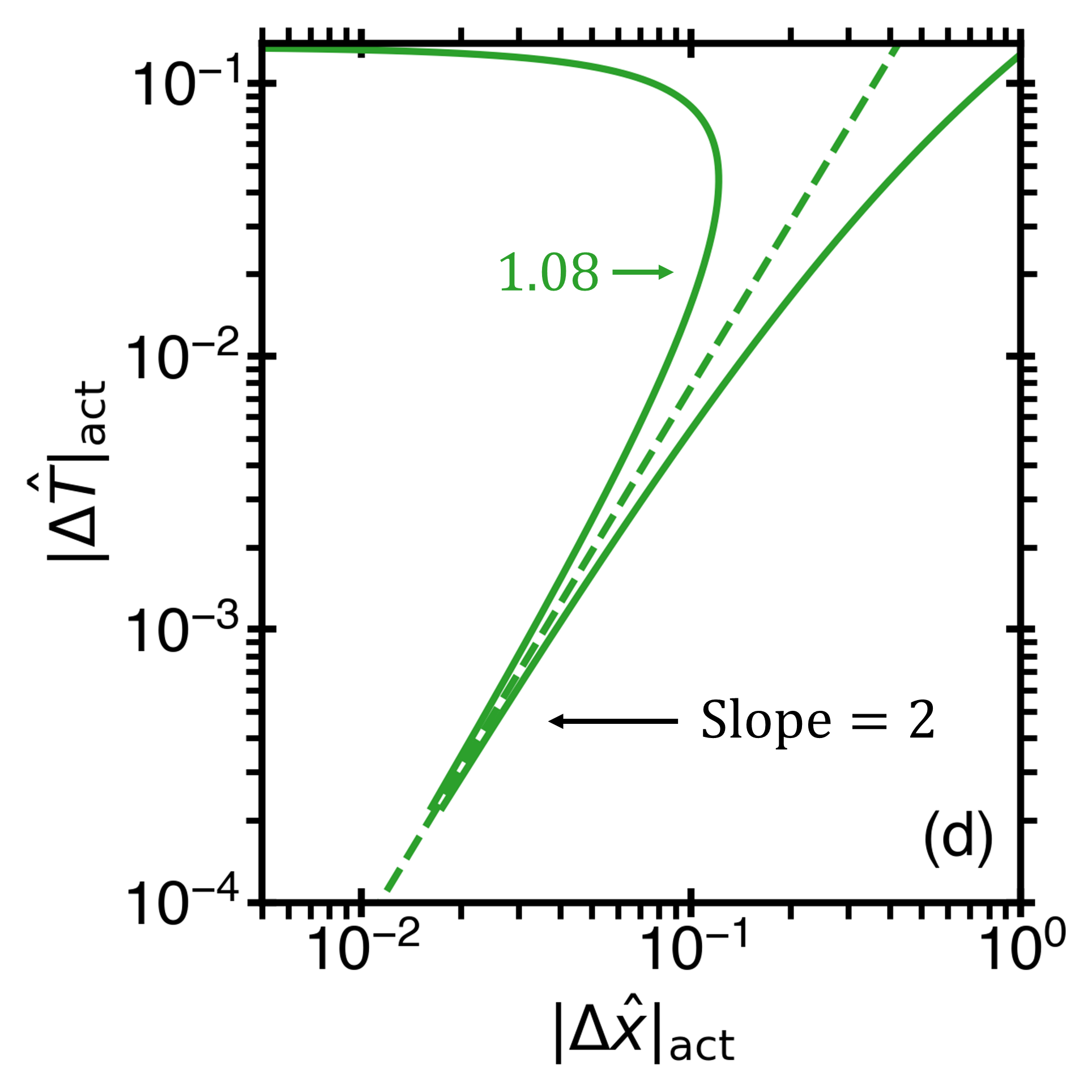}
    \caption{Asymptotic behavior of the liquid vapor coexistence for systems with $\omega_{11}=1.6$, $\omega_{22}=2.0$, $e=3$, $s=4$, and with various values of $\omega_{12}$: $\omega_{12}=1.00$ (blue), $\omega_{12}=1.04$ (orange), $\omega_{12}=1.08$ (green), $\omega_{12}=1.12$ (red), $\omega_{12}=1.16$ (purple), $\omega_{12}=1.20$ (pink), and $\omega_{12}=1.24$ (gray). (a) The temperature-density LV coexistence follows the meanfield power law, $\Delta\hat{T}\sim|\Delta\hat{\rho}|^2$, where $\Delta\hat{T}_\mathrm{act}= 1-T/T_\mathrm{c}^\mathrm{act}$ and $T_\mathrm{c}^\mathrm{act}$ is the actual critical temperature selected by the interconverting path. Likewise, $\Delta\hat{\rho}_\mathrm{act}=1-\rho/\rho_\mathrm{c}^\mathrm{act}$, where $\rho_\mathrm{c}^\mathrm{act}$ is the actual critical density. (b) The temperature-average concentration LV coexistence follows the meanfield power law, $\Delta\hat{T}\sim|\Delta\bar{x}|^2$, where $\Delta\bar{x}_\mathrm{act}=1-\Bar{x}/x_\mathrm{c}^\mathrm{act}$, in which $\Bar{x}=(x_\mathrm{L}+x_\mathrm{V})/2$ and $x_\mathrm{c}^\mathrm{act}$ is the actual critical concentration. (c,d) Illustrate, as an example, the asymptotic behavior of the system with $\omega_{12}=1.08$, in which (d) shows the asymptotic behavior of the each branch of the concentration coexistence. In (a,b) the black arrow indicates the direction of increasing $\omega_{12}$.}
    \label{Fig_Asympt_Coex}
\end{figure}

\clearpage
\newpage

\section{Interfacial Tension of a Sharp Interface}
As indicated by the liquid-vapor coexistence curves presented in Fig.~\ref{Fig_SM_LVcoex}(b), in the low temperature limit ($T\to 0$), the equilibrium fraction of species asymptotically goes to zero as $x_\text{e}\to 0$. Via this feature, we may estimate the liquid-vapor surface tension of a sharp interface. In this limit, the contribution to the free energy from the liquid-vapor interface is given through Eq.~(\ref{Eq_SM_sigmaXeq1}), as $(1/4)\ell^2\omega_{11}|\nabla\rho|^2$, which upon integration over the volume of space gives, $\sigma_\mathrm{LV,shp} \approx \omega_{11}/8 = 0.2$.

Alternatively, from the liquid-liquid coexistence curves presented in Fig.~2(d) in the main text, in the low temperature limit, the density asymptotically goes to $\rho = 1$. We estimate the liquid-liquid interfacial tension for a sharp interface assuming $T\to 0$. In this case, the contribution to the free energy from the liquid-liquid interfacial tension, $\sigma_\mathrm{LL}$, is given through Eq.~(\ref{Eq_SM_sigmaRhoeq1}), as $(1/4)\ell^2\omega|\nabla\rho|^2$, which when integrated over space, gives $\sigma_\mathrm{LL,shp} \approx \omega/8$. As discussed in the main text, we estimate that a sharp interface has surface tension that varies from $\sigma_\text{shp} \approx 0.2$ for the system with $\omega_{12}=1.00$ to $\sigma_\text{shp} \approx 0.17$ for the system with $\omega_{12} = 1.12$.

\section{Correlation Between the Difference in Diameters and the Shift in the Profiles}

\begin{figure}[t!]
    \centering
    \includegraphics[width=0.6\linewidth]{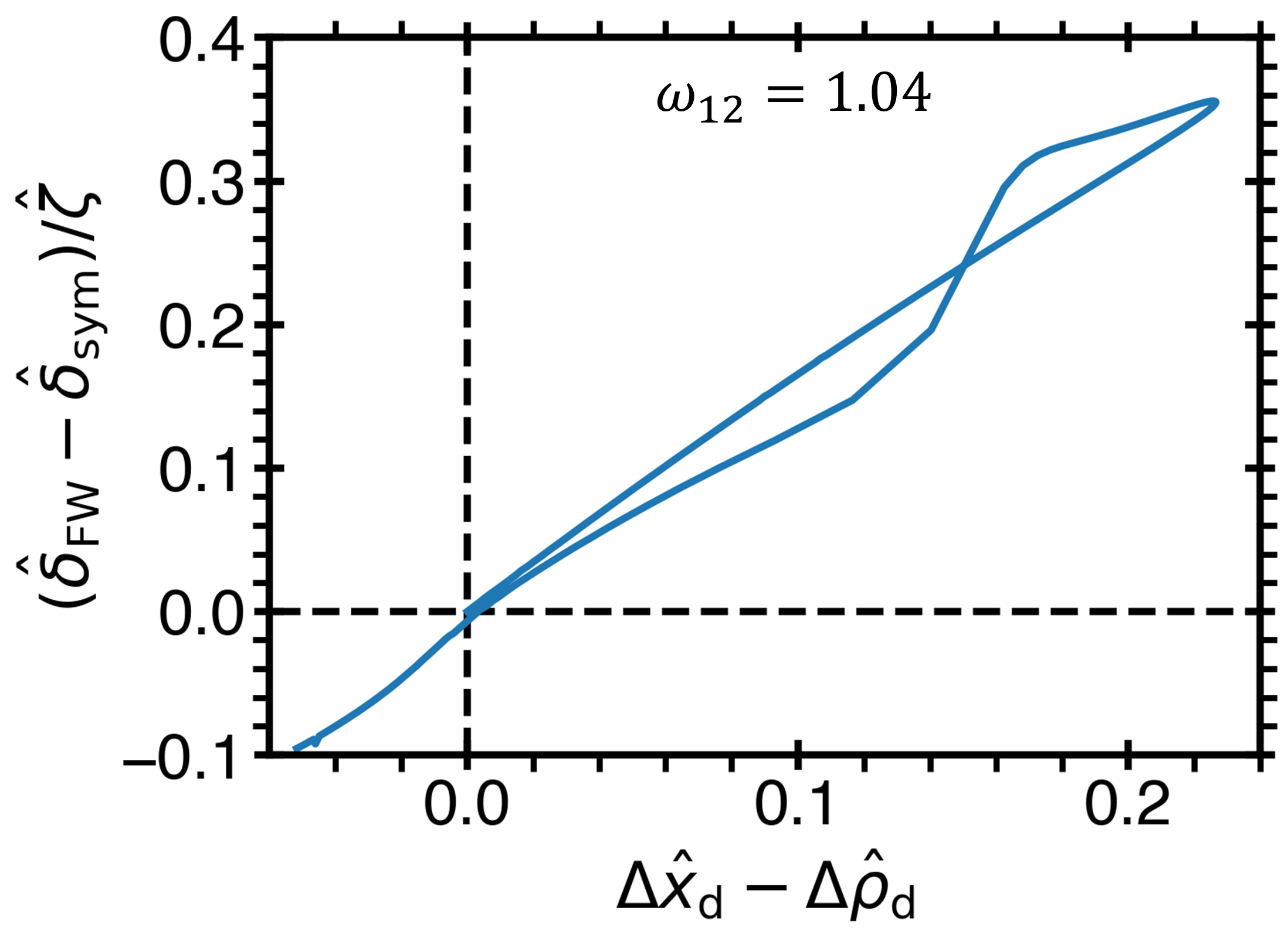}
    \caption{The asymmetric contribution to the shift between the concentration and density profiles for the system with $\omega_{12}=1.04$, $\omega_{11}=1.6$, $\omega_{22}=2.0$, $e=3$ and $s=4$. The asymmetric contribution, $\hat{\delta}_\mathrm{asym} = \hat{\delta}_\mathrm{FW}-\hat{\delta}_\mathrm{sym}$, where $\hat{\delta}_\mathrm{FW}$ is obtained for the Fisher-Wortis ansatzes, while $\hat{\delta}_\mathrm{sym}$ is obtained for the two parameter symmetric ansatzes of Eqs.~(\ref{SM_Eq_rhoAsym}) and (\ref{SM_Eq_xAsym}), is proportional to the difference in the diameters between the concentration, $\Delta\hat{x}_\mathrm{d}$, and the density, $\Delta\hat{\rho}_\mathrm{d}$.}
    \label{SM_Fig_Delta_Understand}
\end{figure}

We observed that the relative shift between the concentration and density profiles, $\hat{\delta}$, may be separated into symmetric, $\hat{\delta}_\mathrm{sym}$, and asymmetric contributions, $\hat{\delta}_\mathrm{asym}$. The asymmetric contribution to the shift may be determined from the Taylor series expansion to first order of the Fisher-Wortis ansatzes, Eqs.~(8) and 9), in the main text, as
\begin{align}
    \hat{\rho}(\hat{z}) &\approx \frac{1}{2}\left(\frac{\hat{z}}{\hat{\zeta}}+1\right) - \Delta\hat{\rho}_\mathrm{d} (\rho_\mathrm{L}-\rho_\mathrm{V})\\
    \hat{x}(\hat{z}) &\approx \frac{1}{2}\left(\frac{\hat{z}+\hat{\delta}}{\hat{\zeta}}+1\right) -\Delta\hat{x}_\mathrm{d}(x_\mathrm{L}-x_\mathrm{V})
\end{align}
Subtracting the concentration from the density profile, at the position $\hat{z}=0$, gives a relationship for the asymmetric contribution to the shift, $\hat{\delta}_\mathrm{asym}$,
\begin{equation}
    \hat{\rho}(\hat{z}=0) - \hat{x}(\hat{z}=0) \approx \Delta\hat{\rho}_\mathrm{d}(\rho_\mathrm{L}-\rho_\mathrm{V}) - \Delta\hat{x}_\mathrm{d}(x_\mathrm{L}-x_\mathrm{V}) + \frac{\hat{\delta}_\mathrm{asym}}{2\hat{\zeta}} = 0
\end{equation}
Therefore, by calculating liquid-vapor interfacial tension for a two-parameter symmetric ansatz of the form
\begin{align}
    \hat{\rho}(\hat{z}) &= \frac{\rho(\hat{z})-\rho_\mathrm{V}}{\rho_\mathrm{L}-\rho_\mathrm{V}} = \frac{1}{2}\left[\tanh\left(\frac{\hat{z}}{\zeta}\right)-1\right]\label{SM_Eq_rhoAsym}\\
    \hat{x}(\hat{z}) &= \frac{x(\hat{z})-x_\mathrm{V}}{x_\mathrm{L}-x_\mathrm{V}} = \frac{1}{2}\left[\tanh\left(\frac{\hat{z}+\hat{\delta}}{\zeta}\right)-1\right]\label{SM_Eq_xAsym}
\end{align}
and subtracting the shift from that of predicted by the Fisher-Wortis ansatzes, we isolate the asymmetric contribution and verify that the asymmetric contribution is proportional to the difference in diameters of the concentration and density. This calculation was performed for the system with $\omega_{12}=1.04$, and is presented in Fig.~\ref{SM_Fig_Delta_Understand}. We note that the asymmetric contribution to the shift, $\hat{\delta}_\mathrm{asym}$, goes to zero at the critical point, while we found that the symmetric contribution, as presented in Fig.~4 of the main text, reaches a finite value at the critical point.

\section{Surface Enrichment Near the Minimum of Liquid-Vapor Interfacial Tension}

\begin{figure}[h!]
	\centering
	\includegraphics[width=0.49\linewidth]{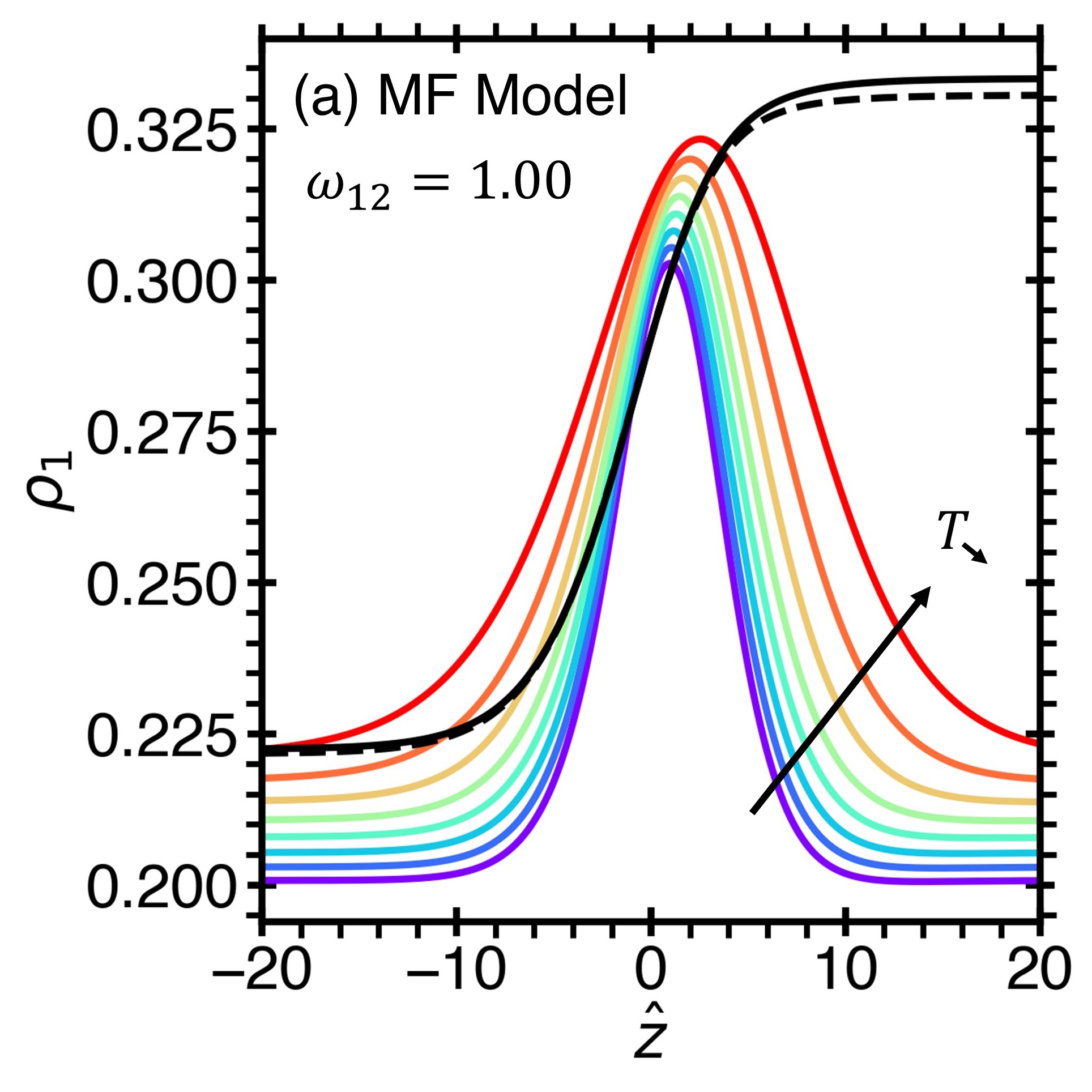}
	\includegraphics[width=0.49\linewidth]{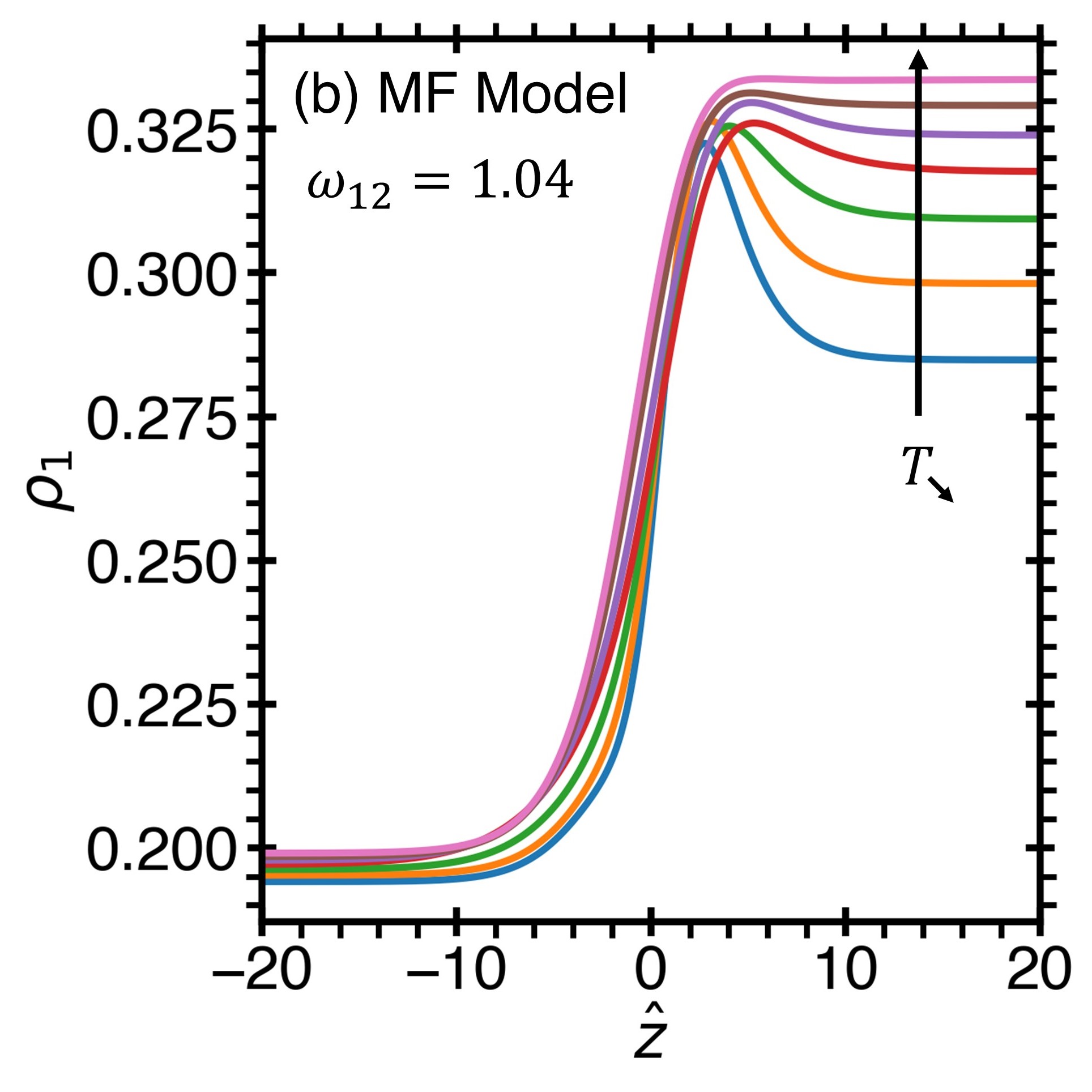}
	\caption{Interfacial profiles of species 1, $\rho_1=\rho x$, in the meanfield blinking-checkers model (MF Model) demonstrate surface enrichment near the TP temperature, $T_\mathrm{TP}=0.68429$. (a) Surface enrichment of species 1 for the system with $\omega_{12}=1.00$. The colored curves indicate temperatures from $T=0.68989$ to $T=T_\text{TP}$ in steps of $\Delta\hat{T}=-0.0008$ in order of purple to red. The black curves are $T=0.68389$ (dashed) and $T=0.68309$ (solid). (b) Surface enrichment of species 1 for the system with $\omega_{12}=1.04$. The curves are $T=0.6882$ to $T=0.6826$ in steps of $\Delta T =-0.0008$ (blue to pink). In (a,b), the black arrows indicate the direction of decreasing temperature. Note that while the transition of a surface enriched profile ($T>T_\text{TP}$) to a smooth profile ($T<T_\text{TP}$) is discontinuous in the system with $\omega_{12}=1.00$, it is continuous in the system with $\omega_{12}=1.04$.}
	\label{SM_Fig_MF_Enrich}
\end{figure}

\begin{figure}[h!]
	\centering
	\includegraphics[width=0.49\linewidth]{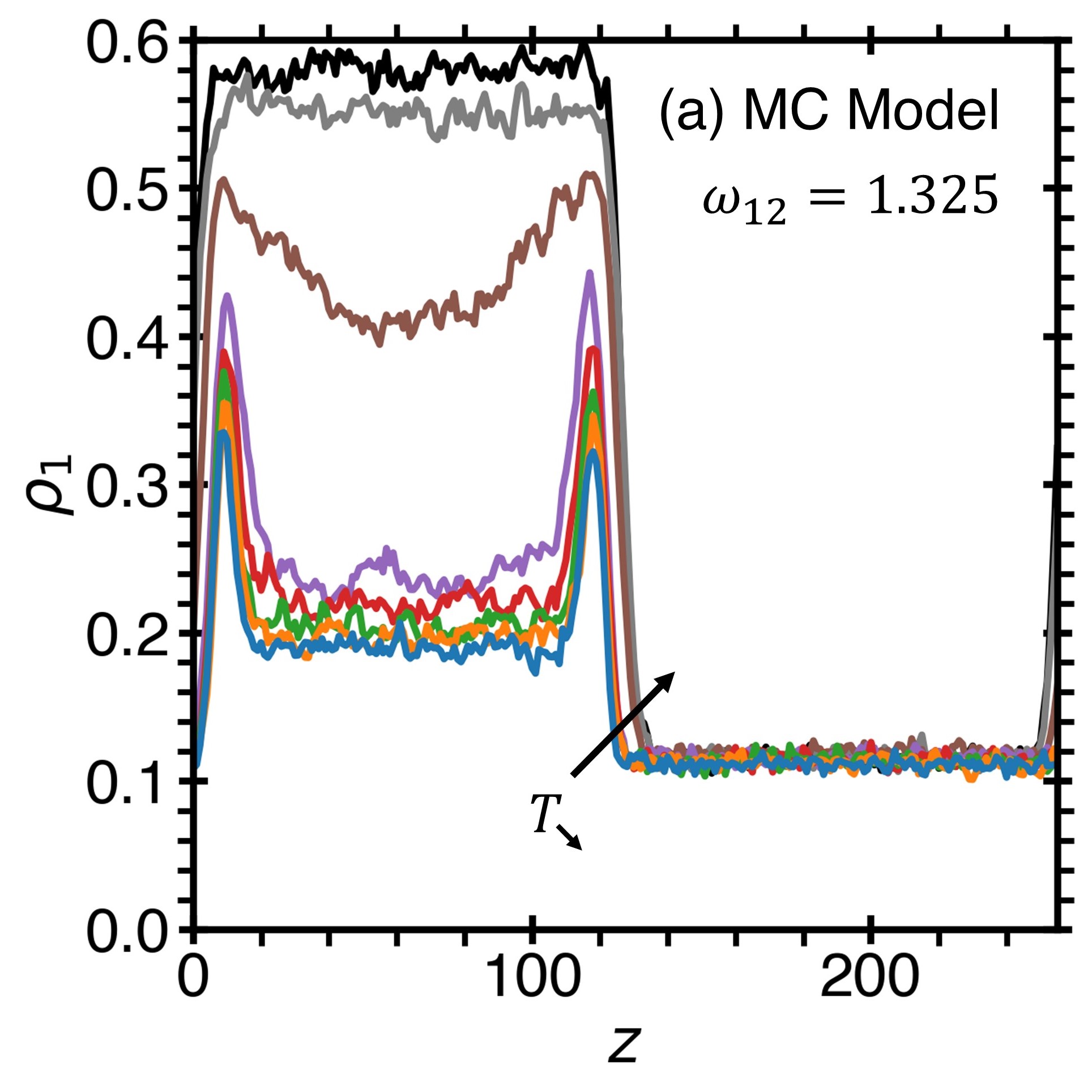}
	\includegraphics[width=0.49\linewidth]{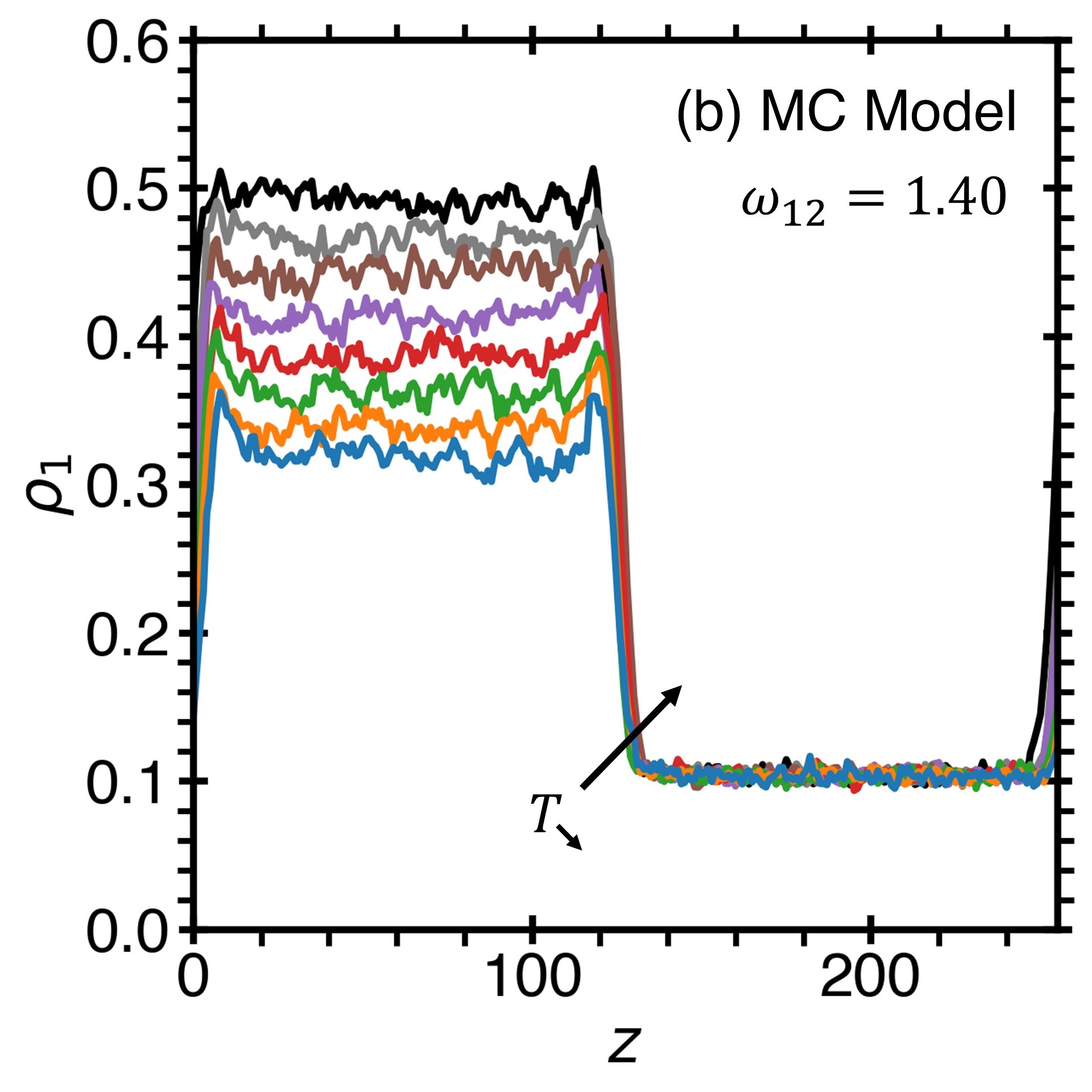}
	\caption{Interfacial profiles of species 1, $\rho_1=\rho x$, in Monte Carlo simulations of the blinking-checkers model (MC Model), for systems with $e=3$, $s=4$, $\omega_{11}=2.0$, and $\omega_{22}=2.5$, demonstrate surface enrichment near the TP temperature, $T_\mathrm{TP}=0.6345$. Surface enrichment of species 1 for (a) a system with a triple point ($\omega_{12}=1.325$) and (b) a system without a triple point ($\omega_{12}=1.40$), see Sec.~\ref{SM_Sec_MC_Results} for details. In (a,b) the curves indicate temperatures from $T=0.640$ (blue) to $T=0.633$ (black) in steps of $\Delta\hat{T}=-0.001$. The black arrows indicate the direction of decreasing temperature. The predictions of the meanfield model (Fig.~\ref{SM_Fig_MF_Enrich}) are verified by the profiles obtained from the MC approach.}
	\label{SM_Fig_MC_Enrich}
\end{figure}

\bibliography{SM_ref}